\documentclass[twocolumn]{aastex62}

\usepackage{url}
\usepackage{amsmath}
\usepackage{amssymb}
\usepackage{natbib}
\usepackage{multirow}
\definecolor{orange}{RGB}{255,127,0}

\newcommand{\project}[1]{\textsf{#1}}
\newcommand{\celerite}{\project{celerite }}

\newcommand{\bvec}[1]{{\ensuremath{\boldsymbol{#1}}}}
\newcommand{\T}{\ensuremath{\mathrm{T}}}
\newcommand{\expandvec}[2]{\left(\begin{array}{ccccc} #1\quad && \cdots\quad && #2 \end{array}\right)}
\newcommand{\mvec}[1]{\ensuremath{\mathrm{vec}({#1})}}

\graphicspath{}

\received{}
\revised{}
\accepted{}
\submitjournal{}

\shorttitle{Celerite2D}
\shortauthors{Gordon et al.}

\begin{document}

\title{A Fast, 2D Gaussian Process Method Based on Celerite: Applications to Transiting Exoplanet Discovery and Characterization}

\correspondingauthor{Tyler A. Gordon}
\email{tagordon@uw.edu}

\author[0000-0001-5253-1987]{Tyler A. Gordon}
\affiliation{Department of Astronomy, University of Washington, Box 351580, U.W., Seattle, WA 98195-1580, USA}

\author[0000-0002-0802-9145]{Eric Agol}
\affiliation{Department of Astronomy, University of Washington, Box 351580, U.W., Seattle, WA 98195-1580, USA}

\author[0000-0002-9328-5652]{Daniel Foreman-Mackey}
\affiliation{Center for Computational Astrophysics, Flatiron Institute, 162 5th Ave, New York, NY 10010}



\begin{abstract}
    Gaussian processes (GPs) are commonly used as a model of stochastic variability in astrophysical time series. In particular, GPs are frequently employed to account for correlated stellar variability in planetary transit light curves. The efficient application of GPs to light curves containing thousands to tens of thousands of datapoints has been made possible by recent advances in GP methods, including the \celerite method. Here we present an extension of the \celerite method to two input dimensions, where, typically, the second dimension is small.  This method scales linearly with the total number of datapoints when the noise in each large dimension is proportional to the same \celerite kernel and only the amplitude of the correlated noise varies in the second dimension. We demonstrate the application of this method to the problem of measuring precise transit parameters from multiwavelength light curves and show that it has the potential to improve transit parameters measurements by orders of magnitude. Applications of this method include transit spectroscopy and exomoon detection, as well  a broader set of astronomical problems.
\end{abstract}

\section{Introduction}
    All exoplanet transit observations must contend with the presence 
    of noise. Light curves can display both uncorrelated, or white, noise 
    and correlated noise. While white noise often results from the 
    the statistics of photon counting, and may only be 
    ameliorated by collecting more photons, correlated noise 
    can arise from a variety of sources. These can be broadly 
    divided into two categories: astrophysical noise, 
    which results from physical processes at the source of the 
    observed photons such as stellar granulation and 
    oscillations \citep{Pereira2019,Barros2020,Sulis2020,Morris2020}, and instrumental noise, which results from 
    imperfections in detectors, errors in spacecraft pointing, 
    or other processes taking place at the location of the observer
    rather than at the source. 
    
    Our ability to detect transits and infer their parameters 
    depends on how well we can model both white and correlated 
    noise. While white noise is straightforward to model as a Gaussian distributed random variable,\footnote{Which is the limit of a Poisson distribution at high photon count rates.} 
    correlated noise can be more challenging to account for. Additionally, as more powerful telescopes 
    yield more precise observations, photon-counting noise 
    will decrease while astrophysical correlated noise 
    (which does not depend on photon counts)
    will not. In fact, correlated noise will become 
    more dominant as decreasing white noise 
    amplitudes reveal previously undetectable 
    variability. 
    
    A number of methods have been used to model, filter, or otherwise account for correlated 
    noise in astrophysics, dating back to work by \citet{Rybicki1992} and \citet{Rybicki1995}.  Among these techniques are wavelet filtering \citep{Carter2008} 
    and Kalman filtering \citep{Kelly2014}. A comprehensive study of various detrending methods is 
    given in \cite{Hippke2019}. These include various sliding filter 
    methods (such as a sliding mean or median), sums of sines or cosines 
    \citep{Kipping2013, Mazeh2010}, and others.
    
    Our work focuses on the Gaussian process method of 
    modeling correlated noise. In this paper we introduce 
    an extension to the popular \celerite code which 
    can be used to model correlated noise in two dimensions. 
    We use this extension to simulate multiwavelength stellar 
    variability in transit observations. We show that by 
    accurately modeling correlation across wavelengths 
    we can improve measurements of transit parameters by 
    orders of magnitude in some common limits. 
    
    While this paper focuses on multiwavelength transit 
    observations with a small number of bands, our method 
    also naturally extends to transit spectroscopy as 
    the number of bands becomes large. In this paper 
    we consider a trapezoidal transit model that has 
    no wavelength dependence, but a wavelength-dependent 
    transit model can easily be incorporated. For transit spectroscopy, the transit
    depth and limb-darkening parameters should be allowed to vary between bands.
    
    \subsection{A short introduction to Gaussian Processes}
    
    While more general definitions of Gaussian processes may be 
    formulated, it is most helpful for our purposes to view Gaussian processes as 
    an ordered collection of random variables along one or more axes often 
    representing time or space. 
    In the case of an exoplanet transit the random variables 
    model a series of observations 
    of the star's flux taken at discrete times. 
    The Gaussian aspect of a 
    Gaussian process describes the relationship between random 
    variables --- we model $N'$ observations with an $N'$-dimensional 
    Gaussian distribution. The covariance of the multi-dimensional 
    Gaussian is described by a kernel function, which gives the 
    covariance between any pair of observations as a function of 
    their separation in time or space. The kernel function then defines the 
    covariance matrix. For a kernel $k(x_i, x_j)$, we have
    \begin{equation}
        K_{i, j} = k(x_i, x_j),
    \end{equation}
    which includes both correlated and white noise components.
    In addition to the kernel function, a GP is characterized by its 
    mean function, $\mu(t)$, which describes the deterministic 
    component of the process. In the case of 
    an exoplanet transit we use a transit model as the mean function. 
    The GP likelihood function, $\mathcal{L}$, describes the likelihood that 
    a set of observations, $\bvec{y}$, is drawn from the GP. It 
    is written as
    \begin{equation}
        \label{eqn:simple_logL}
		\ln\ \mathcal{L} = 
		-\frac{1}{2}(\bvec{y}-\bvec{\mu})^\T K^{-1}(\bvec{y}-\bvec{\mu})  
		-\frac{1}{2}\ln\ \mathrm{det}(K) - \frac{N'}{2}\ln(2\pi)
    \end{equation}
    where $\bvec{\mu}$ is a vector where the entries are given by $\bvec{\mu}_i = \mu(x_i)$.
    
    A typical procedure for measuring exoplanet transit parameters 
    using a Gaussian process noise model (as applied in \citealt{Dawson2014}, \citealt{Barclay2015}, and \citealt{Chakrabarty2019} among 
    others) can be 
    summarized as follows: 
    \begin{enumerate}
        \item Choose a suitable kernel function to describe the 
        correlated noise. 
        \item Choose a transit model to use as the GP mean function. 
        \item Maximize the GP likelihood with respect to the 
        parameters of both the transit model and the kernel function. 
        \item Optionally, use a Monte Carlo method to sample 
        the posterior (defined by the GP likelihood and priors for 
        each parameter) in order to estimate uncertainties for the 
        transit and kernel parameters. 
    \end{enumerate}
    When searching for a previously undetected transit, the results of 
    step 3 will suggest the most likely parameters of the transit. 
    In a Bayesian framework the GP likelihood can then be used to 
    estimate the evidence for a transit with respect to a flat 
    mean. 
    
    In the case of a monochromatic light curve 
    this procedure is effective at 
    identifying transits when the depth or duration of the 
    transit differs sufficiently from the amplitude and 
    characteristic timescales of the noise. For instance, 
    a transit that is much deeper than the noise amplitude 
    is poorly described by the GP noise model and thus the 
    likelihood will be sharply peaked at the location of the 
    correct transit parameters. Similarly, a 
    transit that occurs on a much shorter timescale 
    than the characteristic timescale of the variability 
    will be poorly described by the GP and hence easily 
    detectable via the likelihood. Figures \ref{fig:regimes} and \ref{fig:psd_regimes} 
    illustrate these instances.
    
    A problem occurs when the transit depth and duration are 
    comparable to the noise amplitude and timescale. In this 
    case the GP covariance alone is able to fit the transit 
    without the need for a mean model. The result is that the 
    GP likelihood is not sharply peaked about the location of 
    the correct transit parameters and the transit is thus difficult 
    or impossible to detect. Gathering more photons 
    with a larger telescope does 
    not fix the problem as the correlated noise 
    does not decrease with higher photon count rates as white noise does. 
    One simply obtains a better measurement of the 
    correlated noise, but the transit remains masked 
    by the variability.
    
    One solution to this problem is to gather light in multiple wavebands. With a multiband light curve we can 
    leverage the difference in the spectral dependence of 
    the transit as compared to the correlated variability to disentangle the transit from the noise, and thus detect 
    shallower transits across a broader range in duration 
    than is possible with monochromatic observations.  This approach 
    depends upon the assumption that the correlated noise has the same time dependence for each component of the power spectrum, but varies in amplitude with wavelength.  If, on the other hand, the correlated noise is achromatic, multiple wave bands will not improve upon the monochromatic case.  For the remainder of this paper we will assume that there is in fact a wavelength dependence to the correlated noise which shares a common time dependence, and exploit this dependence to demonstrate an improvement on the inference of transiting planet parameters.  We also ignore instrumental/systematic variations, and assume that the white noise is dominated by Poisson photon counting uncertainty.
    
    In the next section we describe our wavelength-dependent stellar 
    variability model (\S \ref{sec:noise_model}). 
    We then review the one-dimensional version 
    of \celerite before describing our extension to 
    two dimensions (\S \ref{sec:2dcelerite}). Next, we conduct 
    an Information analysis to derive approximate, 
    semi-analytic upper bounds on the precision that can be achieved 
    when inferring transit parameters from multiband light curves 
    with different noise properties, and compare the results of our Information analysis to 
    a full MCMC treatment for select noise parameters (\S \ref{sec:results}). In the discussion 
    (\S \ref{sec:discussion}) we outline additional 
    applications of our method including exomoon detection and 
    transmission spectroscopy. We conclude with a discussion 
    of the limitations of and potential improvements to our 
    method (\S \ref{sec:conclusions}).
    
\section{multiwavelength noise model}\label{sec:noise_model}

Here we describe our model for noise that is correlated across
both time and wavelength.  We start with a description of the
time-dependence of the noise.

    \subsection{Time-correlated variability model}\label{sec:monochromatic_model}
        \cite{Foreman-Mackey2017} describe how 
        \celerite can be used as a physically-motivated 
        model for stellar variability. The following discussion 
        is closely based on the discussion in that paper. 
        
        We follow \cite{Anderson1990} in modeling stellar 
        oscillations as the result of stochastic excitations 
        that are damped by convection and turbulent viscosity 
        in the star. This process is described by the differential 
        equation
        \begin{equation}
            \label{eqn:ornstein}
            \frac{1}{\omega_0^2}\frac{\mathrm{d}^2}{\mathrm{d}t^2}y(t) + \frac{1}{\omega_0Q}\frac{\mathrm{d}}{\mathrm{d}t}y(t) + y(t) = \epsilon(t)
        \end{equation}
        where $\omega_0$ is the characteristic frequency of the 
        oscillator, $Q$ is the quality factor of the oscillator, 
        and $\epsilon(t)$ is a stochastic driving force, and 
        $y(t)$ is the amplitude of the oscillations. 
        If $\epsilon(t)$ is Gaussian distributed 
        then the solution to Equation \ref{eqn:ornstein} is a 
        Gaussian process with the power spectral density
        \begin{equation}
            \label{eqn:sho_psd}
            S(\omega) = \sqrt{\frac{2}{\pi}}\frac{S_0\omega_0^4}{(\omega^2-\omega_0^2)^2 + \omega_0^2\omega^2/Q^2}.
        \end{equation}
        
\begin{figure*}
            \includegraphics[width=0.5\hsize]{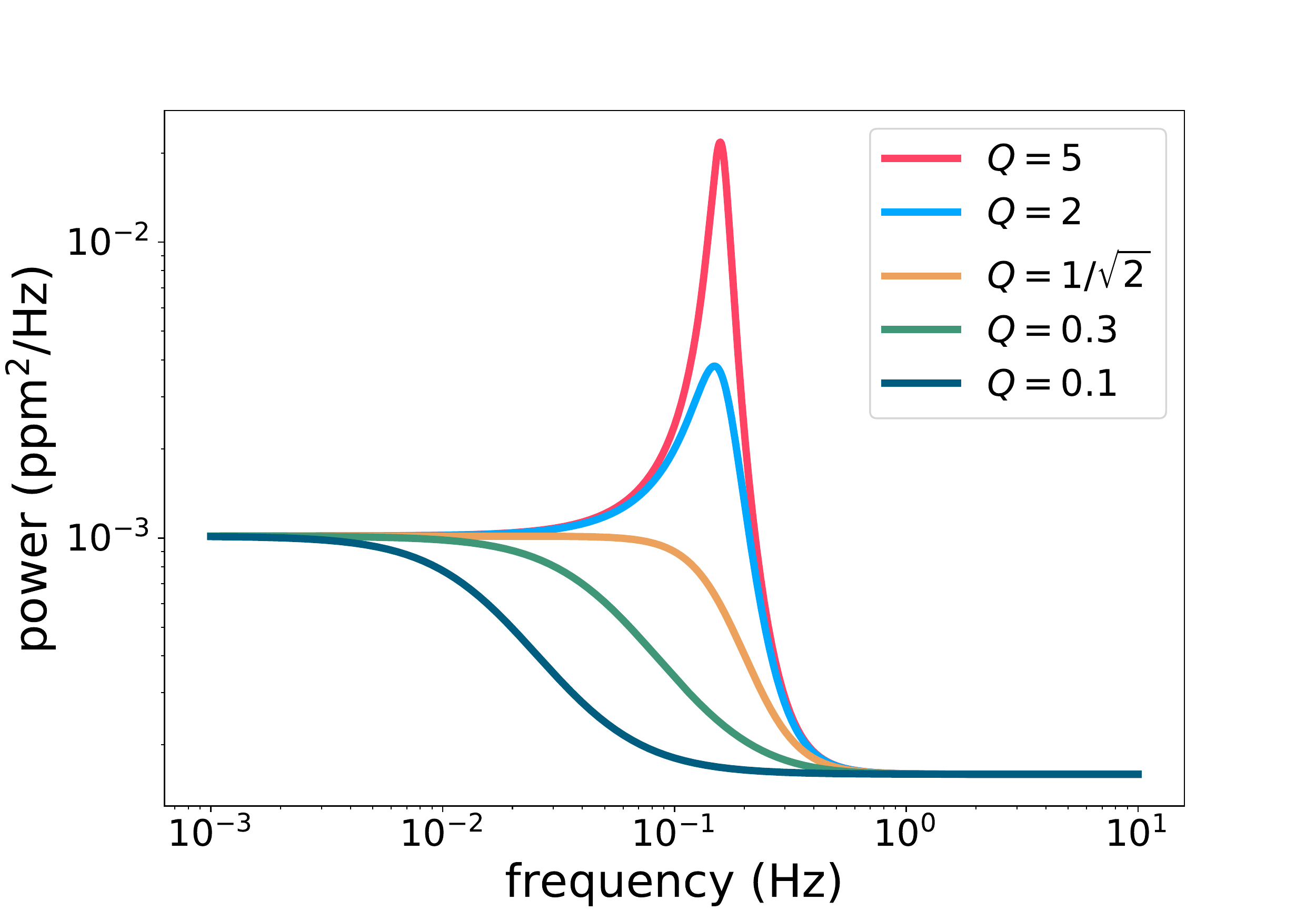}
            \includegraphics[width=0.5\hsize]{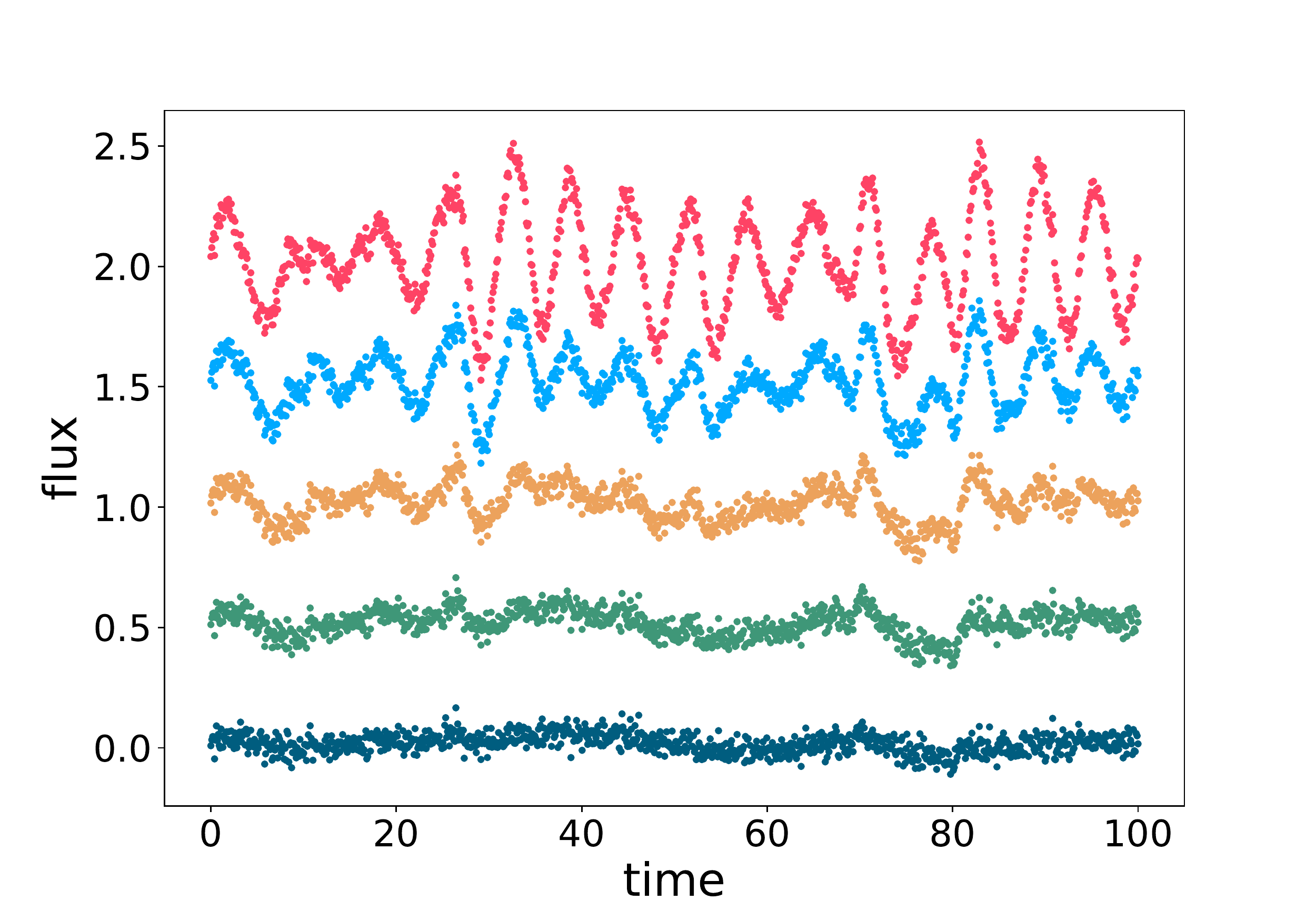}
            \caption{\textbf{Left:} Power spectrum of the 
            SHO kernel for several 
            values of the quality factor $Q$. For $Q < 1/\sqrt{2}$ 
            the system is overdamped. For $Q > 1/\sqrt{2}$ the 
            system is underdamped and the GP shows oscillations 
            at the characteristic frequency. For our simulations 
            we set $Q=1/\sqrt{2}$, in which case the system 
            is critically damped. \textbf{Right:} Noise realizations 
            for each power spectrum on the left. Note the 
            decreasing coherency of the oscillations as we move 
            from high to low values of $Q$. The decreasing 
            noise amplitudes from top to bottom are a result 
            of the fact that the GPs with larger $Q$ values 
            have more total power at constant $S_0$.}
            \label{fig:sho_psd}
        \end{figure*}

        Figure \ref{fig:sho_psd} 
        shows this power spectrum for several values of $Q$. 
        For our modeling we set 
        $Q=1/\sqrt{2}$, in which case the power spectral density 
        simplifies to
        \begin{equation}
            \label{sho_kernel}
            S(\omega) = \sqrt{\frac{2}{\pi}}\frac{S_0}{(\omega/\omega_0)^4 + 1}.
        \end{equation}
        This power spectrum has been used to 
        describe granulation-driven stellar variability \cite{Kallinger2014}. The corresponding kernel function is 
        \begin{equation}\label{eqn:1dcov}
            k(\tau) = S_0\omega_0e^{-\omega_0\tau/\sqrt{2}}\cos{\left(\frac{\omega_0\tau}{\sqrt{2}} - \frac{\pi}{4}\right)},
        \end{equation}
        where $\tau = |t_i-t_j|$. 
        
    \subsection{Wavelength dependence of variability}
    
        We are now interested in constructing a simple model 
        for the wavelength dependence of stellar variability based upon 
        our time-dependent correlated variability model. To begin, we 
        consider a two-component photosphere where each 
        component has a unique spectrum and covering fraction. 
        The star's variability is then a result of variations 
        in the covering fraction of these components, and 
        the covering fractions vary according to the 
        stochastic process described in Section 
        \ref{sec:monochromatic_model}. 
        
        We label the two components ``hot'' and ``cold.''
        Their spectra are given by $S_\mathrm{h}(\lambda)$ 
        and $S_\mathrm{c}(\lambda)$ and their 
        covering fractions are given by $x_\mathrm{h}$ and 
        $x_\mathrm{c} = 1-x_\mathrm{h}$. In the 
        absence of limb-darkening the flux observed 
        in a band $B_1$ is given by 
        \begin{equation}
            F_{B_1} = \frac{\pi R_*^2}{d^2}\int (x_\mathrm{c}S_\mathrm{c}(\lambda) + x_\mathrm{h}S_\mathrm{h}(\lambda))\mathcal{R}_{B_1}(\lambda)\mathrm{d}\lambda
        \end{equation}
        where $\mathcal{R}_{B_1}(\lambda)$ is the response curve for the filter and the integral 
        is taken over all wavelengths and $d$ is the distance from the 
        observer to the star. Substituting $x_\mathrm{h} = 1-x_\mathrm{c}$ 
        allows us to rewrite this expression as 
        \begin{eqnarray}
            \label{eqn:flux_B1}
            F_{B_1}&=& \frac{\pi R_*^2}{d^2}\left(\int S_\mathrm{h}(\lambda)\mathcal{R}_{B_1}(\lambda)\mathrm{d}\lambda\right) \\
            &-&  \frac{\pi R_*^2}{d^2}x_\mathrm{c}\left(\int (S_\mathrm{h}(\lambda) - S_\mathrm{c}(\lambda))\mathcal{R}_{B_1}(\lambda)\mathrm{d}\lambda\right) \nonumber.
        \end{eqnarray} The first term of Equation \ref{eqn:flux_B1} is 
        the total flux for a photosphere completely covered 
        by the hot component, and the second term is a 
        correction dependent on the contrast between the 
        hot and cold components. For simplicity, we 
        define:
        \begin{equation}
            F_\mathrm{B_1, hot} = \frac{\pi R_*^2}{d^2}\int S_\mathrm{h}(\lambda)\mathcal{R}_{B_1}(\lambda)\mathrm{d}\lambda
        \end{equation} 
        and 
        \begin{equation}
            \alpha_\mathrm{1} = \frac{\pi R_*^2}{d^2} \sigma_c \int (S_\mathrm{h}(\lambda) - S_\mathrm{c}(\lambda))\mathcal{R}_{B_1}(\lambda)\mathrm{d}\lambda,
        \end{equation}
        where $\sigma_c^2 = \mathrm{var}(x_c)$.
        With these definitions we have
        \begin{equation}\label{eqn:FB1}
            F_{B_1} = F_\mathrm{{B_1}, hot} - \frac{x_\mathrm{c}}{\sigma_c}\alpha_1.
        \end{equation}
        We can do the same for a second hypothetical band $B_2$, giving 
        us 
        \begin{equation}\label{eqn:FB2}
            F_{B_2} = F_{B_2, \mathrm{hot}} - \frac{x_\mathrm{c}}{\sigma_c}\alpha_2.
        \end{equation}
        Since the only time-dependent quantity in Equations \ref{eqn:FB1} and 
        \ref{eqn:FB2} is the covering fraction of the cold component $x_\mathrm{c}$, 
        we see that the flux in each band will vary coherently with the same 
        power spectral density and the amplitude of the variability will 
        be set by the contrast between the hot and cold components of 
        the photosphere in each band. 
        
        The covariance between two bands can now be computed: 
        \begin{eqnarray}
            \mathrm{cov}(F_{B_1}, F_{B_2}) &=& \sigma_c^{-2}\mathrm{cov}(x_c\alpha_1, x_c\alpha_2) \\ 
            &=& \alpha_1\alpha_2\mathrm{corr}(x_c, x_c). \nonumber
        \end{eqnarray}
    
        Now we let $x_c$ be a function of time, $x_c(t)$, and assert that it is drawn from a one-dimensional Gaussian process evaluated at times $t_i$ for $i=1,\dots,N$ (i.e.\ a correlated time series) with a kernel which can be described with the \celerite formalism.
        Then the full covariance matrix for the time and 
        wavelength dimensions is given by the block matrix
        \begin{equation}
            K = \begin{bmatrix}
                \Sigma_1 + T_{1,1}R & T_{1, 2}R & \dots & T_{1, N}R \\
                T_{2, 1}R & \ddots & & \\
                \vdots & & & \\
                T_{N, 1}R & & & \Sigma_N + T_{N, N}R
            \end{bmatrix},
        \end{equation}
        where
        \begin{equation}
            \Sigma_i = \begin{pmatrix}
                \sigma_{i, 1}^2 & 0 \\
                0 & \sigma_{i, 2}^2
            \end{pmatrix}
        \end{equation}
        is a diagonal matrix containing the 
        white noise components for each band at 
        time $i$; $T_{i,j}=\mathrm{corr}(x_c(t_i), x_c(t_j))$ is the 
        time covariance matrix for the process described in 
        Section \ref{sec:monochromatic_model} normalized by 
        the variance of $x_c$; and $R$ is the covariance 
        matrix across bands, defined: 
        \begin{equation}
            \label{eqn:R_outer}
            R = \begin{pmatrix}
                \alpha_1^2 & \alpha_1\alpha_2 \\
                \alpha_2\alpha_1 & \alpha_2^2
            \end{pmatrix}.
        \end{equation}
        
        For $M$ bands $B_1, B_2, \dots B_M$ with amplitudes given by $\alpha_1, \alpha_2, \dots \alpha_M$, $R$ becomes
        \begin{eqnarray}
            \label{eqn:wavelength_cov}
            R &=& \begin{pmatrix}
                \alpha_1^2 & \alpha_1\alpha_2 & \dots & \alpha_1\alpha_M \\
                \alpha_2\alpha_1 & \ddots & & \alpha_2 \alpha_M \\
                \vdots & & & \vdots\\
                \alpha_M\alpha_1 & \alpha_M \alpha_2 & \dots & \alpha_M^2
            \end{pmatrix} \cr 
            \cr
            &=& \bvec{\alpha}\bvec{\alpha}^T
        \end{eqnarray}
        where $\bvec{\alpha} = (\alpha_1, \alpha_2, \dots, \alpha_M)^T$. 
        The covariance matrix can now be written
        \begin{equation}
            \label{eqn:kronecker_cov}
            K = \Sigma + T \otimes R,
        \end{equation}
        where $\Sigma$ is the block matrix
        \begin{equation}
            \Sigma = \begin{bmatrix}
                \Sigma_1 & \dots & 0 \\
                0 & \ddots & \\
                \vdots & & \Sigma_N
            \end{bmatrix},
        \end{equation}
        and where $\otimes$ 
        denotes the Kronecker product. The Kronecker product 
        is defined for two matrices $A$ and $B$ with dimensions 
        $N {\times} M$ and $P{\times} Q$, respectively, as the 
        $NP {\times} MQ$ block matrix
        \begin{equation}
            A\otimes B = \begin{bmatrix}
                a_{1, 1}\bvec{B} & a_{2, 1}\bvec{B} & \dots & a_{1, N}\bvec{B}\\
                a_{2, 1}\bvec{B} & a_{2, 2}\bvec{B} & \\
                \vdots &  & \ddots & \\
                a_{N, 1}\bvec{B} & & & a_{N, N}\bvec{B}
            \end{bmatrix}.
        \end{equation}

        When the number of bands, $M$, is small, this covariance 
        matrix can be used to model multiband observations. We 
        can also allow $M$ to become arbitrarily large, in 
        which case the resultant covariance matrix can be used 
        to model spectral observations. Here each entry in $\bvec{\alpha}$ 
        would represent the amplitude of the correlated variability in one 
        wavelength bin of the spectrum. The linear scaling of our method 
        with respect to both the time and wavelength dimension 
        makes it feasible to model high spectral resolution 
        time series this way. We include additional discussion 
        on the subject of modeling transmission spectra in section 
        \ref{sec:discussion}.
        
\begin{figure*}
            \epsscale{1.15}
            \plottwo{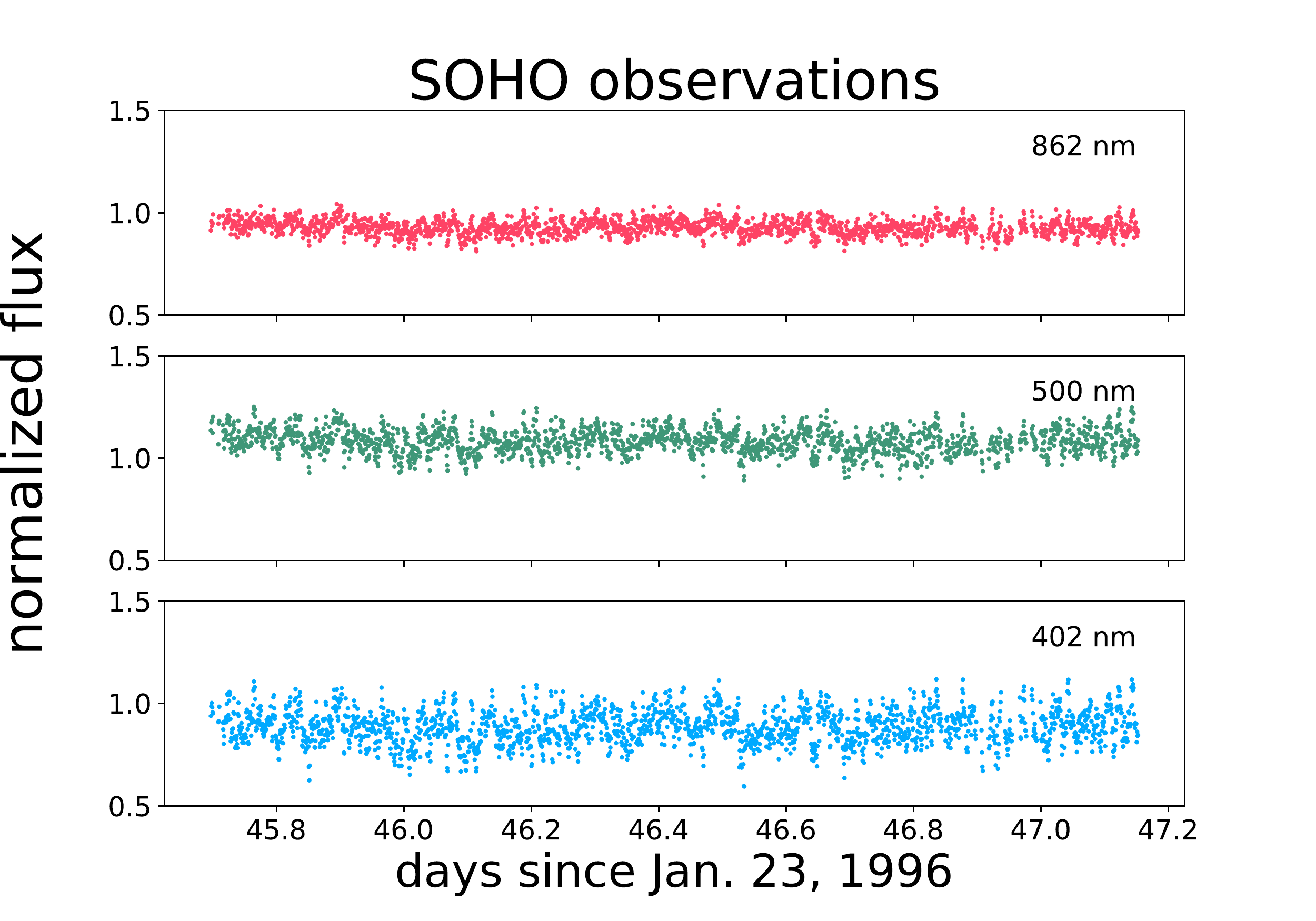}{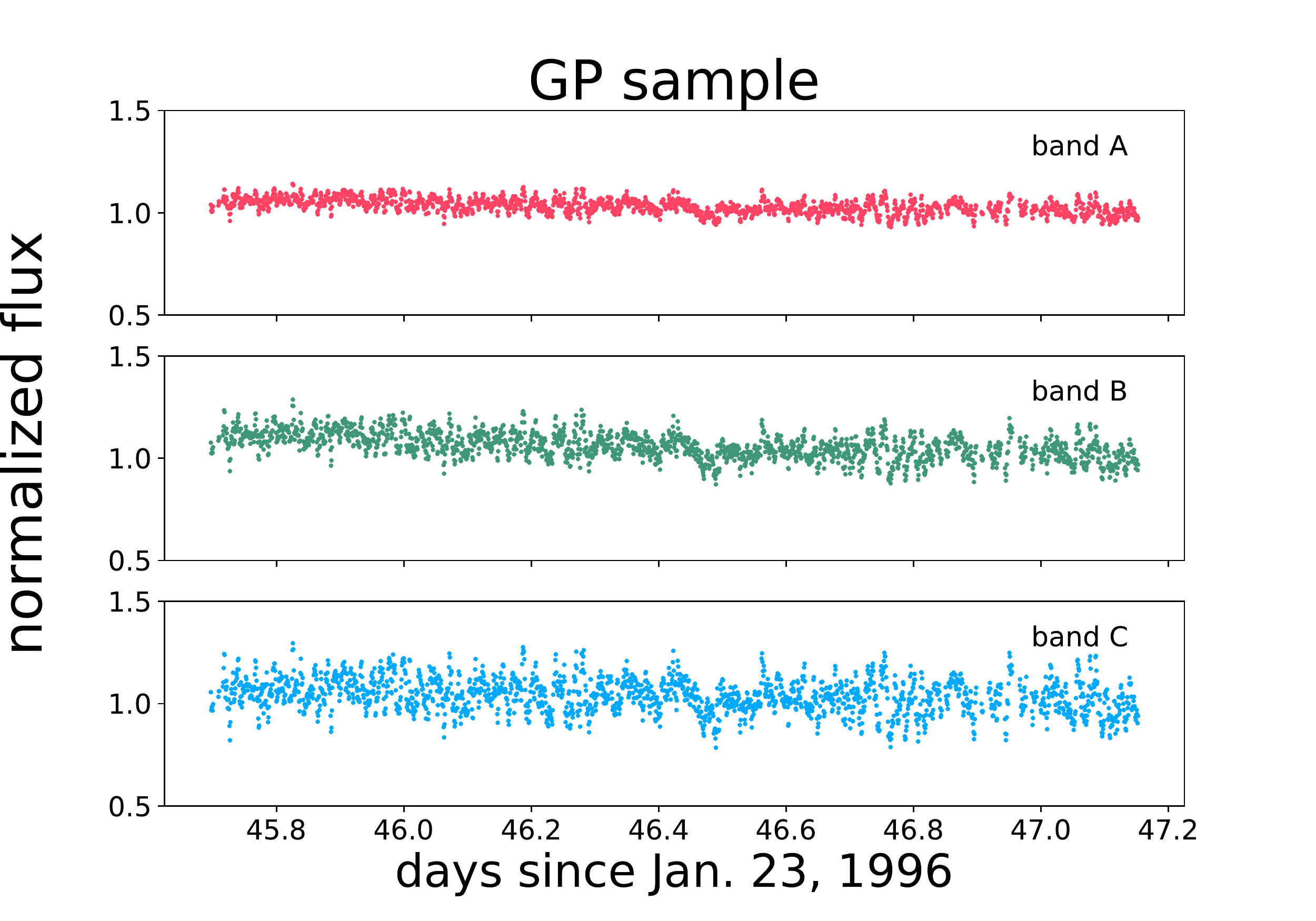}
            \caption{\textbf{Left:} SOHO three-channel 
            sunphotometer time series of the Sun. \textbf{Right:} 
            A three-band light curve simulated 
            from a GP with a kernel consisting 
            of three Kronecker-product terms (see equation \ref{eqn:kronecker_sum}), each 
            term having the covariance described 
            by equation \ref{eqn:wavelength_cov}. 
            The GP hyperparameters were obtained 
            by optimizing the GP likelihood with 
            respect to the data in the left panel.}
            \label{fig:comparison}
        \end{figure*}

        To validate this model of multiwavelength stellar variability, we compare with observed Solar variability in Figure \ref{fig:comparison}. 
        This figure shows a time series from the SOHO VIRGO three-channel sun photometer \citep[SPM;][]{Frohlich1995}.  The SPM monitors the Sun's variability in three visible light wave bands at one-minute cadence, and each of these bands exhibits a power spectrum which has the same shape, but with amplitude which increase from red (862 nm) to blue (402 nm) as shown in \citet{Sulis2020}.  Alongside the SOHO SPM data we show a Gaussian Process drawn from our two-dimensional celerite algorithm in which the amplitudes in each band have been scaled to match the SOHO SPM multiband data.  The qualitative agreement between the observed and simulated data is remarkable, and indicates that our model contains the necessary properties to capture high-precision multiwavelength stellar variability.
        
        The algorithm used to simulate multiwavelength stellar variability and to compute the likelihood model is described in \S \ref{sec:outer_product_method}. Our implementation of 
        the multiband GP, which is based on the \celerite 
        GP method,  achieves $\mathcal{O}(NMJ^2)$ scaling where 
        $N$ is the size of $T$ corresponding to 
        the length of the vector $x_c$ and
        $M$ is the number of bands and 
        corresponds to the size of the vector $\bvec{\alpha}$.
        Appendix \ref{sec:arbitrary_covariance_method} introduces 
        a more general form of the two-dimensional 
        GP which scales as $\mathcal{O}(NJ^2M^3)$ 
        for arbitrary covariance in the second 
        dimension.  The remaining component of the likelihood function is the mean model, which for this paper we take to be a transit model, described next.
        
        Similarly, \citet{Loper2020} recently derived a multivariate generalization of \celerite\ with linear scaling for a class of covariance functions called Latent Exponentially Generated (LEG) kernels.
        These LEG kernel functions are presented for multivariate \emph{outputs} instead of multivariate \emph{inputs} as described here, but it should be possible to express the kernels described here as members of the LEG family.
        However, for our restricted application, the computational cost and scaling of our method is better, since LEG GPs will scale as $\mathcal{O}(NJ^2M^3)$, in the notation above.
        
\subsection{Transit model}

\begin{figure}
            \epsscale{1.2}
            \plotone{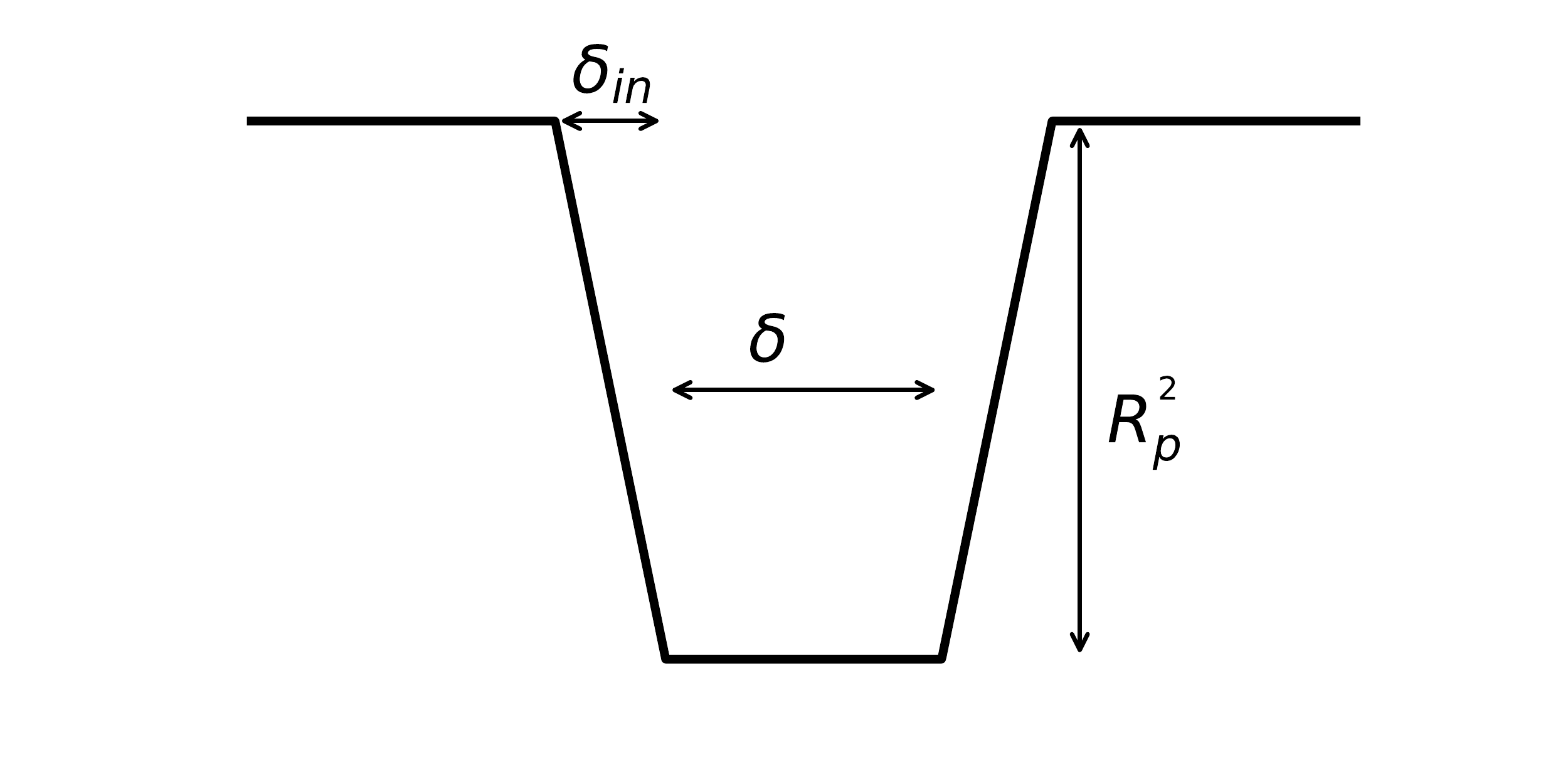}
            \caption{Schematic of the trapezoidal 
            transit model. The center of transit 
            $t_0$ is the midpoint of the transit.}
            \label{fig:trapezoid}
        \end{figure}

To simplify and sharpen our simulated light curves, we use a trapezoidal transit model \citep{Carter2008};  this is the mean model whose parameters we wish to infer. For all our simulations the out-of-transit flux is normalized to unity in order to reduce the number of parameters to be inferred, though we note that this would represent an additional free parameter when modeling real observations. A schematic of this transit model is shown in Figure \ref{fig:trapezoid}.  For the purposes of this paper, we ignore limb-darkening (which can have a wavelength dependence), and we ignore the slight curvature which occurs during ingress and egress.

The model is described by the function $\mu_\mathrm{trap}(t,\bvec{\theta})$ with 
        $\bvec\theta = (R_p, t_0, \delta, \delta_\mathrm{in})$ where 
        $R_p$ is the planet's radius in units of the star's radius, 
        $t_0$ is the time at 
        center of transit, $\delta$ is the transit duration, and $\delta_\mathrm{in}$ 
        is the ingress/egress duration.  Note that we set the normalization of this model to one under the assumption that the out-of-transit data will be sufficiently lengthy to constrain the unocculted stellar flux. 
        
        With the noise and mean models specified, we next describe our simulated data.
     
    \subsection{Simulations}
    
\begin{figure}
            \epsscale{1.15}
            \plotone{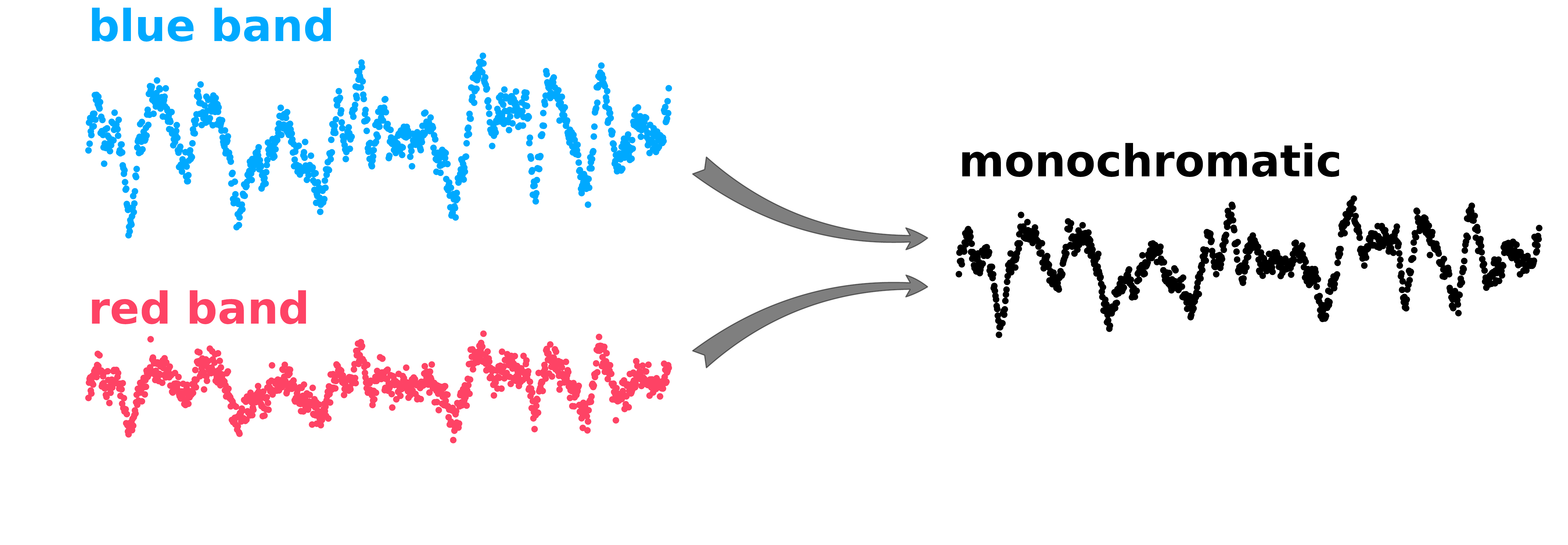}
            \caption{Two bands from a multiband simulation 
            are combined to simulate a monochromatic 
            light curve with the same noise realization. 
            Note that the white noise amplitude is 
            smaller in the monochromatic light curve 
            than for either individual band, while the 
            amplitude of the correlated noise is the photon-weighted mean 
            of the amplitude in the two bands. Here the 
            blue band has a correlated noise amplitude 
            twice that of the red band.}
            \label{fig:schematic}
        \end{figure}

        We simulate 
        a suite of multiband light curves and construct 
        a parallel set of monochromatic light 
        curves by summing the flux between the bands 
        of our multiband light curves. Figure 
        \ref{fig:schematic} shows 
        schematically how we produce a monochromatic 
        light curve from the simulated multiband 
        light curve. We compute the 
        Information matrix (see \ref{sec:fisher}) and 
        run MCMC analysis on each light curve using our 
        multiband GP model. The 
        Information matrix tells us the theoretical 
        lower limit for the uncertainty of each 
        parameter, while the MCMC analysis gives us 
        an estimate of the uncertainty 
        on the parameters. 
        
\begin{figure*}
            \epsscale{1.2}
            \plotone{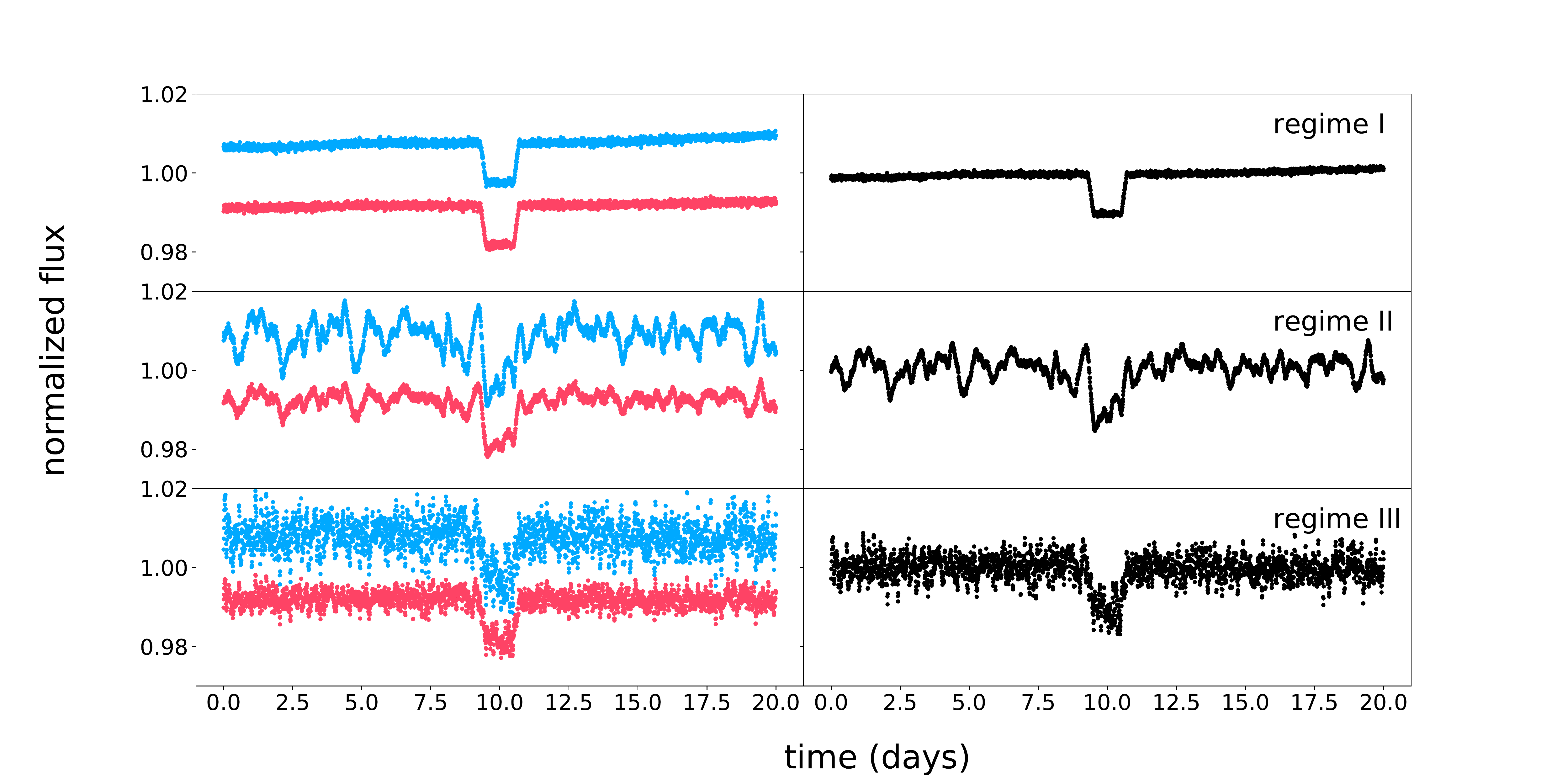}
            \caption{Representative light curves 
            for the three noise regimes. The 
            left panels show the two bands separately 
            and the right panels show the monochromatic 
            light curve resulting from the summation of 
            the two bands. \textbf{Top:} 
            In regime I the variability timescale is 
            much longer than the transit duration. \textbf{Middle:} 
            In regime II the variability timescale is between 
            the transit duration and ingress/egress duration. 
            \textbf{Bottom:} In regime II the variability timescale 
            is shorter than the ingress/egress duration. Figure \ref{fig:psd_regimes} 
            shows power spectra corresponding to each of these regimes (but not to the light curves pictured here).}
            \label{fig:regimes}
        \end{figure*}

\begin{figure}
            \plotone{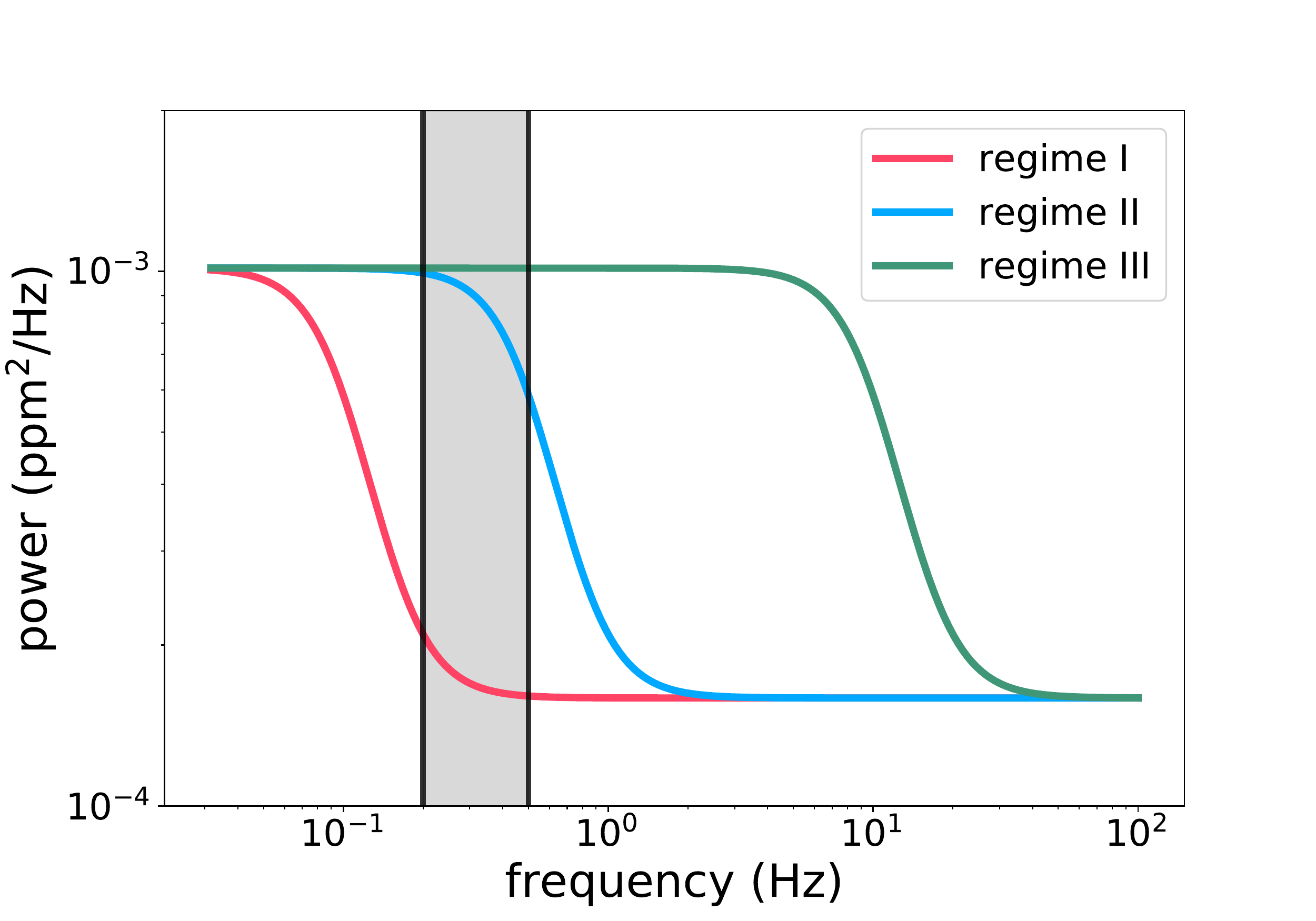}
            \caption{Power spectral densities for 
            the three regimes. The shaded region spans 
            from the inverse transit duration on the left 
            to the inverse ingress/egress 
            duration on the right. Note that the 
            densities plotted here are only meant 
            to be illustrative, and do not correspond to the power spectra of the light curves in figure \ref{fig:regimes}}
            \label{fig:psd_regimes}
        \end{figure}

        We split our simulations into three noise 
        regimes based on the ratio between the 
        characteristic variability timescale,  and the 
        ingress/egress and total duration of the transit. 
        The characteristic variability timescale is 
        given by $2\pi\omega_0^{-1}$ where 
        $\omega_0$ is the characteristic 
        frequency of the variability 
        appearing in equation 
        \ref{eqn:sho_psd}. We define the three regimes 
        as follows:
        \begin{itemize}
            \item Regime I: $1/f_0 > \delta$
            \item Regime II: $\delta_\mathrm{in} < 1/f_0 < \delta$
            \item Regime III: $1/f_0 < \delta_\mathrm{in}$
        \end{itemize}
        where $f_0 = \omega_0/(2\pi)$ is the 
        characteristic frequency of the 
        variability. 
        Figure \ref{fig:regimes} contains representative 
        light curves from each regime, chosen where the 
        white and correlated noise amplitudes are comparable. 
        In regime I the transit 
        signal is distinguishable from the noise by its 
        duration --- all of the power in the correlated 
        variability is on longer timescales than the transit 
        duration. In regime II the characteristic 
        timescale of the noise is smaller than the transit 
        duration, but longer than the ingress/egress 
        timescale. The transit still stands out from 
        the noise because the transition into and 
        out of transit is sharper than is characteristic 
        for the SHO variability. In regime III 
        the variability timescale is shorter 
        than all of the relevant transit durations. We 
        can see from Figures \ref{fig:sho_psd} and 
        \ref{fig:psd_regimes} that 
        the SHO power spectrum allocates equal power to 
        all oscillations on timescales longer than the 
        characteristic timescale. The transit durations 
        are thus swamped by correlated noise.
        As a result in the monochromatic case it is difficult to
        differentiate between the transit signal and noise, both by 
        eye and with the GP. Fortunately the multiband 
        GP is able to make use of additional information in the 
        correlation between bands to disentangle the 
        transit signal from the variability. 
        
        Amongst all of our simulations we hold 
        constant the total noise, $\bar\alpha^2 + \sigma^2$ 
        where $\bar\alpha^2$ is the weighted variance of the 
        correlated noise over all bands and $\sigma^2$ is the 
        variance of the white noise summed over all bands. We then vary 
        the ratio between the noise amplitudes in 
        order to analyze the simulations as a 
        function of $\bar\alpha/\sigma$. For the 
        multiband simulations, $\bar\alpha$ is the 
        weighted mean of the amplitudes of variability in 
        the individual bands, given by $\alpha_i$. 
        For all of our simulations, unless otherwise 
        specified, we use a two-band model with 
        $\alpha_2 = 2\alpha_1$ to represent the multiband 
        case. 
        
        We hold the transit duration and ingress/egress 
        duration constant so that the 
        value of $\omega_0$ changes to determine 
        which noise regime we fall under. 
        
        Into all of our simulations we inject a transit 
        signal with a fractional depth of 1\% of the star's 
        flux.  We use a 
        transit duration of 12 hours in the middle of 
        a 10 day baseline. The ingress/egress duration 
        is set to 1.2 hours. 
        
        \subsection{Information matrix analysis}\label{sec:fisher}
        The Information matrix encodes 
        the amount of information about a signal that can be 
        determined from observations taken in the presence of 
        noise with a given covariance. For a model made up of 
        a mean function $\bvec{\mu_\theta}$ with $N_\theta$ parameters $\theta_1, \theta_2, 
        \dots \theta_{N_\theta}$ obscured by noise 
        drawn from a multivariate Gaussian with covariance 
        $K$, the Information matrix is the $N_\theta{\times} N_\theta$ 
        matrix with entries given by 
        \begin{equation}
            \left[\mathcal{I}_\bvec\theta\right]_{i,j} = \left(\frac{d \bvec{\mu}}{d\theta_i}\right)^\T 
            K^{-1} \left(\frac{d\bvec{\mu}}{d\theta_j}\right).
        \end{equation}
        The covariance between parameters of the mean 
        are then approximated by 
        \begin{equation}
            \left[\mathcal{I}_\bvec{\theta}^{-1}\right]_{i,j} \approx \mathrm{cov}(\theta_i, \theta_j).
        \end{equation}
        This approximation represents a lower limit on the 
        covariance that can be estimated in practice via methods 
        such as MCMC simulation. It is valid in the limit that the posterior 
        probability is a multi-dimensional Gaussian distribution 
        near the maximum likelihood solution. This corresponds to 
        the limit in which a signal may be approximated as 
        linear with respect to its parameters, known as the 
        linear signal approximation (LSA). \citet{Vallisneri2008} 
        shows that in order for LSA to apply we must be 
        in the high signal-to-noise limit. Accordingly, the following 
        analysis should be taken to apply only to a transit with a depth 
        much larger than both the correlated and white noise 
        components of the noise. While the approximation 
        may continue to be 
        accurate for smaller signal-to-noise, a full quantification 
        of the uncertainty in the low SNR limit should rely on 
        sampling the posterior directly via MCMC analysis. 
        
        We compute the Information 
        matrix for the transit 
        parameters assuming that the hyperparameters of the 
        GP are known exactly. In practice the GP hyperparameters 
        will be unknown, and should be fit simultaneously 
        with the transit parameters. Our results thus represent 
        a scenario in which there are sufficient out-of-transit 
        observations to determine the covariance of the 
        noise to arbitrary precision. 
        
        We adopt a semi-analytic approach to computing the 
        Information matrix by using exact derivatives 
        of the trapezoidal transit model and using \celerite 
        to compute products of the inverted covariance matrix 
        with the transit model's derivatives. This approach 
        is necessary because the covariance matrix for our 
        GP model cannot be inverted analytically except 
        in special cases. 

    \subsection{Analytical Estimates For Parameter Uncertainties}
        \label{sec:analytical_estimates}
        The Information matrix approach can yield analytic results 
        for the depth uncertainty in the limit that limb-darkening 
        is ignored, the ingress/egress duration is short, $\delta_{in} \approx 0$, and 
        all other parameters are assumed to have no uncertainty. In 
        particular we make the approximation that 
        the out-of-transit flux is measured to high precision from extensive monitoring.
        In this limit the transit model has a derivative of 
        \begin{equation}
            \frac{\partial \bvec{\mu}_\mathrm{trap}}{\partial R_p^2} = 
            \begin{cases}
                0 & \mathrm{out-of-transit}\\
                -1 & \mathrm{in-transit}
            \end{cases},
        \end{equation}
        where $R_p^2$ is the 
        depth of the transit. If we assume that the transit duration matches exactly 
        a single observation cadence then the covariance matrix 
        may be written in the two-band case as:
        \begin{equation}
            K = \begin{pmatrix}
            \sigma_1^2 + \alpha_1^2 & \alpha_1 \alpha_2\\
            \alpha_1\alpha_2 & \sigma_2^2 + \alpha_2^2
            \end{pmatrix},
        \end{equation}
        where $\sigma_{1, 2}$ are the white noise 
        components on the timescale of the transit and 
        $\alpha_{1, 2}$ are the correlated noise amplitudes 
        on the timescale of the transit in the two bands. 
        
        For this covariance matrix, the Information matrix
        gives an uncertainty on the depth 
        of the transit, $\sigma_{R_p^2}$, of 
        \begin{equation}
            \sigma_{R_p^2,poly}^2 = \left(\frac{1}{\sigma_1^2}+
            \frac{1}{\sigma_2^2}\right)^{-1}
            \left(\frac{1 + \left(\frac{\alpha_1}{\sigma_1}\right)^2 + \left(\frac{\alpha_2}{\sigma_2}\right)^2}{1 + \frac{(\alpha_1-\alpha_2)^2}{\sigma_1^2+\sigma_2^2}}\right).
        \end{equation}
        Note that the prefactor equals the noise in the limit 
        of no correlated noise component $(\alpha_1 = \alpha_2 = 0)$. 
        
        In the monochromatic case we can compute the uncertainty 
        assuming that the noise is Poisson, in which case the  mean 
        amplitude of correlated noise is given by 
        \begin{equation}
            \sigma_{R_p^2,mono}^2 =  \left(\frac{1}{\sigma_1^2}+
            \frac{1}{\sigma_2^2}\right)^{-1}
            + \bar \alpha^2,
        \end{equation} 
        where we have assumed the noise to be Poisson and 
        $\bar\alpha$ is defined to be the weighted mean amplitude of 
        the correlated noise in both bands, given by 
        \begin{equation}
            \bar \alpha = \left(\frac{1}{\sigma_1^2}+
            \frac{1}{\sigma_2^2}\right)^{-1}\left(\frac{\alpha_1}{\sigma_1^2}+\frac{\alpha_2}{\sigma_2^2}\right).
        \end{equation}
        
\begin{figure}
            \centering
            \includegraphics[width=\hsize]{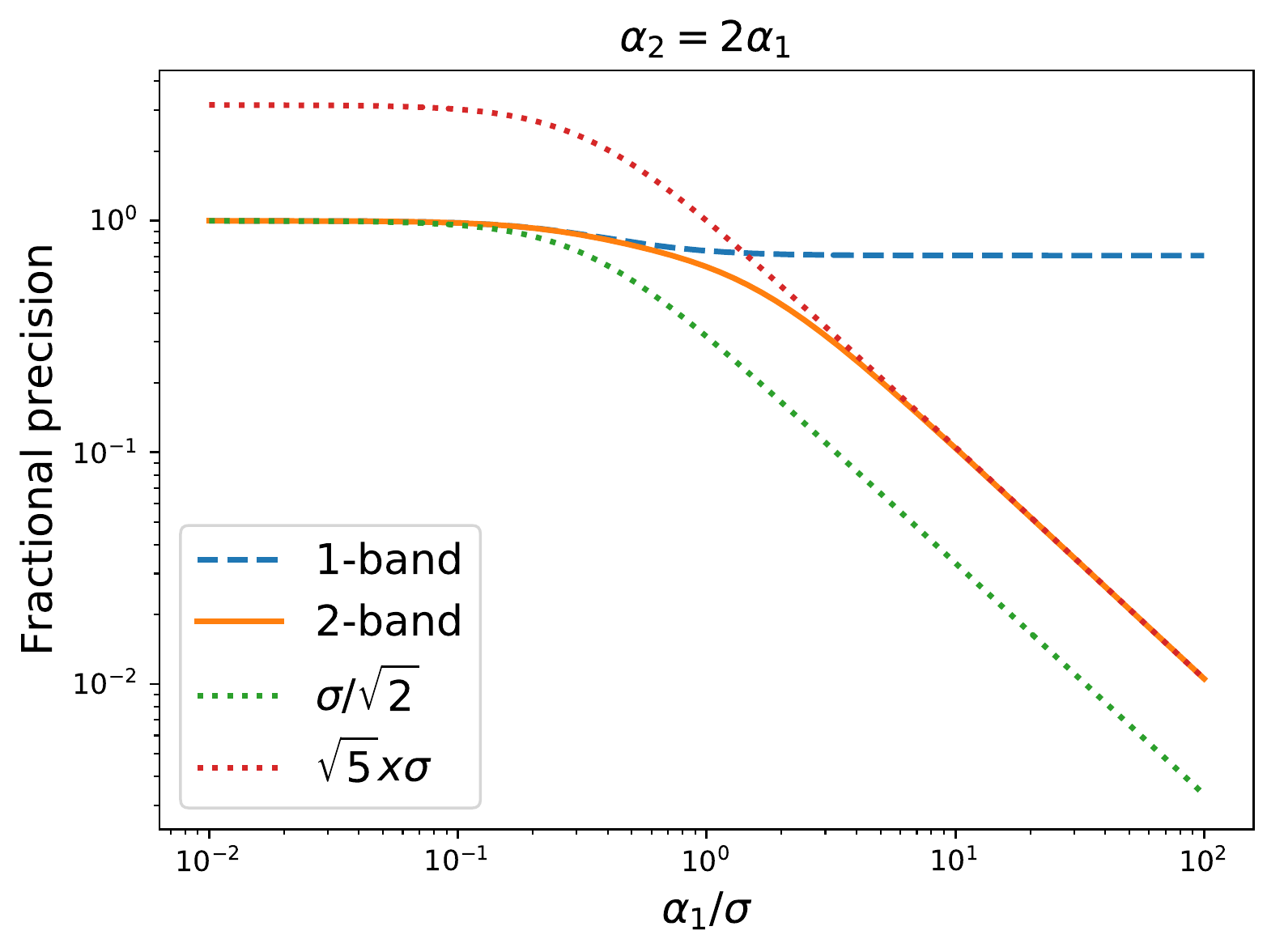}
            \caption{Analytic approximation for the fractional uncertainty on depth for two bands versus the ratio of the correlated noise to white noise in first band, $\alpha_1/\sigma$, in the limit of a constant amplitude of the sum of correlated and white noise (so that the white noise declines as the correlated noise increases).  The ratio of the correlated noise in the two bands is two, i.e.\ $\alpha_2 = 2 \alpha_1$.  Plotted are the single-band case (blue dashed), two-band case (orange solid), and the white noise in each band, $\sigma$, times $1/\sqrt{2}$ and $\sqrt{5}$ (dotted).  The fractional precision is normalized to the case $\alpha_1=0$.}
            \label{fig:depth_noise}
        \end{figure}

        The relations for the polychromatic and monochromatic 
        cases are plotted in Figure \ref{fig:depth_noise} in the case 
        $\alpha_2 = 2\alpha_1$ and $\sigma_1 = \sigma_2 = \sigma$ 
        in which the sum of the white noise and correlated noise 
        is held fixed. Compare with Figure 
        \ref{fig:white_noise_comparison} to see the 
        similarity of this analytic approximation with the 
        Information matrix results for the full trapezoidal 
        model. 
        
        We can generalize these expressions for the depth 
        uncertainty in the monochromatic and two-band 
        case to an arbitrary number of bands when 
        the white noise is identical for each band (i.e. 
        $\sigma_i = \sigma$ for $M$ bands indexed by $i$). In this 
        case the uncertainties are given by 
        \begin{equation}
            \label{eqn:sig_poly}
            \frac{\sigma_{R_p^2,M,poly}^2}{\sigma^2}
            = \frac{1 + \sigma^{-2}\sum_{i=1}^M \alpha_i^2}{M\left(1+\sigma^{-2}\sum_{i=1}^M \alpha_i^2\right) - \left(\sigma^{-1}\sum_{i=1}^M \alpha_i\right)^2},
        \end{equation}
        for the $M$-band case, 
        and 
        \begin{equation}
            \frac{\sigma_{R_p^2,M,mono}^2}{\sigma^2}
            = \frac{1}{M} + \left(\frac{1}{M\sigma}\sum_{i=1}^M \alpha_i\right)^2.
        \end{equation}
        for the corresponding monochromatic case. Similar 
        expressions may likely be found for the other transit 
        parameters as well as for non-uniform noise in $M$ bands, 
        which we leave to future work. 
        
        While the uncertainties predicted by these 
        equations differ from those found by 
        a full Information matrix analysis of the trapezoidal 
        transit, we find that they correctly predict the 
        relationship between the monochromatic and multiband 
        uncertainties in the limits $\alpha \gg \sigma$ and 
        $\alpha \ll \sigma$ not only for the depth, but for 
        the other parameters of the trapezoidal 
        transit as well. 
        
        This is illustrated by Figure 
        \ref{fig:white_noise_comparison} which shows the 
        Information uncertainties for each parameter of the 
        trapezoidal transit model in the presence of 
        correlated noise. We use 
        a two band noise model with $\alpha_2 = 2*\alpha_1$. 
        When the white noise 
        dominates over the correlated noise 
        ($\sigma \gg \alpha_{1, 2})$, the Information uncertainties
        for the model with correlated noise is identical 
        to a white noise-only model with the same white 
        noise component, as we expect given that the 
        correlated noise component is insignificant in this 
        limit. We can use equation \ref{eqn:sig_poly} to 
        predict the Information matrix for the two band 
        model in the limit that the correlated noise 
        component dominates over the white noise 
        component ($\sigma \ll \alpha_{1, 2}$). Taking 
        this limit equation \ref{eqn:sig_poly} becomes 
        \begin{equation}
            \label{eqn:sig_poly_limit}
            \lim_{\sigma \ll \alpha_{1, 2}} \frac{\sigma_{R_p^2, M, \mathrm{poly}}^2}{\sigma^2} = 
            \frac{\sum_{i=1}^M \alpha_i^2}{M\sum_{i=1}^M \alpha_i^2 - \left(\sum_{i=1}^M \alpha_i\right)^2}.
        \end{equation}
        Setting $\alpha_1 = 1$ and $\alpha_2 = 2$, 
        we find 
        $\sigma_{R_p^2, M, \mathrm{poly}}/\sigma = \sqrt{10}$ 
        which explains the scaling of the Information uncertainty 
        at large $\alpha/\sigma$ in Figure
        \ref{fig:white_noise_comparison}.

\begin{figure}
            \plotone{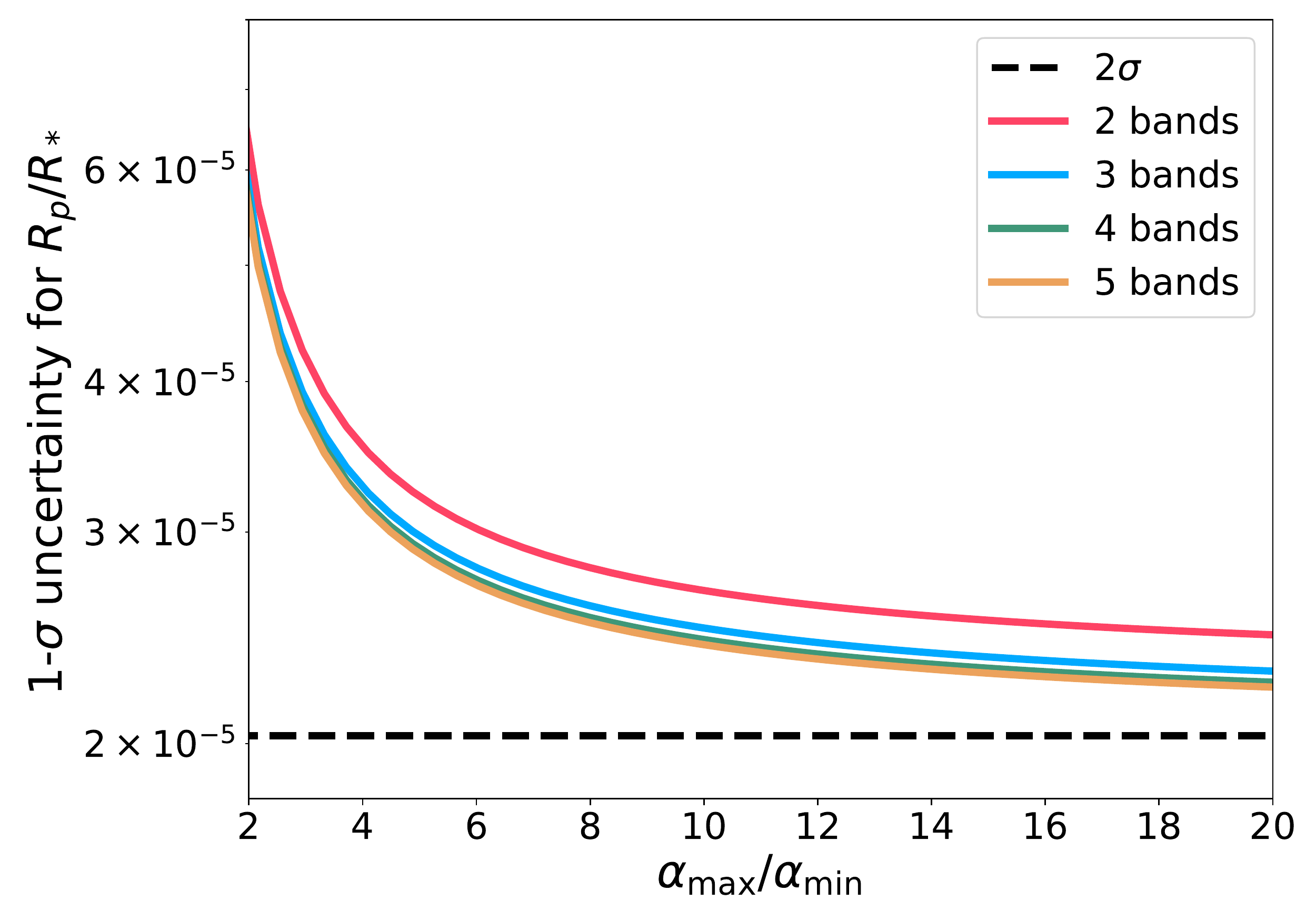}
            \caption{Information uncertainty curves for the planet/star radius ratio as 
            a function of contrast ratio for a 
            spectrum that increases linearly with photon flux 
            from $\alpha_\mathrm{min}$ to $\alpha_\mathrm{max}$. We plot the 
            Information uncertainty for different values of $M$, the number of bands into which 
            the spectrum is binned for modeling. 
            The dashed line is the minimum uncertainty achievable as the contrast 
            ratio becomes infinite which, for the two-band case, is equal to $2\sigma$ where $\sigma$ is the Information uncertainty in the absence of correlated noise.}
            \label{fig:fisher_nbands}
        \end{figure}

        We also examine the Information uncertainties as a function of 
        number of bands.  We consider a photon spectrum for which the 
        variability increases from a value of $\alpha_\mathrm{min}$ 
        to $\alpha_\mathrm{max}$.  We assume that the photon spectrum variability is split into $M$ bands with an equal photon count rate in each band to give equivalent Poisson noise across all bands.  In addition, we assume that $\alpha$ varies linearly with the photon count rate across all bands, so that the $i$th band has a correlated noise amplitude of $\alpha_i = \alpha_\mathrm{min} + (\alpha_\mathrm{max}-\alpha_\mathrm{min}) (i - 1/2)/M$. For example, in the case of two bands with $\alpha_\mathrm{max}/\alpha_\mathrm{min} = 5$, we have $\alpha_2 = 2\alpha_1$, as in Figure \ref{fig:depth_noise}.  
        Figure 
        \ref{fig:fisher_nbands} shows the uncertainty for 
        the planet-star radius ratio as a function of
        the ratio between the minimum and 
        maximum variability, $\alpha_\mathrm{max}/\alpha_\mathrm{min}$, for several 
        values of $M$. The minimum achievable uncertainty as $M$ 
        approaches infinity and $\alpha_\mathrm{max} \gg \alpha_\mathrm{min}$, which 
        can be arrived at by taking the appropriate limits 
        of Equation \ref{eqn:sig_poly} and transforming the 
        sums into integrals as $M$ approaches 
        infinity. In these limits the 
        minimum achievable uncertainty is twice the 
        uncertainty for the white noise-only case, which is 
        represented by the dashed line in Figure 
        \ref{fig:fisher_nbands}. 
        
        The same calculation may be performed for alternative 
        spectra. For a blackbody spectrum we arrive at a 
        limit of 2.2 times the white noise-only 
        case when the number of bands and the contrast ratio 
        is large. For arbitrary spectra the integrals 
        can be computed numerically to yield the 
        minimum achievable uncertainties for realistic 
        stellar spectra and spot models. 
        
        The Information matrix and analytic approaches describe approximations to the parameter uncertainties.  We next summarize our MCMC analysis to check and validate these approximations.
        
        \subsection{MCMC analysis}
        
        We use the \texttt{exoplanet} package \citep{exoplanet:exoplanet} 
        which interfaces with PyMC3 to conduct our MCMC simulations. 
        Each simulation is initialized with the true parameters. During 
        MCMC we hold the GP hyperparameters constant as we did 
        for the Information matrix analysis, and vary only the 
        parameters of the trapezoidal transit model. We 
        use PyMC3's implementation of No U-Turn sampling \citep[NUTS;][]{Hoffman2014}, which requires the derivatives of the log likelihood to carry out the Hamiltonian markov chain integration.   The NUTS sampler is initialized by 
        tuning 
        each simulation for 2000 steps.  Subsequently,  the simulation is run another 
        2000 steps to sample the posterior. This procedure results 
        in about $10^3$ effective samples for each parameter of the 
        model for each simulation as the autocorrelation length of the chains is extremely short (one of the advantages of using the NUTS sampler).  
        
        The final ingredient needed for our Information matrix and MCMC simulations involves our novel 2D version of \celerite, which we describe next. 

\section{Implementation of the multiwavelength variability model} \label{sec:2dcelerite}

    We implement our multiwavelength variability 
    model as an extension of the \celerite 
    GP method to two dimensions. The \celerite algorithm
    \citep{Foreman-Mackey2017} is a method for 
    computing Gaussian processes in one 
    dimension that scales 
    as $\mathcal{O}(NJ^2)$ where $N$ is the number 
    of datapoints being modeled and $J$ is the 
    number of terms used to represent the 
    covariance matrix. While one-dimensional 
    GPs are suitable for a wide range of 
    applications, there are many problems for 
    which we need to model covariance between 
    datapoints in two or more dimensions. Here we 
    describe a method for computing a two 
    dimensional GP when the covariance in the 
    second dimension can be written as the outer 
    product of a vector with itself. This 
    covariance matrix is relevant to the common task 
    of modeling time-variable spectra, as in our multiband transit model application. Our method 
    is scalable, with computational time increasing 
    linearly with the number of datapoints. In this 
    section we introduce the method and revisit the 
    \celerite algorithm for Cholesky 
    decomposition of the covariance matrix as it 
    applies to a two-dimensional dataset. In appendices 
    \ref{sec:likelihood}, \ref{sec:prediction}, 
    and \ref{sec:sampling} we discuss 
    the algorithms for computing the likelihood, 
    predicting 
    or extrapolating from the GP, and 
    sampling from the GP. 
    
    For problems where the covariance cannot 
    be modeled as an outer product 
    we offer a more general extension 
    of \celerite where the covariance matrix
    for the second dimension can be arbitrary. We 
    discuss our implementation of the arbitrary 
    covariance method in Appendix 
    \ref{sec:algorithm}. 
    
    \subsection{The one-dimensional \celerite method}
    
        Until recently, adoption of GP methods has been limited by 
        computational expense. As a reminder, the log-likelihood function for a GP 
        model for a series of $N$ flux measurements, $\bvec{y}=(y_1, y_2, \dots y_{N})$, so that the total number of datapoints $N' = N$,
         taken at times 
        $\bvec{t} = (t_1, t_2, \dots t_{N})$ is given by 
        \begin{equation}
	        \label{eqn:simple_logL_v2}
		    \ln\ \mathcal{L} = 
		    -\frac{1}{2}(\bvec{y}-\bvec{\mu})^\T K^{-1}(\bvec{y}-\bvec{\mu})  
		    -\frac{1}{2}\ln\ \mathrm{det}(K) - \frac{N}{2}\ln(2\pi)
	    \end{equation}
	    where $\bvec{\mu} = (\mu(t_1), \mu(t_2), \dots \mu(t_{N}))$ and 
	    $K$ is the covariance matrix of the GP. This equation involves 
	    the inverse and determinant of the $N{\times} N$ matrix $K$. In general, computing 
	    the inverse and determinant of an $N{\times} N$ matrix requires $\mathcal{O}({N}^3)$ 
	    operations. Thus computing the likelihood for a GP by directly 
	    inverting $K$ becomes prohibitively expensive for datasets larger 
	    than about $10^4$ observations \citep{Deisenroth2015}. This is especially true for 
	    applications that require repeated calls to the likelihood function 
	    as is the case for minimization and MCMC. 
	
	    The computational expense of GPs can be mitigated in two ways. 
        The first is by employing inexact methods in which the 
        full GP covariance matrix is approximated by a matrix for 
        which the relevant matrix operations (primarily inversion 
        and computation of the determinant) can be computed more 
        efficiently than $\mathcal{O}({N}^3)$ \citep{Rasmussen2006}. 
        The second is by restricting 
        the user to covariance matrices of a specific form for which, 
        again, the relevant matrix operations are quicker to compute; 
        \celerite \citep{Foreman-Mackey2017} is a member of this 
        second class of methods. \celerite is a fast,     
        one-dimensional 
        GP method which takes $\mathcal{O}(NJ^2)$ operations to 
        compute where $J$ is the number of \celerite terms that make 
        up the kernel function. For commonly-used kernel models the number of 
        terms will be very small compared to $N$. 
        
        \celerite works by representing the GP covariance matrix 
        as the sum of a diagonal matrix and $J$ semi-separable matrices. The Cholesky 
        factorization of $J$ semi-separable matrices plus a diagonal matrix 
        can be computed in $\mathcal{O}(NJ^2)$ rather than 
        the $\mathcal{O}(N^3/3)$ required for an ordinary 
        matrix. Once the Cholesky factors are in hand, the 
        inverse and determinant of the covariance matrix 
	    can be computed in $\mathcal{O}(NJ)$ and $\mathcal{O}(N)$ 
	    respectively. Here we briefly describe the 
	    \celerite algorithm, referring the reader to 
	    \citet{Foreman-Mackey2017} for a more detailed exposition of the method. 
		
		Consider a one-dimensional Gaussian process evaluated at the coordinates
		\begin{equation}
			\bvec{x} = \expandvec{t_1}{t_{N}}
		\end{equation}
		The \celerite kernel is given by
		\begin{eqnarray}
		    \label{eqn:k_alpha}
			k_\bvec{\beta}(t_n, t_m) &=& \sigma_n^2 \delta_{nm} + \sum_{j=1}^J 
			\frac{1}{2}\left[(a_j+ib_j) e^{-(c_j+id_j)\tau_{nm}} \right. \cr &+& \left.  (a_j-ib_j)e^{(c_j-id_j)\tau_{nm}}\right]   
		\end{eqnarray}
		where $\bvec{\beta} = (a_1...a_J, b_1...b_J, c_1...c_J, d_1...d_J)$, $\sigma_n^2$ is the variance of the Gaussian-distributed white noise, and $\tau_{nm} = |t_n-t_m|$ with $n,m \in {1,\dots,N}$. 
		This kernel defines a \celerite model with $J$ terms. 
		
		For a kernel function of this form, the covariance matrix is a symmetric, semiseparable matrix with semiseparability rank $P=2J$. A matrix of this type can 
		be written in terms of two generator matrices $U$ and $V$, both of size $(N{\times} P)$, along with a diagonal matrix $A$:
		\begin{equation}
		\label{eqn:KUV}
			K = A + \mathrm{tril}(UV^\T) + \mathrm{triu}(VU^\T),
		\end{equation}	
		where tril is the lower-triangular operator which, when applied to 
		a square matrix, preserves the entries below the diagonal and 
		replaces all entries on and above the diagonal with zeros. The triu operator 
		does the same for the upper-triangular entries in the matrix.
		In the case of our covariance matrix, the generator matrices are specified by:	
		\begin{eqnarray}
		    \label{eqn:UV_defs}
			U_{n, 2j-1} &=& a_je^{-c_jt_n}\cos(d_jt_n) + b_je^{-c_jt_n}\sin(d_jt_n), \cr
			U_{n, 2j} &=& a_je^{-c_jt_n}\sin(d_jt_n) - b_je^{-c_jt_n}\cos(d_jt_n), \cr
			V_{m, 2j-1} &=& e^{c_jt_m}\cos(d_jt_m), \cr
			V_{m, 2j} &=& e^{c_jt_m}\sin(d_jt_m),
		\end{eqnarray}
		and A is given by:
		\begin{equation}
		    \label{eqn:A_matrix}
			A_{n,n} = \sigma_n^2 + \sum_{j=1}^Ja_j.
		\end{equation}
		We will soon see that the Cholesky decomposition for this covariance matrix can be computed 
		in $\mathcal{O}(NJ^2)$ operations, allowing for the 
		fast evaluation of the GP likelihood function. 
		
\begin{figure}
		    \centering
		    \includegraphics[width=\hsize]{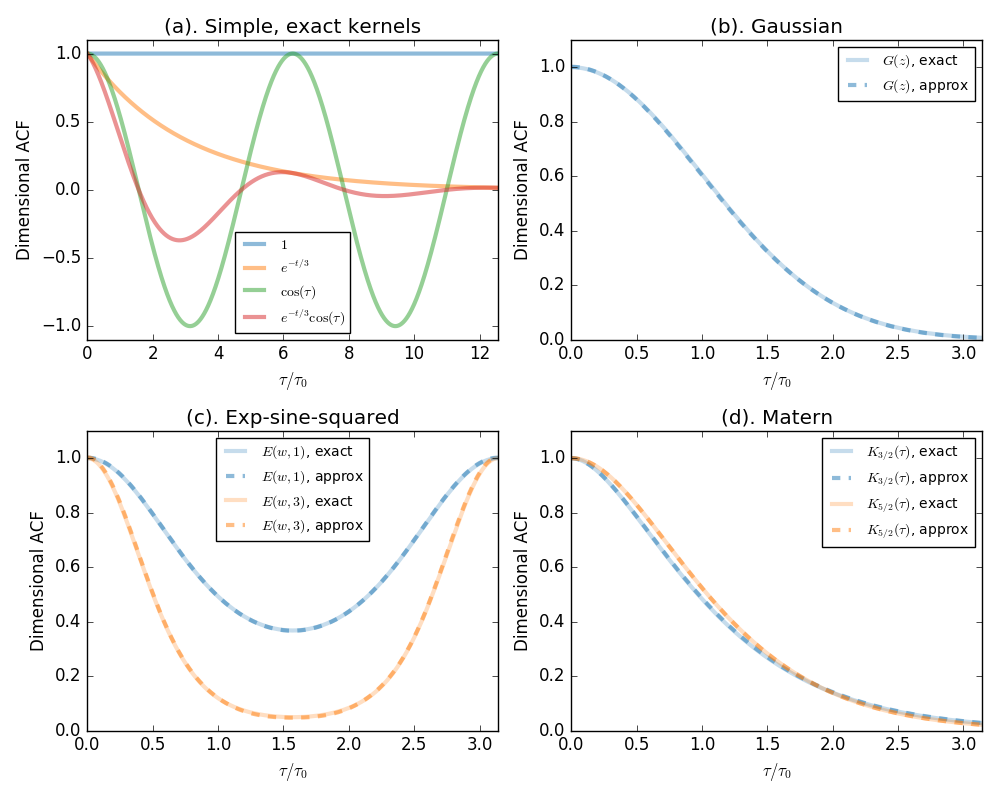}
		    \caption{Approximation to various commonly used GP kernels 
		    (a) Simple kernels with an exact \celerite representation: 
		    cosine, or exponential times cosine. (b) Approximation of a 
		    referred to as ``exponential-squared" to distinguish it from 
		    sine-squared kernels. (d) Matern kernels.}
		    \label{fig:approx_kerns}
		\end{figure}

		The kernel function implemented by \celerite is versatile in that 
	    by choosing appropriate coefficients it can be made to approximate a 
	    wide range of other kernel functions.
	    Furthermore, \citet{Loper2020} demonstrated that \celerite\ kernels provide a complete basis for one dimensional stationary covariance functions, meaning that these methods can, in principle, be used to approximate \emph{any} stationary kernel, though there might be issues with numerical precision and computational cost when a large number of terms are required for accuracy.
	    This versatility is demonstrated qualitatively in 
	    Fig.\ \ref{fig:approx_kerns} which shows approximations of several popular 
	    kernels achieved by carefully choosing $\left\{a_j\right\}$, $\left\{b_j\right\}$, $\left\{c_j\right\}$ and $\left\{d_j\right\}$. 
		Since each of these kernels may be approximated well by a \celerite kernel,
		the products and sums of these component kernels are also \celerite kernels, meaning that
		complex kernels can still be approximated within the \celerite kernel formalism.
		
		The Cholesky factorization of the 
		covariance matrix $K$ is given by 
		\begin{equation}
		    K = LDL^\T
		\end{equation}
		where $L$ is the lower-triangular Cholesky 
		factor and $D$ is a diagonal matrix. 
		\cite{Foreman-Mackey2017} begin their 
		derivation of the Cholesky factorization 
		algorithm with the ansatz that $L$ can 
		be represented in terms of $U$ and a new 
		(at this point unknown) matrix $W$ with 
		the same dimensions as $U$, as
		\begin{equation}
		    L = I + \mathrm{tril}(UW^\T).
		\end{equation}
		Then $W$ and $D$ can be found via the 
		recursion relations
		\begin{eqnarray}
    	     \nonumber S_{n, j, k} = S_{n-1, j, k} + D_{n-1, n-1}W_{n-1, j}W_{n-1, k} \\
    	    \nonumber D_{n, n} = A_{n, n} - \sum_{j=1}^P\sum_{k=1}^PU_{n,j}S_{n, j, k}U_{n, k} \\
    	    W_{n, j} = \frac{1}{D_{n, n}}\left[V_{n, j} - \sum_{k=1}^PU_{n, k}S_{n, j, k}\right],
    	\end{eqnarray}
    	where $S_{1, j, k}$ is a matrix of zeros and $P$ is both the rank 
    	of the semiseparable covariance matrix and the number of columns 
    	in $U$ and $V$, here equal to $2J$. In the original 
    	\celerite paper it was found that, in order to 
    	avoid numerical stability issues caused by the 
    	exponential factors in equation \ref{eqn:UV_defs}, it was 
    	necessary to redefine the generator matrices $U$ and $V$ 
    	and to define an additional matrix $\phi$ of the same 
    	dimensions as $U$ and $V$. The generators become
    	\begin{eqnarray}
    	    \label{eqn:tildes}
    	    \tilde U_{n, 2j-1} &=& a_j\cos(d_jt_n) + b_j\sin(d_jt_n) \cr 
			\tilde U_{n, 2j} &=& a_j\sin(d_jt_n) - b_j\cos(d_jt_n) \cr
			\tilde V_{m, 2j-1} &=& \cos(d_jt_m) \\
			\tilde V_{m, 2j} &=& \sin(d_jt_m). \nonumber
		\end{eqnarray}
		The unknown matrix $W$ becomes 
		\begin{eqnarray}
			\tilde W_{n, 2j-1} &=& e^{-c_jt_n}W_{n, 2j-1} \cr
			\tilde W_{n, 2j} &=& e^{-c_jt_n}W_{n, 2j}.
    	\end{eqnarray}
    	And the new matrix $\phi$ is defined
    	\begin{eqnarray}
    	    \phi_{n, 2j-1} = \phi_{n, 2j} = e^{-c_j(t_n-t_{n-1})}.
    	\end{eqnarray}
    	The algorithm for decomposing the covariance matrix 
    	becomes
    	\begin{eqnarray}
    	    \label{eqn:cholesky_algorithm1}
    	    S_{n,j,k} &=& \phi_{n,j}\phi_{n,k}\left[S_{n-1, j, k} + D_{n-1, n-1}\tilde{W}_{n-1, j}\tilde{W}_{n-1,k}\right] \\
    	    D_{n, n} &=& A_{n, n} - \sum_{j=1}^P\sum_{k=1}^P\tilde U_{n, j}S_{n, j, k}\tilde U_{n, k} \\
    	    \label{eqn:cholesky_algorithm2}
    	    \tilde W_{n, j} &=& \frac{1}{D_{n, n}}\left[\tilde V_{n, j}-\sum_{k=1}^P\tilde U_{n, k}S_{n, j, k}\right].
    	\end{eqnarray}
    	
    	This completes our recap of the one-dimensional version of \celerite; next we describe our novel 2D version.
    	
    \subsection{Computing the two-dimensional GP} \label{sec:outer_product_method}
        
        We now consider the Cholesky decomposition 
        of the covariance matrix for a two-dimensional 
        GP when the covariance in the second 
        dimension can be written as the outer 
        product of a vector with itself.  This form of the covariance applies when the correlated component of the noise has the same shape along the first large dimension (of size $N$) and varies proportionally in amplitude along the second small dimension (of size $M$), as is the case for the multiwavelength stellar variability problem discussed above.
        
        This 
        covariance matrix is given by equation 
        \ref{eqn:kronecker_cov}, reproduced here: 
        \begin{equation}
            K = \Sigma + T\otimes R,
        \end{equation}
        which has  size $N'{\times} N' = N M {\times} NM$.
        Here $\Sigma$ is a diagonal matrix 
        containing the white noise components for 
        each datapoint, which may be heteroskedastic, 
        $T$ is the covariance matrix in the 
        first dimension, which must be defined 
        by a \celerite kernel, and 
        $R$ is the covariance matrix for the 
        second dimension which must be an outer 
        product of the form 
        \begin{equation}
            R = \bvec{\alpha}\bvec{\alpha^\T}, 
        \end{equation}
        where $\bvec{\alpha} = (\alpha_1, \alpha_2, \dots, \alpha_M)^\T$ is a vector of length $M$. 
        
        Writing $K$ in terms of the celerite 
        generator matrices from 
	        equation \ref{eqn:KUV}:
	        \begin{eqnarray}
	            K &=& \Sigma + \left[A_0 + \mathrm{tril}(UV^\T) 
	            + \mathrm{triu}(VU^\T)\right] \otimes R \cr
	            &=& \Sigma + \mathrm{diag}(A_0\otimes R) \cr
	            &+& \mathrm{tril}(UV^\T\otimes R) 
	            + \mathrm{triu}(VU^\T\otimes R),
	        \end{eqnarray}
	        where $A_0$ is the 
	        diagonal component of $T$ obtained by setting 
	        $\sigma_n = 0$ for all $n \in {1,\dots,N}$ in equation \ref{eqn:A_matrix}.
	        Substituting the outer product $\bvec{\alpha}\bvec{\alpha}^\T$ 
	        for $R$ inside the upper and lower triangular operators we 
	        have
	        \begin{eqnarray}
	            K = \Sigma &+& \mathrm{diag}(A_0\otimes R) \cr
	            &+& \mathrm{tril}(UV^\T\otimes \bvec{\alpha}\bvec{\alpha}^\T) \cr
	            &+& \mathrm{triu}(VU^\T\otimes \bvec{\alpha}\bvec{\alpha}^\T).
	        \end{eqnarray}
	        Applying the formula for mixed Kronecker and 
	        matrix products,
	        \begin{equation}
	            \label{eqn:mixed_products}
	            (AB)\otimes(CD) = (A\otimes C)(B\otimes D),
	        \end{equation}
	        we can rewrite the covariance matrix as 
	        \begin{eqnarray}
	            K = \Sigma &+& \mathrm{diag}(A_0\otimes R) \cr
	            &+& \mathrm{tril}((U\otimes \bvec{\alpha})(V\otimes \bvec{\alpha})^\T) \cr
	            &+& \mathrm{triu}((V\otimes \bvec{\alpha})(U\otimes \bvec{\alpha})^\T).
	        \end{eqnarray}
	        We now see that the two-dimensional covariance matrix 
	        has exactly the same semi-separable structure as 
	        the one-dimensional covariance matrix with new 
	        definitions of the generator matrices in terms of 
	        their Kronecker products with $\bvec{\alpha}$:
	        \begin{eqnarray}
	            \label{eqn:kronecker_generators_outer}
	            A' &=& \Sigma + \mathrm{diag}(A_0\otimes R) \cr
	            U' &=& U\otimes \bvec{\alpha} \\
	            V' &=& V\otimes \bvec{\alpha} \nonumber
	        \end{eqnarray}
	        
	        The components of the refactored generator matrices, 
	        corresponding to equation \ref{eqn:tildes}, are now 
	        given by 
	        \begin{eqnarray}
    	        \label{eqn:kronecker_tildes}
    	        \tilde U'_{M(n-1)+p, 2j-1} &=& \alpha_p(a_j\cos(d_jt_n) + b_j\sin(d_jt_n)) \cr 
			    \tilde U'_{M(n-1)+p, 2j} &=& \alpha_p(a_j\sin(d_jt_n) - b_j\cos(d_jt_n)) \cr
			    \tilde V'_{M(m-1)+p, 2j-1} &=& \alpha_p\cos(d_jt_m) \\
			    \tilde V'_{M(m-1)+p, 2j} &=& \alpha_p\sin(d_jt_m), \nonumber
		    \end{eqnarray}
		    
		    and $\phi'$ is given by 
		    \begin{eqnarray}\label{eqn:phi_2d}
    	        \phi'_{M(n-1)+p, :} = \left\{\begin{matrix}
    	            e^{-c_j(t_n-t_{n-1})} & p=1 \\
    	            1 & p > 1
    	        \end{matrix}\right. , 
    	    \end{eqnarray}
        	with $n,m \in {1,\dots,N}$, $p\in 1,\dots, M$, and the colon indicating 
    	    that the element is identical for every entry of that 
    	    row. For these definitions of the generator matrices the 
    	    recursive Cholesky decomposition algorithm becomes 
    	    \begin{eqnarray}
    	        \label{eqn:cholesky_algorithm_final}
    	        S_{n,j,k} &=& \phi'_{n,j}\phi'_{n,k}\left[S_{n-1, j, k} + D_{n-1, n-1}\tilde{W}_{n-1, j}\tilde{W}_{n-1,k}\right], \cr
    	        D_{n, n} &=& A'_{n, n} - \sum_{j=1}^{P}\sum_{k=1}^{P}\tilde U'_{n, j}S_{n, j, k}\tilde U'_{n, k}, \\
    	        \tilde W_{n, j} &=& \frac{1}{D_{n, n}}\left[\tilde V'_{n, j}-\sum_{k=1}^{P}\tilde U'_{n, k}S_{n, j, k}\right], \nonumber
    	    \end{eqnarray}
    	    where again $P$ is the number of columns in $U'$ and $V'$.
    	    
    	    The recursive algorithm defined above 
    	    requires one pass through each of the $N' = NM$ rows 
    	    of $\tilde U'$ and $\tilde V'$. At each step we 
    	    compute a double sum over the $P$ columns of these matrices. 
    	    The resultant scaling is thus $\mathcal{O}(NMP^2)$. For 
    	    the outer-product definition of $R$ we have $P=2J$ and the method 
    	    scales as $\mathcal{O}(NMJ^2)$ (see figure \ref{fig:benchmarks} for benchmarks). 
    	    
    	    As shown in 
    	    appendix \ref{sec:algorithm}, we can 
    	    come up with similar definitions of $\tilde U'$, $\tilde V'$, 
    	    and $\phi'$ for arbitrarily defined $R$ which yield $P=2JM$, 
    	    allowing us to compute the Cholesky decomposition 
    	    in $\mathcal{O}(NJ^2M^3)$.
    	    Algorithms for computing the likelihood function, computing predictions 
    	    or extrapolations from the GP, 
    	    and sampling the GP are given in appendices \ref{sec:likelihood}, 
    	    \ref{sec:prediction}, and \ref{sec:sampling} respectively for 
    	    both outer-product and arbitrary 
    	    definitions of $R$.
    	    
    	    For this two-dimensional GP the set of observations used to 
    	    compute the GP likelihood is also two-dimensional. Actual 
    	    computation of the likelihood however requires that the 
    	    input be reduced to one dimension. The Kronecker structure 
    	    of the covariance matrix determines the form of the 
    	    vector of observations. For input defined on 
    	    a grid of size $\bvec{t}{\times} \bvec{r}$ where $\bvec{t}$ 
    	    represents the dimension along which the covariance is 
    	    described by a celerite kernel and $\bvec{r}$ represents the second small dimension, we have a 2D matrix of 
    	    observations:
    	    \begin{equation}
    	        Y_{i, j} = y(r_i, t_j).
    	    \end{equation}
    	    We define the observation vector to be 
    	    \begin{equation}
    	        \bvec{y} = \mathrm{vec}(Y),
    	    \end{equation}
    	    where $\mathrm{vec}(Y)$ is the concatenation of the 
    	    rows of $Y$. In other words, 
    	    \begin{equation}
    	        \bvec{y} = \left(Y_{:, 1}, Y_{:, 2},\dots Y_{:, N}\right).
    	    \end{equation}

    With the description of our computational methods completed, we now turn to the results of transit simulations.

\section{Results}\label{sec:results}
    
    We have carried out an analysis of simulated transit light curves with a wide
    range of noise amplitudes, timescales, and ratios of correlated to
    white noise, which we summarize the results of here.  We start with a discussion of the results from a case study of seven examples with different ratios of correlated to white noise (\S \ref{sec:casestudies}), and then expand the discussion to a wider range of simulations for which we compare the Information matrix, analytic, and MCMC error analyses (\S \ref{sec:noisecomparison}).
    
     \begin{figure*}
            \epsscale{1.1}
            \includegraphics[width=\textwidth]{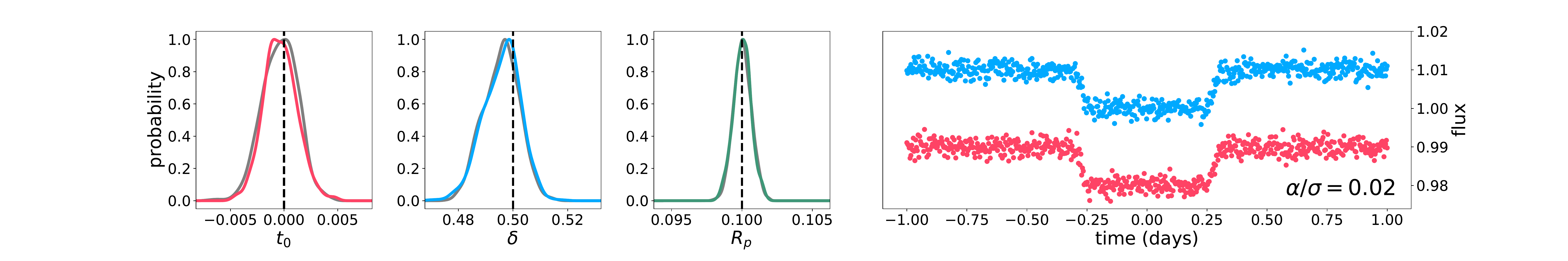}
            \includegraphics[width=\textwidth]{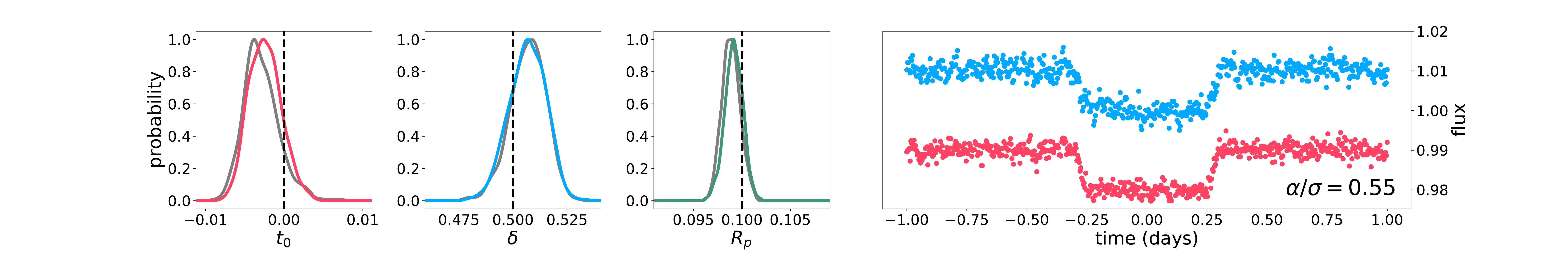}
            \includegraphics[width=\textwidth]{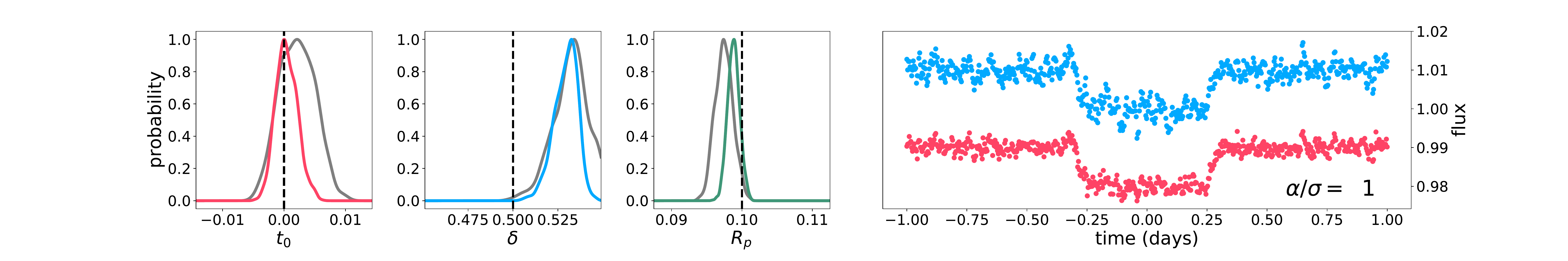}
            \includegraphics[width=\textwidth]{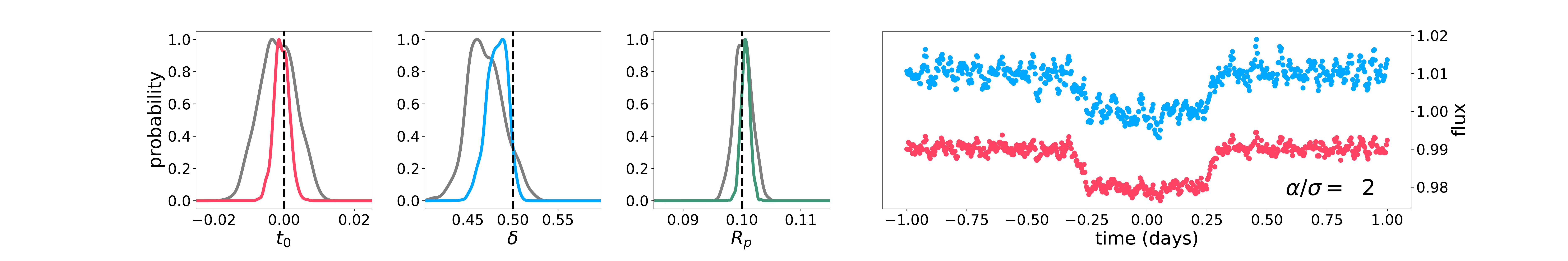}
            \includegraphics[width=\textwidth]{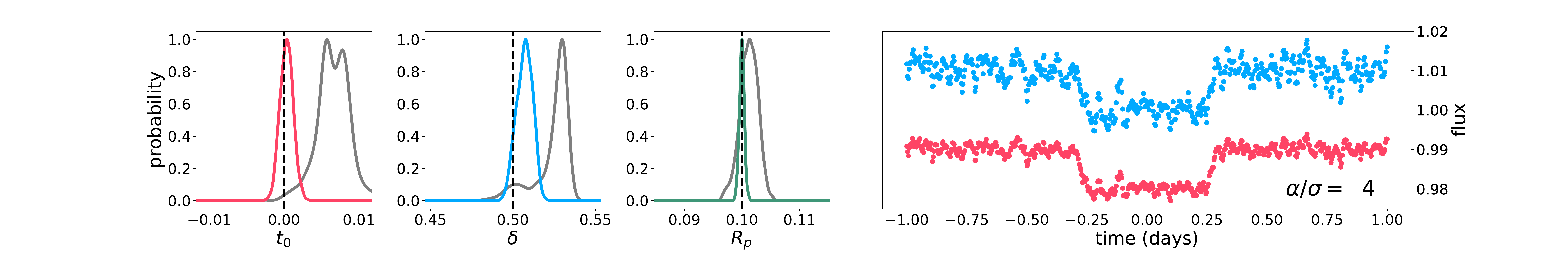}
            \includegraphics[width=\textwidth]{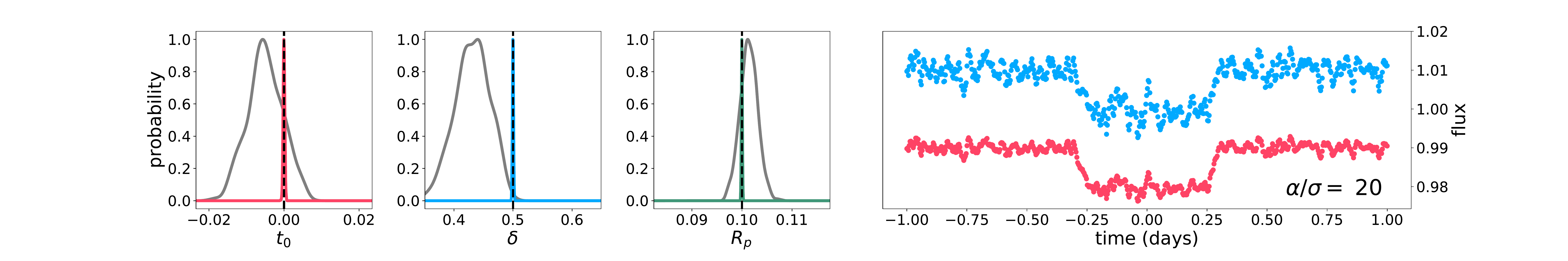}
            \includegraphics[width=\textwidth]{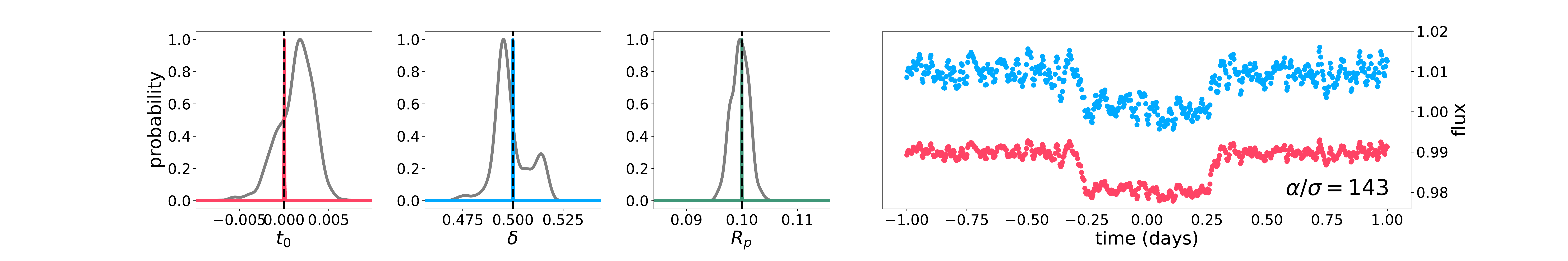}
            \caption{\textbf{Left:} Posteriors 
            for three transit parameters estimated 
            by MCMC analysis on the two band (colored) and 
            single band (gray) data. Posteriors are smoothed 
            using Gaussian kernel density estimation for $\omega_0 \delta = 100$ 
            (corresponding to the final panel of Figure \ref{fig:fisher}). 
            From left to right: the 
            center of transit $t_0$, transit 
            duration $\delta$, and radius 
            ratio $R_p/R_*$. For 
            $\alpha/\sigma = 20$ and $\alpha/\sigma = 143$ 
            the posterior distributions for the 
            two-band case 
            are too sharply peaked to be visible. \textbf{Right:} 
            Representative light curves for 
            each value of the noise amplitude 
            ratio $\alpha/\sigma$ zoomed in 
            on the transit signal (the input light 
            curves have a duration of 
            10 days). }
            \label{fig:posteriors}
        \end{figure*}
        
    \subsection{Case studies}\label{sec:casestudies}
    
    To start with, Figure 
    \ref{fig:posteriors} shows seven examples of our simulations for two bands with correlated noise amplitudes which differ by a ratio of two.  These were made with moderate signal-to-noise and with
    $\omega_0 \delta = 100$, which corresponds to a characteristic timescale of the correlated noise which is shorter than the transit duration and 
    the ingress/egress timescales (regime III). In this case we held the white noise in the two bands to be identical in amplitude (corresponding to an identical photon count rate in both bands), and we compared a joint analysis of the two bands (we refer to this as ``polychromatic") with an analysis of a single band consisting of the sum of the same simulated light curves from the two bands (this analysis we refer to as ``monochromatic"). Across these simulations we have varied the ratio of the total correlated noise to the white noise, $\alpha/\sigma$, over seven values,
    $\{0.02, 0.55, 1, 2, 4, 20, 143\}$, to examine the precision of the two-band analysis compared with a monochromatic analysis.

    For the first two simulations, $\alpha = 0.02\sigma$ and $\alpha = 0.55\sigma$, the 
    variance of the correlated noise is smaller than 
    that of the white noise. At this low ratio of 
    $\alpha/\sigma$ we find that the measurement of the transit 
    depth and timing parameters is about the 
    same in the two-band case as in the monochromatic 
    case (top panel, 
    Figure \ref{fig:posteriors}). In the 
    third panel where the white and correlated 
    noise amplitudes are equal, we see a 
    slight improvement in the measurement of 
    the transit time and depth. In the 
    remaining panels (bottom four 
    panels of \ref{fig:posteriors}), we find 
    an increasing degree of improvement 
    in all the measured parameters as 
    $\alpha$ increases relative to $\sigma$. 
    As we approach the 
    small white noise limit the improvement 
    in all parameters between 
    the single-band and two-band analyses 
    is dramatic, with the transit depth 
    improving by a factor of $18$ at 
    $\alpha = 20\sigma$ and by a factor 
    of $118$ at $\alpha = 143\sigma$. The 
    transit time measurement improves by 
    a factor of $21$ and $65$ respectively 
    for these simulations. 
    This improvement results from the ability to distinguish correlated noise variations from the transit signal when two bands are utilized, thanks to the different amplitudes of the correlated noise in the two bands; the correlated noise variations are measured to high precision in this case due to the small photon noise.  Even so, the precision of the transit parameters is worse than it would be if there were no correlated noise by a factor of $\sqrt{10}$.  This is an astrophysical limitation, and yet it still demonstrates a dramatic improvement in the analysis which splits the the photons into two bands versus a single summed band.  
    
    The intermediate values of $\alpha/\sigma = \{1, 2, 4\}$ shown in Figure \ref{fig:posteriors} have a behavior which is intermediate between the high white noise and low white noise limits that we discuss above: a monotonic improvement in all of the measurements with the increase in $\alpha/\sigma$.
    
    The general trends of these simulations hold over a broader range of parameters.  To examine a larger number of cases, we summarize the uncertainties of the monochromatic cases and polychromatic cases based on the measurement precision as a function of the noise parameters, which amounts to measuring the breadth of the posterior distributions inferred for each parameter (left-hand panels of Figure \ref{fig:posteriors}).  We also compare these to the uncertainty estimates using the  Information matrix approach and the analytic estimates given in \S \ref{sec:fisher} and  \S \ref{sec:analytical_estimates}, which we discuss next.
    
    \subsection{Noise comparison}\label{sec:noisecomparison}
    
\begin{figure}
            \includegraphics[width=\hsize]{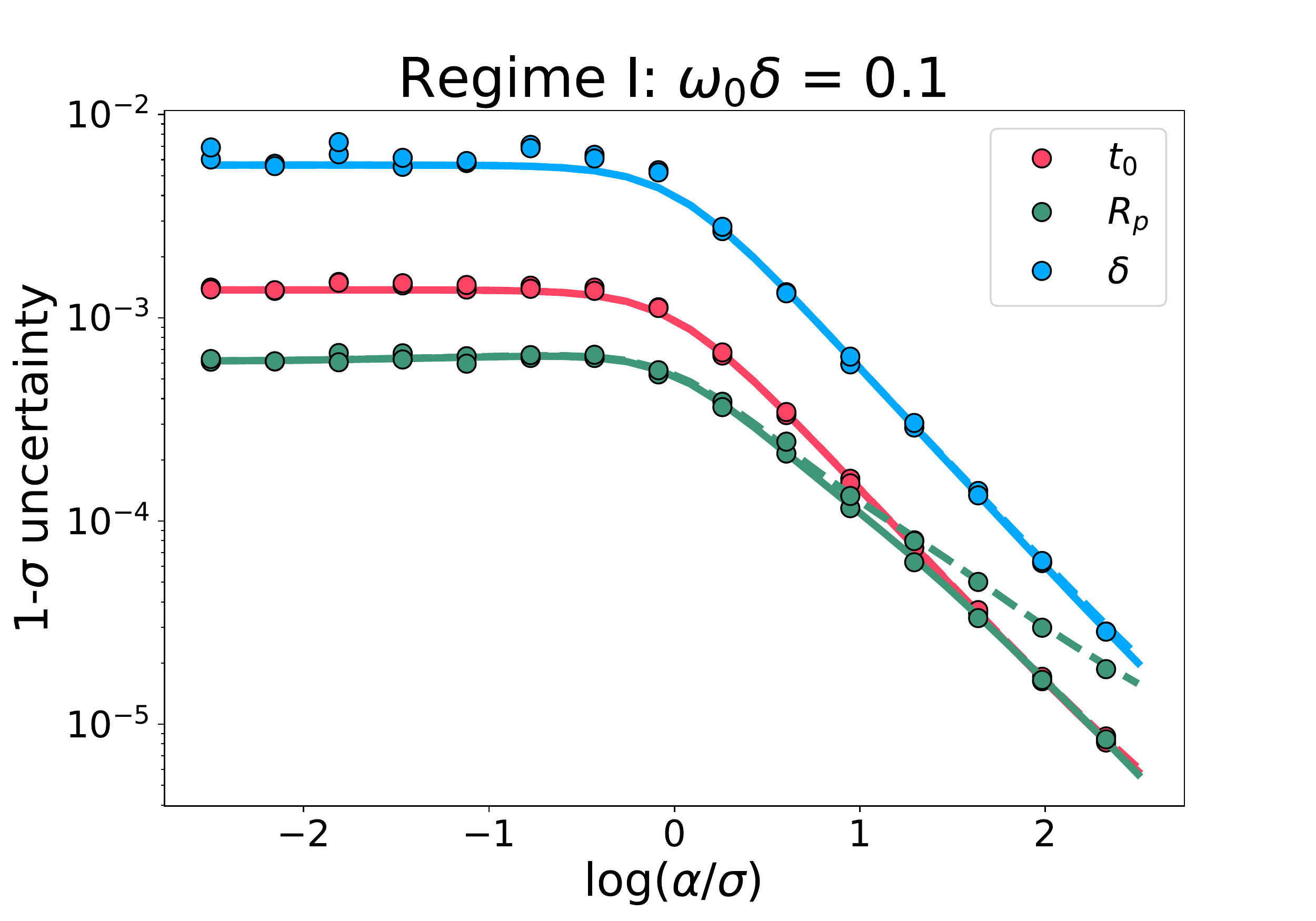}
            \includegraphics[width=\hsize]{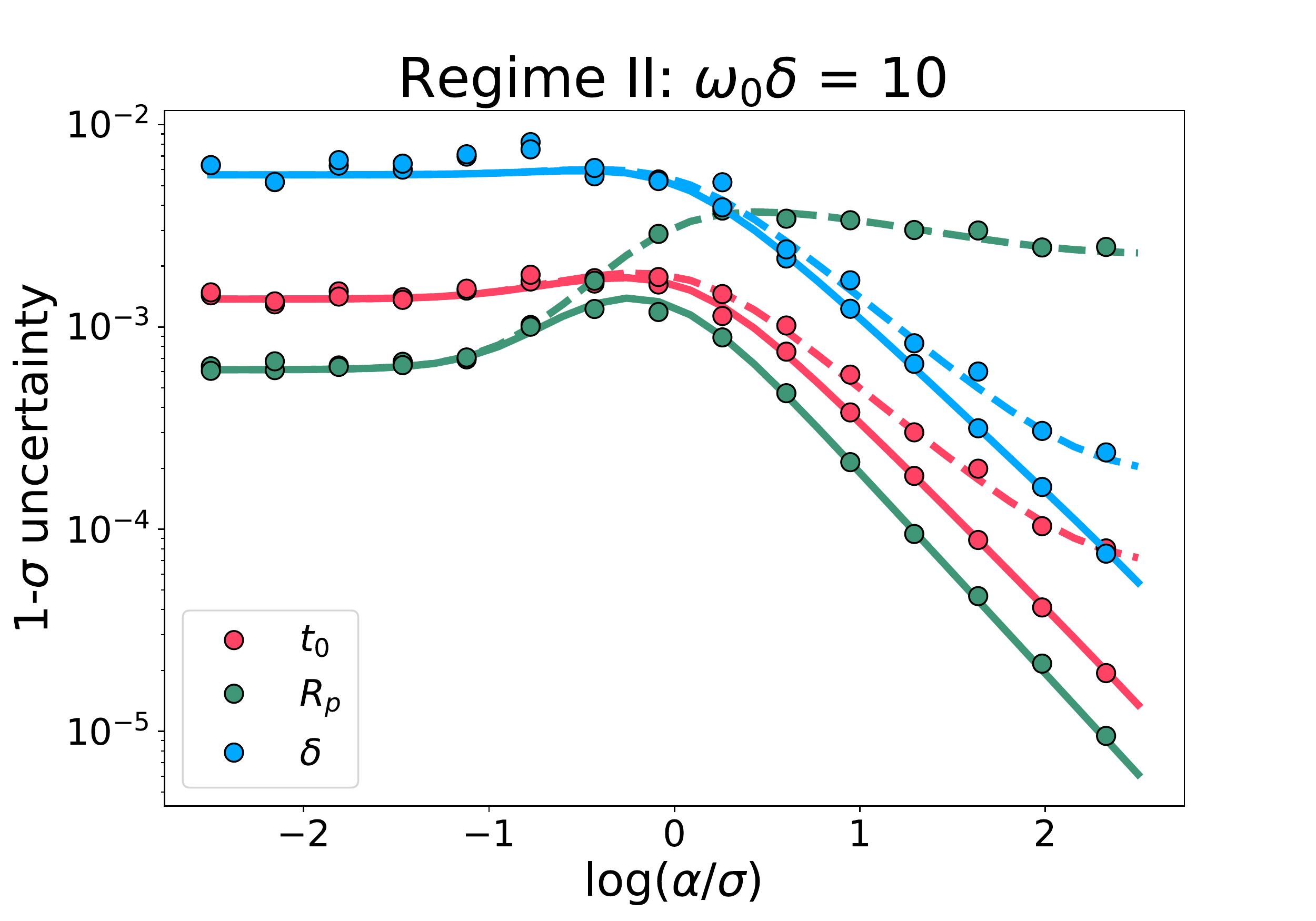}
            \includegraphics[width=\hsize]{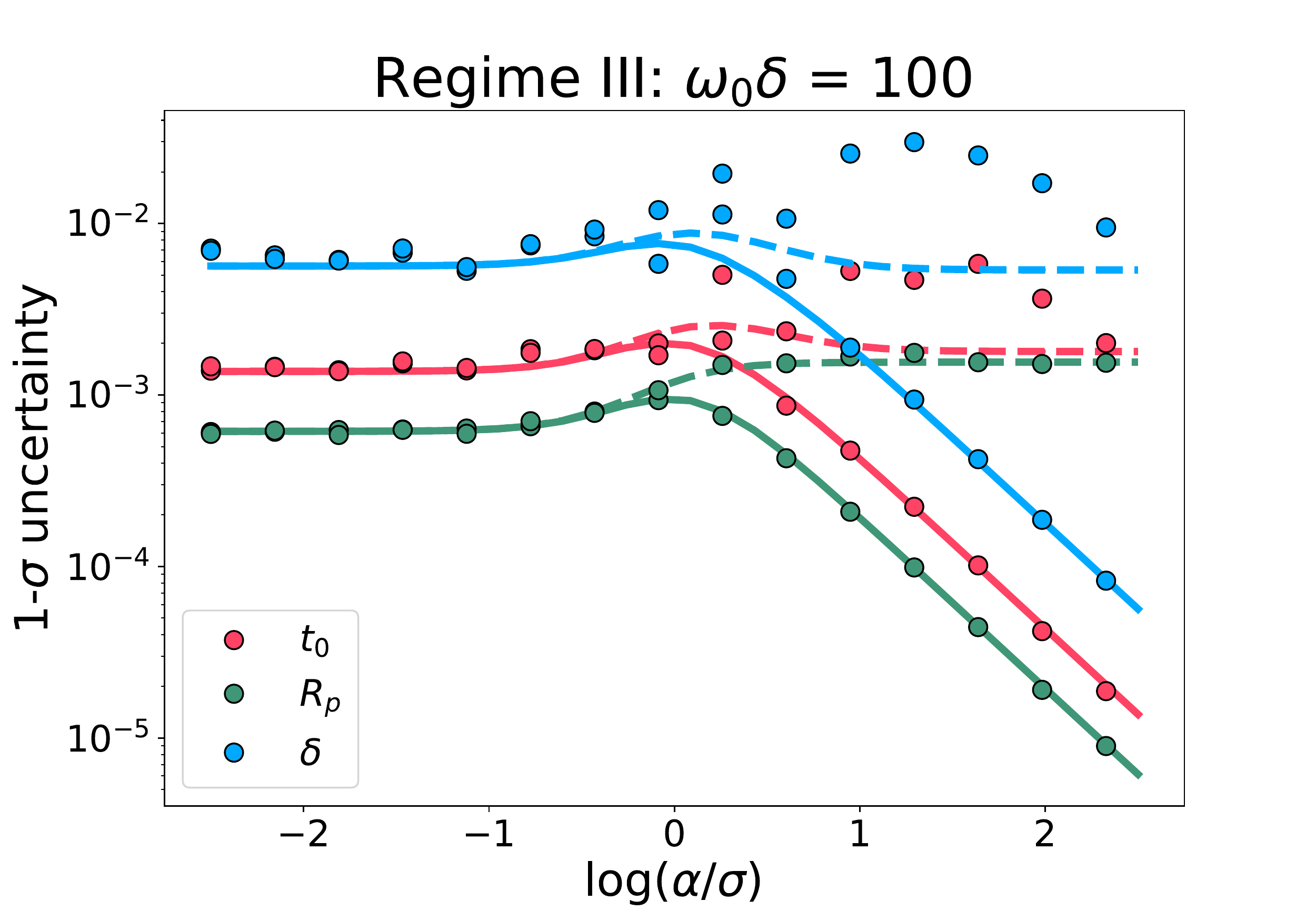}
            \caption{Information uncertainty curves overlayed with 
            MCMC uncertainty estimates for trapezoidal 
            transit parameters. Dashed lines show results for 
            the monochromatic noise model and solid lines show 
            results for the two-band noise model. Circles represent 
            the MCMC uncertainty for distinct realizations of 
            the noise and transit.}
            \label{fig:fisher}
        \end{figure}

    We have carried out a much broader parameter study, varying the ratio of $\alpha/\sigma$ over a wide range of values for three values of the timescale:  $\omega_0 \delta = 0.1, 10$, and 100.  We compare the Information matrix analysis against the MCMC anlysis in the monochromatic case with the two-band case, also with $\alpha_2=2\alpha_1$, in Figure \ref{fig:fisher}.  
    The MCMC uncertainty estimates agree closely with 
    the Information uncertainty curves for almost all of our simulations, 
    as demonstrated by Figure \ref{fig:fisher} for moderate signal-to-noise.
    
    In regime I, $\omega_0\delta = 0.1$, in which the characteristic variability 
    timescale is longer than the transit duration, the 
    uncertainties on the transit parameters are nearly 
    identical between the monochromatic and multiband 
    simulations up to $\alpha/\sigma \approx 10$, where 
    the multiband uncertainties begin to 
    diverge slightly from the monochromatic uncertainties. 
    Since the transit signal is distinguishable from the 
    noise on the basis of its duration alone, the amount of 
    additional information contained in the inter-band 
    correlation is insignificant and both models 
    perform similarly well. 
    
    We now skip to regime III, with $\omega_0\delta = 100$ (the same as the case studies in the prior subsection), in which the characteristic 
    variability timescale is smaller than the 
    transit duration.  
    Because the SHO power spectrum 
    allocates equal power to all oscillations on 
    timescales longer than $1/\omega_0$, the 
    transit signal is not distinguishable from 
    the variability on the basis of its duration. 
    In this case the inter-band correlation 
    contains the additional information necessary to 
    correctly infer transit parameters. 
    Both models perform similarly when the 
    correlated noise amplitude is small compared to 
    the white noise, but when the correlated noise amplitude 
    $\alpha$ begins to dominate over the white noise 
    $\sigma$ the monochromatic model does a poor 
    job of inferring parameters (as evidenced by the 
    large uncertainties) while the multiband model 
    infers more and more precise values as the white 
    noise decreases relative to the correlated noise. 
    
    The results for regime II, here represented 
    by $\omega_0\delta = 10$, fall intermediately 
    between regimes I and III. In regime II, the 
    characteristic timescale of the 
    variability falls between the transit duration 
    and the ingress/egress timescale so 
    that measurements of the transit 
    duration must contend with correlated 
    noise on the same timescale, whereas 
    measurements of the ingress and egress 
    are affected primarily by white noise 
    rather than correlated noise. Since the 
    transit time is constrained 
    by the ingress and egress times rather than 
    by the transit duration, measurements of 
    $t_0$ are also primarily affected by white 
    noise. This is why we see significant 
    improvement in the measurement of 
    the transit depth at high $\alpha$ and 
    low $\sigma$ between the single-band 
    and two-band simulations, while the timing parameters show much less improvement until 
    we reach the low white noise limit. At this 
    point the white noise amplitude is small 
    enough compared to the correlated noise 
    amplitude that the relatively low 
    correlated noise on the timescale of the 
    ingress/egress duration does begin to 
    interfere with timing measurements in the 
    single-band case.
    
\begin{figure*}
            \centering
            \includegraphics[width=18cm]{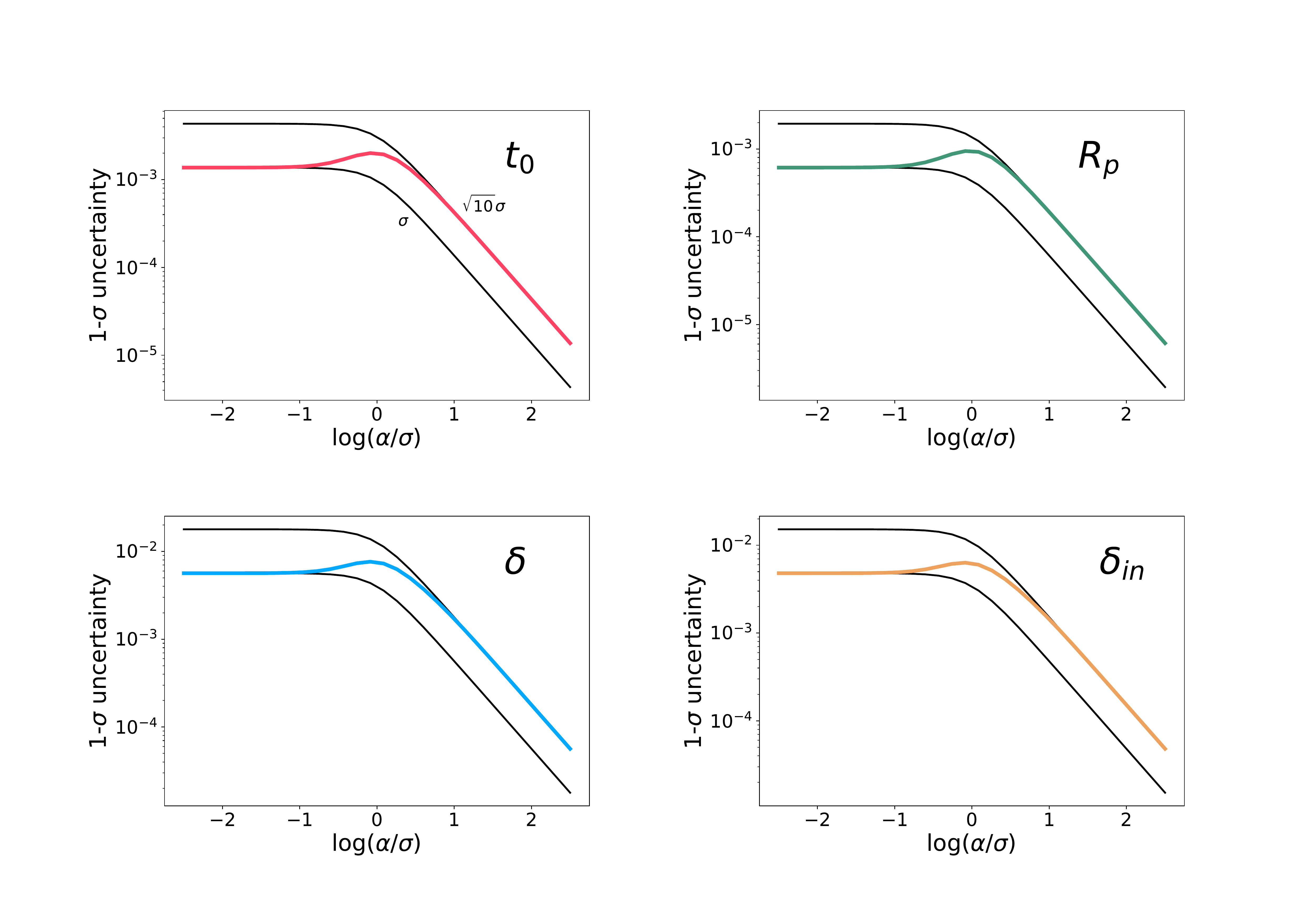}
            \caption{Information uncertainty curves (colored lines) 
            for the two-band 
            model compared to the white noise-only versions 
            of the corresponding monochromatic noise model (black lines) in 
            regime III. 
            For the white noise-only 
            models we set the correlated noise amplitude to 
            zero and leave all other parameters the same as 
            the monochromatic model. As we transition from 
            the white noise dominated to the correlated noise 
            dominated regimes the Information uncertainty curves 
            for the two-band model transition from following 
            the white noise model with $\sigma'=\sigma$ to 
            the white noise model with $\sigma'=\sqrt{10}\sigma$. 
            In effect perfect knowledge of the two-band 
            correlated noise hyperparameters allows us to 
            recover transit parameters at the same precision 
            as if the correlated noise were simply white noise 
            with a $\sqrt{10}$ larger amplitude.}
            \label{fig:white_noise_comparison}
        \end{figure*}

    In Figure \ref{fig:white_noise_comparison} we plot 
    Information uncertainty curves in regime III for the 
    multiband model against 
    Information uncertainty curves for the monochromatic 
    model having the same transit 
    parameters but with only a white noise 
    component --- the correlated noise amplitude 
    is set to zero. The colored curves representing 
    the Information uncertainties for the full noise 
    model (white and correlated noise) match 
    the white noise-only uncertainty in the 
    limit that the correlated noise component 
    is very small, as expected. As we increase the 
    relative amplitude of the correlated noise component 
    the uncertainty for the full model jumps from 
    the white noise-only curve with the same white noise 
    amplitude to the white noise-only curve with 
    $\sqrt{10}$ times greater amplitude. As the 
    correlated noise amplitude further increases, 
    the Information uncertainty for the full model 
    behaves as though we're doing inference on 
    an equivalent model with the correlated noise 
    component exchanged for a larger white noise 
    amplitude. 
    
    The behavior seen here is explained by the analytical 
    model outlined in Section \ref{sec:analytical_estimates}. 
    In particular, Equation \ref{eqn:sig_poly_limit} explains 
    why the uncertainty scales as the white noise-only 
    uncertainty with $\sqrt{10}\sigma$ in the large 
    correlated noise limit for two bands with amplitudes 
    related by $\alpha_2 = 2\alpha_1$. 

This completes our description of the simulated light curves and the results from these simulations.  We next discuss the implications of these results. 
    
\section{Discussion}\label{sec:discussion}

        We have demonstrated the application of our method to 
        the problem of fitting a transit observed 
        in multiple bands in the presence of 
        correlated noise. We now revisit and summarize 
        the results of that demonstration before 
        outlining some other potential applications of 
        our method. 
        
        Monochromatic transit observations are ill-equipped 
        to deal with correlated noise, as the 
        wavelength-integrated flux does not provide 
        enough information to distinguish between 
        transits and noise features except when 
        the correlated noise amplitude is low on the 
        timescale of the transit duration. 
        When transits occur on timescales 
        similar to or longer than the variability timescale 
        we must rely on the spectral dimension to 
        provide the information necessary to 
        distinguish between the two. 
        
        We use the Information matrix to explore 
        the difference between inference on a monochromatic 
        noise model and a multiband model 
        with wavelength-dependent variability. 
        We construct sets of 
        monochromatic and multiband models with identical 
        noise properties by splitting a given number of 
        photons per wavelength into different spectral 
        bins. We find that our results depend strongly 
        on the timescale of the noise with respect to the 
        transit duration. When the timescale of the 
        correlated variability is much longer than the 
        transit duration the monochromatic and multiband 
        models perform similarly, though the multiband 
        model still allows us to infer slightly more 
        precise parameters in the limit that the correlated 
        noise amplitude is much larger than the white noise 
        amplitude (see Figure \ref{fig:fisher}). 
        
        For the noise regime in which the correlated 
        variability timescale is similar to or shorter 
        than the transit duration we summarize our results 
        as follows:
        
        \begin{itemize}
            \item As the white noise amplitude 
            decreases and the correlated noise amplitude 
            increases, the precision inferred by the 
            monochromatic noise model stays approximately 
            constant, getting slightly worse for 
            the radius ratio but improving slightly 
            for the timing parameters $\delta$ and 
            $t_0$. In contrast, the precision inferred 
            by the multiband noise model improves as 
            the white noise amplitude decreases even 
            with increasing correlated noise amplitude. 
            The increase in precision scales the same 
            as if the correlated noise were held constant. 
            The presence of correlated noise simply decreases 
            the precision of the parameters by a constant 
            factor which is related to the form of the 
            variability as a function of wavelength. 
            
            \item Most of the benefits of the multiband 
            noise model can be realized by splitting the 
            monochromatic variability into just two bands, 
            but more bands achieve slightly better precision (see 
            Figure \ref{fig:fisher_nbands})
            
            \item In the limit that 
            we approach an infinitely high-resolution 
            spectrum we can derive the factor by which 
            the precision of the transit parameters 
            is worse than the case where there is 
            no correlated variability. Using equation 
            \ref{eqn:sig_poly} we find that the precision 
            inferred in the presence of correlated noise 
            is worse than in the white noise-only case by 
            a factor of 2 when the variability amplitude 
            scales linearly with cumulative photon counts with wavelength and 2.2 when 
            the variability amplitude is distributed according 
            the the blackbody distribution. In other words, 
            in the presence of linearly scaling correlated 
            variability amplitudes, we need four times 
            as many photons to achieve the same precision in 
            the presence of correlated noise as can be 
            achieved when there is only white noise, 
            provided we use a multiband noise model to 
            do our inference. 
            
        \end{itemize}
        
        \subsection{Low Transit SNR Limit}
            
            The limit where the transit depth is small 
            compared to the correlated noise amplitude 
            is important if we are interested in detecting 
            planets with small radii, or rocky planets 
            around sun-like stars. The Information matrix
            analysis above was done in the high SNR limit, 
            because that is the limit in which the  
            Information matrix can be shown to approximate the 
            uncertainty on model parameters. We now include 
            results on the correspondence between the 
            Information matrix and MCMC uncertainties in 
            the low SNR limit. Since we are primarily 
            interested in the correlated noise component, 
            we use SNR to refer to the ratio of the 
            transit depth to the correlated noise amplitude.
            
\begin{figure*}
                \plottwo{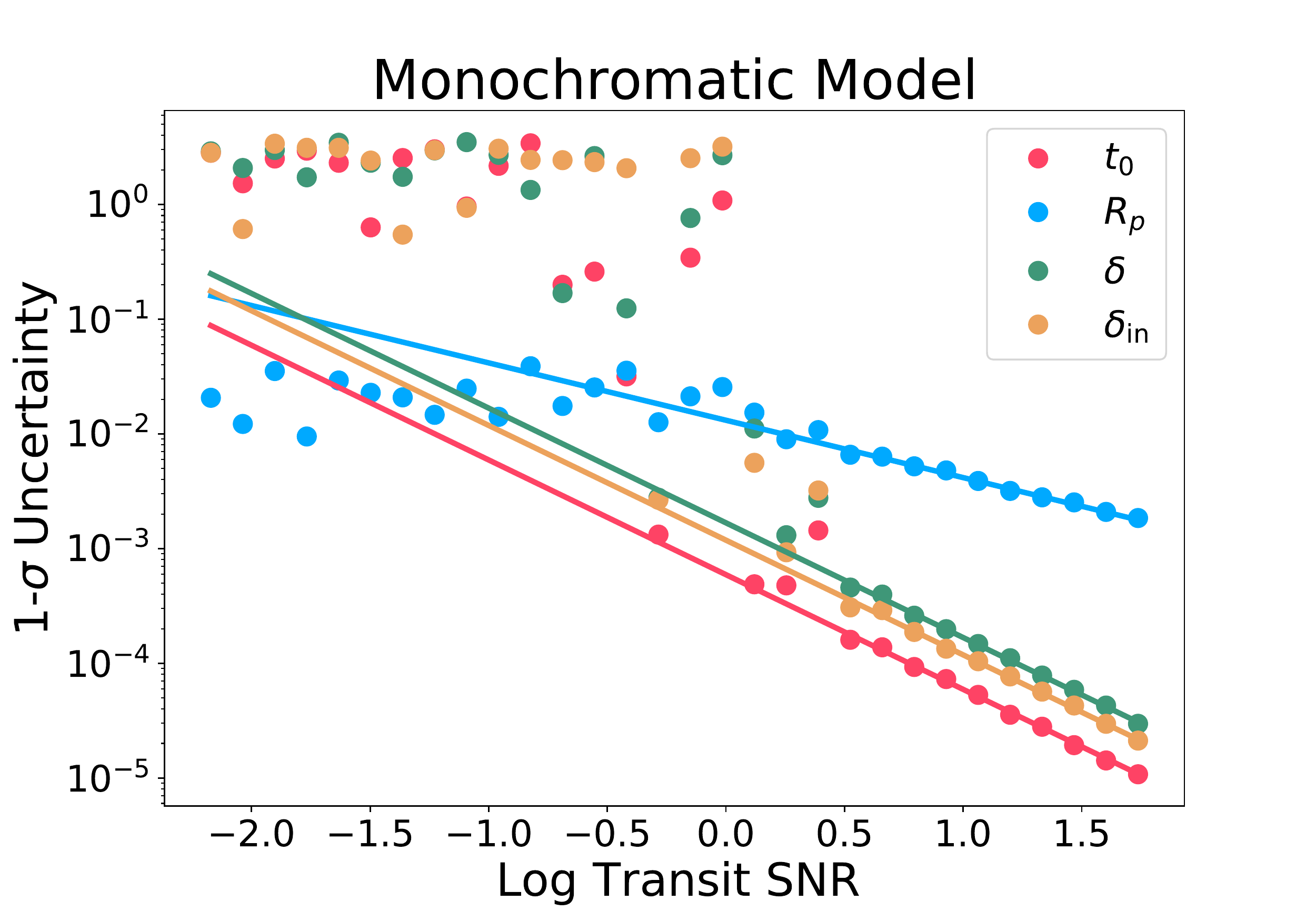}{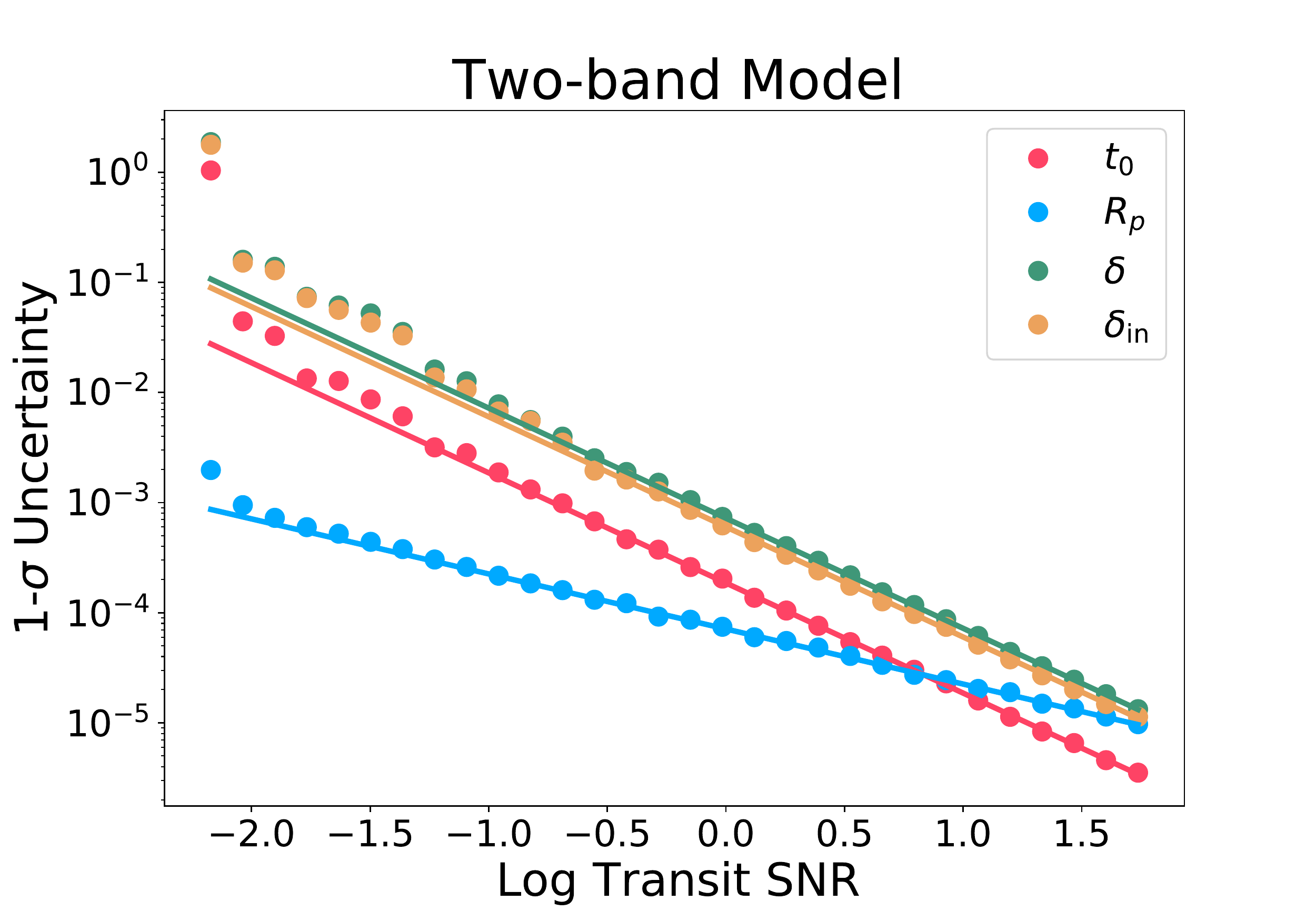}
                \caption{MCMC uncertainties (dots) and Information
                matrix
                uncertainties (lines) for monochromatic and two-band 
                noise models as a function of the transit SNR with $\alpha_2 = 2 \alpha_1$ for 
                the two-band simulations. 
                For these   simulations the correlated 
                noise is held constant 
                at 150 times the amplitude of the 
                white noise component and the total 
                noise defined to be the sum in quadrature 
                of the white noise and correlated 
                noise amplitudes 
                is conserved. The variability timescale $1/\omega_0 = \delta / 10$, placing these simulations in regime II. For the monochromatic model, 
                the Information and 
                MCMC uncertainties correspond down to an SNR of 
                about 10, which is the point at which the MCMC 
                simulations no longer converge to 
                the correct transit 
                solution, as evidenced by the scatter in MCMC 
                uncertainties at lower SNR. For the two-band 
                simulations the Information and MCMC uncertainties     correspond 
                down to an SNR of 1/100.}
                \label{fig:snr}
        \end{figure*}

            Figure \ref{fig:snr} shows the MCMC-derived 
            uncertainties and the Information uncertainties 
            for our four trapezoidal transit parameters 
            in both the monochromatic and two-band cases.
            We use a correlated noise to white noise 
            amplitude ratio $(\alpha/\sigma)$ of 150 
            for this portion of the analysis. 
            
            When we use a monochromatic model the Information 
            uncertanties diverge from the MCMC uncertainties 
            at a SNR of about 10. This corresponds to the 
            point at which the MCMC uncertainties jump 
            to very high values for the timing parameters, 
            indicating that the MCMC fails to converge 
            to the correct solution. 
            
            This contrasts strongly with the two-band 
            model. Using two bands the  Information 
            analysis finds the same uncertainty as the 
            MCMC analysis down to an SNR of about 
            $1/100$, for which the ratio of the transit depth to  the white noise is near unity.
            
\begin{figure*}
                \plottwo{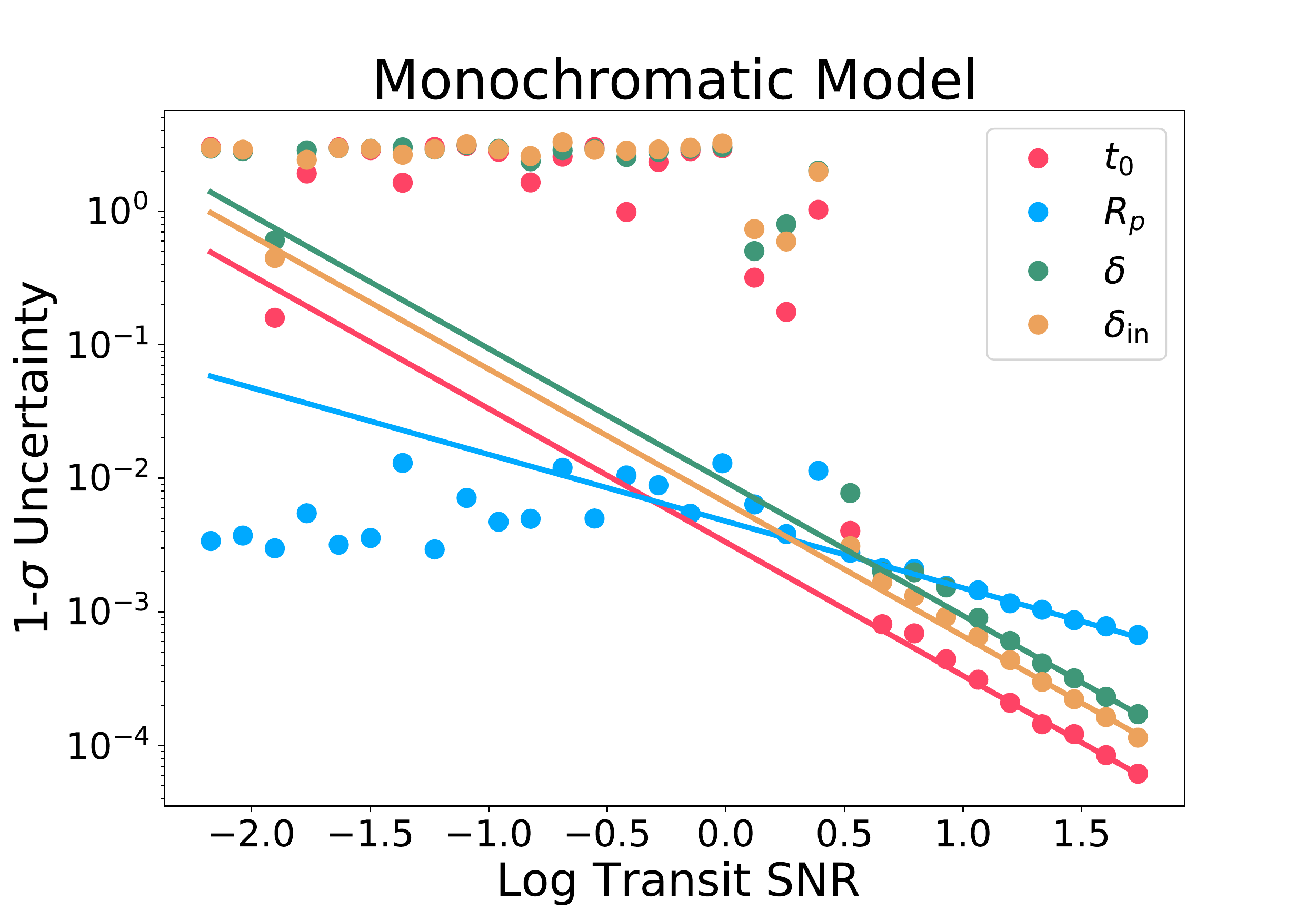}{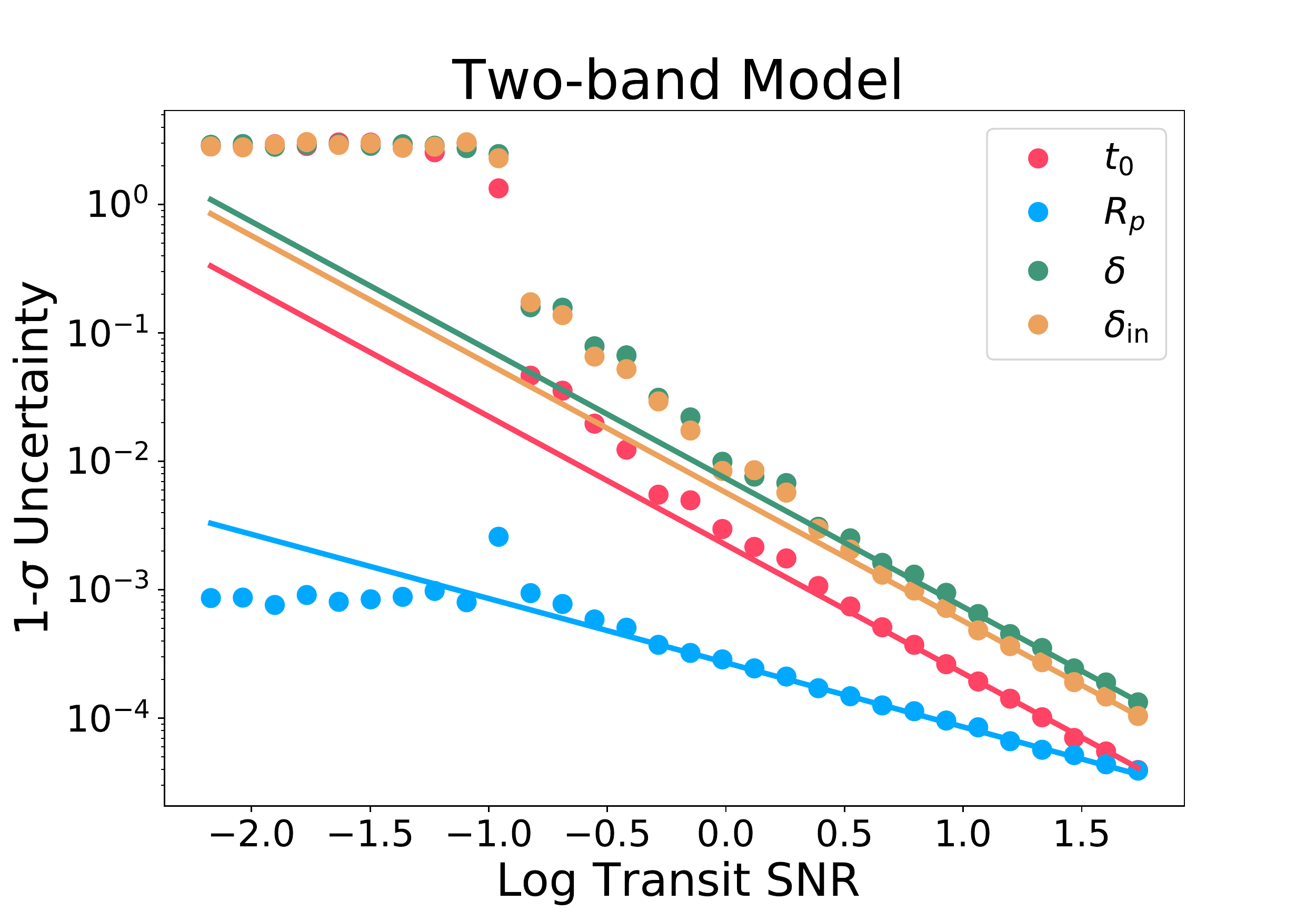}
                \caption{MCMC uncertainties (dots) and Information 
                uncertainties (lines) for monochromatic and two-band 
                noise models as a function of the transit SNR with wavelength 
                dependence specified $\alpha_2 = 2\alpha_1$ for the two-band simulations. 
                For these simulations the correlated 
                noise is held constant 
                at 10 times the amplitude of the 
                white noise component and the total 
                noise defined to be the sum in quadrature 
                of the white noise and correlated 
                noise amplitudes 
                is conserved. The variability timescale $1/\omega = \delta / 10$, placing these simulations in regime II. The larger white noise 
                component compared to figure \ref{fig:snr} 
                pushes the SNR limit below which the MCMC 
                and Information uncertainties diverge to higher 
                SNR. As before, there is an abrupt transition 
                at this limiting SNR where the MCMC suddenly 
                fails to converge to the correct transit 
                solution.}
                \label{fig:snr_wn}
        \end{figure*}

            In Figure \ref{fig:snr_wn} we repeat the 
            analysis for $\alpha/\sigma = 10$. With a 
            larger white noise component the MCMC 
            uncertainties diverge from the Information 
            uncertainties at a higher SNR. However, 
            the two-band model still outperforms the 
            monochromatic model with the MCMC 
            corresponding to the Information and 
            converging to the correct solution down 
            to an SNR of about $1/10$, where again the transit depth is comparable to the white noise.
            
            These results imply that the improvement resulting from multiple bands applies only when the signal is larger than the white noise, and in this limit, the Information matrix provides an adequate estimate of the uncertainties on the model parameters, assuming that the Gaussian Process parameters are well constrained as these were not varied in our analysis.  This approach may be used to estimate sensitivity and detection of transiting bodies, such as exomoons, discussed next.
            
        \subsection{Other applications}

        \ \\ \\
        \textbf{Exomoons}, or moons of exoplanets, are 
        are an oft-theorized 
        but thus far undetected object of interest both for their 
        ability to inform understandings of planetary formation 
        and for their potential habitability. While one 
        candidate exomoon, Kepler-1625b-i \citep{Teachey2018}, has 
        been identified it remains unconfirmed \citep{Teachey2019, Kreidberg2019}. 
        The saga of Kepler-1625bi illustrates one of the 
        primary barriers to observing exomoons: their small size and 
        correspondingly shallow transits. An additional complication is 
        that exomoon transits will not be strictly periodic, due 
        to orbital motion about their planets. This means that folding 
        the light curve on the planet's orbital period to increase 
        the signal-to-noise for a detection will not be effective. 
        
        Observations 
        designed for detecting transiting exomoons may likely need to consist of 
        very high signal-to-noise photometry of more than one transit 
        of a known exoplanet. In the near future JWST 
        will be the observatory best suited to 
        these observations \citep{Beichman2014}. It has the ability to 
        observe time series spectra of bright objects via 
        the NIRSpec instrument \citep{Bagnasco2007}. Our method is well-suited 
        to model these observations and we believe it 
        may end up being the optimal method of identifying an 
        exomoon transit signal.  Simulating JWST observations of transiting planet systems with realistic noise \citep{Sarkar2019}, while applying our multiwavelength GP model to the results, would reveal what sensitivity JWST would have to shallow transiting bodies such as exomoons.
        
        \ \\ \\
        \textbf{Transit transmission spectroscopy} aims to measure the transmission 
        spectrum of an exoplanet by measuring the effective 
        radius of the planet as a function of wavelength. This is 
        typically accomplished by varying the transit depth in the fit 
        to the time series photometry at each wavelength as in \citep{Berta2012} and
        \citet{Mandell2013}. In studies like these
        the effects of stellar variability have been minimal 
        and largely ignored. However in the future 
        high precision observations of 
        bright stars at optical and NIR wavelengths 
        will likely have to contend with variability 
        resulting from stellar granulation and/or   
        pulsations \citep{Sarkar2018}. 
        
        Our method offers an elegant means of measuring the transmission 
        spectrum. Given a 
        sufficiently long time baseline, the wavelength dependence
        of a star's variability can be arbitrarily well-determined. 
        In this case any ``leftover" variability --- variations in 
        transit depth that aren't explained by the wavelength-dependence 
        of the star's variability --- can be attributed to the 
        planet's transmission spectrum. By allowing the GP mean function 
        to vary in transit depth across wavelength during MCMC 
        analysis we can recover an estimate of the transmission 
        spectrum with uncertainties in the presence of 
        stellar variability.  As such, this is a straightforward extension of our model as the only change involves varying the depth and limb-darkening as a function of wavelength, while the covariance remains the same as in the examples we have already shown.
         
        \ \\ \\
        \textbf{Transit timing variations} occur when the gravitational 
        interaction between planets in a multi-planet system 
        perturbs a transiting planet away from a Keplerian 
        orbit \citep{Agol2005,Holman2005}. The perturbed planet will transit 
        earlier or later than the Keplerian solution would 
        dictate based on the relative position of the 
        transiting planet and perturbing planet. Observations 
        of these transit timing variations over the course 
        of many orbits help to constrain the orbital 
        parameters of the perturber as well as the masses 
        of both the perturber and the transiting planet. 
        A notable application of this technique is to 
        the seven-planet TRAPPIST-1 system \citep{Gillon2017, Grimm2018}.
        Correlated noise on timescales similar to the 
        ingress/egress time of a transit can 
        substantially affect measurements of the transit 
        time \citep{Agol2018}. 
        
        At present correlated noise 
        is observable for transiting planets around 
        evolved stars. A notable example is Kepler-91b 
        \citep{Barclay2015}, a hot Jupiter orbiting 
        a red giant. Individual transits of Kepler-91b 
        are nearly undetectable due to correlated noise 
        on similar timescales and amplitudes to the transit 
        signal. While most main-sequence Kepler targets 
        don't show significant correlated variability, 
        we expect that this variability will become 
        observable in the near future with the advent 
        of larger space-based telescopes such as JWST. 
        This means that accurate transit timing measurements 
        for small planets transiting main-sequence 
        stars will require the use of methods like 
        ours to overcome the effects of correlated noise. 
        
        \ \\ \\
        \textbf{Variable phenomena:}
        While we are primarily interested in the transiting planet 
        problem, our multiwavelength GP implementation 
        is likely to be useful for studies of other astronomical objects displaying time-correlated, stochastic variations. Many 
        subfields in astronomy make use of GP variability models, or
        stochastic models which are equivalent to a Gaussian process,
        including the study of eclipsing binaries \citep[e.g.][]{Mahadevan2019}, pulsating
        binaries \citep[e.g.][]{Hey2020}, X-ray variability of the logarithm
        of the flux of X-ray binaries and AGN \citep[e.g.][]{Uttley2005,Kelly2014}, 
        the study of transient phenomena such as supernovae \citep[e.g.][]{Kim2013}, 
        quasar variability \citep{Kelly2009,MacLeod2010}, reverberation mapping \citep{Zu2011,Pancoast2014}, and 
        gravitational lensing time delays \citep{Press1992,Hojjati2013,Hojjati2014}.  multiwavelength data may be exploited
        to better characterize these systems.  For example, \citet{Boone2019}
        found much better characterization of transients with multiwavelength
        Gaussian process modeling, while \citet{Peters2015} use the color-dependence
        of the time-correlation of quasar variability to better characterize their
        physical properties.  It is our hope that 
        some of these fields may benefit from applying our 
        new multiwavelength GP implementation to study the 
        wavelength-dependence of these various phenomena. 

    \subsection{Limitations of the method}
    
        When the second dimension's covariance matrix can 
        be represented in terms of an outer product 
        between a vector and itself, our method has a fast scaling with the number of data points.
        If the second dimension cannot be described as an outer product, then we obtain a poor scaling with the size of this 
        dimension cubed. For the method to be computationally 
        efficient in this case, the non-\celerite 
        dimension should be small compared to the size of 
        the dimension along which the covariance is specified 
        by a \celerite kernel function. For 
        problems where the second dimension is comparable in 
        size to the first and 
        where $R$ must be arbitrarily defined, approximate 
        methods such as the \project{HODLR} method \citep{Ambikasaran2015}, \project{KISS-GP} \citep{Wilson2015},
        or the black box methods implemented in \project{GPyTorch} \citep{Gardner2018}
        may be more efficient. 
        
        
        Another limitation is fundamental to the 
        Gaussian process framework --- our 
        method, like all GP methods, does a poor 
        job of modeling outliers. When analyzing 
        observational data, outliers are often dealt with by 
        discarding them prior to analysis. However, 
        in some cases outliers may represent 
        useful information and should be 
        included in a model. A Student-t process (TP) 
        may thus be a better model for datasets containing 
        outliers \citep{Tracey2018}. We leave to future investigation 
        the prospects for implementing a 
        TP version of \celerite and evaluating the 
        performance of a TP regression on transit 
        photometry. 
    
    \subsection{Limitations of the multiband photometric noise model}
    
        We make several assumptions in the construction of 
        our multiband noise model which likely do not 
        hold in all cases.  First and foremost, a Gaussian process assumes that the noise is stationary and Gaussian.  This does not apply to some sources of noise, such as stellar flares, or sources which undergo outbursts in which the amplitude and/or shape of the power spectrum change dramatically.  Likewise our method does not apply if there is a significant time delay between the bands, if one band involves a time convolution of the other, nor if the correlated components of the bands have no correlation with one another.
        
        Second, the specific form we've chosen for 
        the wavelength covariance assumes that the wavelength-dependence 
        of the flux is due to varying covering fractions of a 
        hot and cold component in a two-component photosphere.  We expect that this model will work under different assumptions;  for instance, small-amplitude temperature variations should have a similar behavior as area fluctuations.  However, 
        different sources of variability will result in different forms 
        for the covariance in the wavelength dimension.  
        
        Additionally, if there are more than two components to 
        the photosphere, then we must consider the possibility that 
        each component's covering fraction varies with a different 
        characteristic timescale. In this case rather than pairing 
        a single wavelength covariance matrix $T$ with a single 
        time covariance matrix $R$ to form the full 
        covariance matrix $K = T\otimes R$, we should 
        pair multiple wavelength covariance matrices with 
        corresponding time covariance matrices, each having 
        different characteristic timescales:
        \begin{equation}
            \label{eqn:kronecker_sum}
            K = \sum_{i=1}^{N} T_i\otimes R_i
        \end{equation}
        Our code accepts multiple kernel components, each 
        with a unique $T$ matrix. While we have limited 
        ourselves to the case of a single kernel component 
        in this paper for the sake of clarity and 
        simplicity, we plan to introduce this extension 
        in more detail in a future paper.
        
        In the examples in this paper 
        we chose to fix the kernel parameters. 
        In practice the kernel parameters will 
        need to be measured alongside the parameters 
        of the mean model. This brings up the question of 
        how long of a time series is required to produce 
        a sufficiently strong constraint on the kernel 
        parameters that the inference of a transit 
        is unambiguous. We also defer this question to future work.
        
        Finally, our formulation assumes that the observations are complete; i.e.\ in the multiband times series example, every time of observation contains data in every band.  In principle this assumption could be relaxed, and in equation (\ref{eqn:kronecker_generators_outer}) the Kronecker products with $\bvec{\alpha}$ could be replaced with an $\bvec{\alpha}$ (and corresponding $R$ matrix) which varies with time stamp, and only contains the amplitudes of the bands observed at each time stamp.  This would also require modifying the indexing in equations (\ref{eqn:kronecker_tildes}) and (\ref{eqn:phi_2d}), but the rest of the method would remain the same.
        
\section{Conclusions} \label{sec:conclusions}

    We have extended the \celerite method for 
    fast one-dimensional GP computations to two 
    dimensions. Our method inherits the $\mathcal{O}(N)$ scaling 
    of \celerite in one of the two dimensions 
    while incurring a computational cost of $\mathcal{O}(M)$ 
    for a grid with size $M$ in the second dimension. Computing 
    the 2D GP on an $N{\times} M$ grid thus costs $\mathcal{O}(NM)$ 
    using our method, compared to $\mathcal{O}(N^3M^3)$ for the 
    direct solution (i.e. inverting the full $NM{\times} NM$ 
    covariance matrix).  This scaling applies only when the amplitude of correlated noise varies across the bands; a more general dependence on the second dimension has a poorer scaling, yet still improves upon direct solution.
    
    This extension may have many possible applications, among them 
    simultaneous modeling of stellar variability across wavelength. 
    This application is of particular interest to us, as we 
    would like to mitigate the effects of stellar variability 
    on detecting transiting exoplanets and measuring their 
    properties. We demonstrate that we can improve the precision 
    of transit depth, time, and duration measurements by 
    modeling the transit in multiple wavelengths when compared 
    to the monochromatic case.   
    
    When the signal-to-noise is high, we have shown that a precision which is proportional to the photon noise limit is achievable. For instance, in the two-band case in which the correlated noise in one band is twice that in the second band, one can achieve $\sqrt{10}$ of the photon-noise limit. This means that to reach the same precision as the no correlated noise case requires 10 times as many photons, or a telescope which has a collecting area ten times larger.  In the limit of a blackbody which is photon-noise dominated, with a large number of bands, one can reach 2.2 times the photon-noise limit in which the correlated noise is absent.  Hence, one needs to use a telescope which has $2.2^2 = 4.8$ times the collecting area.  Thus, in general one can achieve a precision of measurement which is comparable to the pure photon-noise limit, but this requires about an order of magnitude more photons to do so.
    
    In future work, we plan to extend our variability model 
    to model more realistic stellar variability by 
    including terms in the covariance kernel function that 
    capture variability on different timescales with 
    different wavelength dependencies. We suggest that 
    the SOHO spacecraft's three-channel sunphotometer 
    data may be a useful starting point for exploring the 
    wavelength dependence of variability in sun-like stars. 
    This dataset consists of measurements of the Sun's 
    irradiance in three visible-light bands 
    at one-minute cadence \citep{Frohlich1995}. 
    
    We are additionally interested in applying our 
    method to RV observations of exoplanet host 
    stars, following the method demonstrated by 
    \citeauthor{Rajpaul2015}. This requires us to 
    compute linear combinations of the GP 
    and its time derivatives, which in principle should be feasible.
    
    Our code is available in the form of a \texttt{pip} 
    installable python package called \texttt{specgp}. 
    \texttt{specgp} extends \texttt{exoplanet}\footnote{\url{https://github.com/exoplanet-dev/exoplanet}} to enable 2D Gaussian process computations. 
    Interested users can find instructions and 
    tutorials at \url{https://github.com/tagordon/specgp}.

    \acknowledgments
    
    We acknowledge support from NSF grant AST-1907342.
    We thank Jackson Loper for useful conversations about LEG GPs.
    EA was supported by a Guggenheim Fellowship and NSF grant AST-1615315.
    We also acknowledge support from NASA's NExSS Virtual Planetary Laboratory, funded under
    NASA Astrobiology Institute Cooperative Agreement Number NNA13AA93A, and the NASA Astrobiology
    Program grant 80NSSC18K0829.
    This research was partially conducted during the Exostar19 program at the Kavli Institute
    for Theoretical Physics at UC Santa Barbara, which was supported in part by the National
    Science Foundation under Grant No.\ NSF PHY-1748958.
    
    This work was facilitated though the use of the advanced computational, storage, and
    networking infrastructure provided by the Hyak supercomputer system at the University of
    Washington.



\acknowledgments

%






\appendix
    \section{\texttt{Celerite} algorithm for arbitrary covariance matrix in 
second dimension}
    \label{sec:algorithm}

            \label{sec:arbitrary_covariance_method}
	        In this section we assume that the covariance in 
	        the second dimension, defined by the covariance matrix 
	        $R$, is arbitrary, subject to 
	        the constraint that the full covariance matrix $K$ 
	        must be positive definite \footnote{Positive definiteness 
	        can be defined by the requirement that all eigenvalues 
	        of the matrix are positive. For a 1D 
	        \celerite model positive definiteness can be guaranteed 
	        by the methods outlined in Appendix A of \cite{Foreman-Mackey2017}. For a 2D \celerite model 
	        guaranteeing positive definiteness is more complicated, and 
	        we recommend investigating kernels on a case-by-case basis.}. 
	        
	        We start by rewriting $T$ in terms of the 
	        celerite generator matrices $A$, $U$, and $V$ 
	        from equation \ref{eqn:KUV}:
	        \begin{eqnarray}
	            K &=& \Sigma + \left[A_0 + \mathrm{tril}(UV^\T) 
	            + \mathrm{triu}(VU^\T)\right] \otimes R \cr
	            &=& \Sigma + \mathrm{diag}(A_0\otimes R) \cr
	            &+& \mathrm{tril}(UV^\T\otimes R) 
	            + \mathrm{triu}(VU^\T\otimes R) 
	        \end{eqnarray}
	        We rewrite $R$ as $R I_M$ where $I_M$ is the $M{\times} M$ 
	        identity matrix, which allows us to write $K$ as 
	        \begin{eqnarray}
	            K = \Sigma &+& \mathrm{diag}(A_0\otimes R) \cr
	            &+& \mathrm{tril}((U\otimes R)(V\otimes I_M)^\T) \\
	            &+& \mathrm{triu}((V\otimes I_M)(U\otimes R)^\T) \nonumber 
	        \end{eqnarray}
	        where we have again applied equation \ref{eqn:mixed_products}. 
	        As for the outer product case, we now have a semi-separable 
	        matrix defined by a new set of generators: 
	        \begin{eqnarray}
	            \label{eqn:kronecker_generators}
	    	    \nonumber A' &=& \Sigma + \mathrm{diag}(A_0\otimes T) \\
    		    \nonumber U' &=& U\otimes T \\
	    	    V' &=& V\otimes I_M.
	        \end{eqnarray}
	        In terms of the \celerite coefficients the 
	        refactored generator matrices 
	        are defined element-wise as follows:
	        \begin{eqnarray}
    	    \label{eqn:UV_prime_defs}
    	        A'_{(n-1)M+p, (n-1)M+p} &=& \sigma_{(n-1)M+1}^2 + R_{p, p}\sum_{j=1}^J a_j \cr
    	        \tilde U'_{(n-1)M+p,(2j-1)M+q} &=& R_{p, q}\tilde U_{n, 2j-1} \\
    	        \tilde U'_{(n-1)M+p, 2jM+q} &=& R_{p, q}\tilde U_{n, 2j} \cr
    	        \tilde V'_{(n-1)M+p,(2j-1)M+q} &=& \delta_{p, q}\tilde V_{n, 2j-1} \cr
    	        \tilde V'_{(n-1)M+p, 2jM+q} &=& \delta_{p, q}\tilde V_{n, 2j}, \nonumber
        	\end{eqnarray}
        	where $\tilde U$ and $\tilde V$ are 
        	the refactored generator matrices defined in equation 
        	\ref{eqn:tildes}, $n$ ranges over $(1,N)$, $p$ and $q$ range over $(1, M)$, 
        	and $\delta_{p, q}$ is the 
        	Kronecker delta function:
        	\begin{equation}
    	        \delta_{p, q} = \left\{\begin{matrix}
    	            1 & p=q \\
    	            0 & p\neq q
    	        \end{matrix}\right. .
	        \end{equation}
	        
	        The recursive algorithm for carrying out the Cholesky 
	        decomposition is identical to the outer-product case.  Starting with $D_{1,1} = A_{1,1}^\prime$ and $\tilde W_{1,j} = \tilde V_{1,j}/D_{1,1}$, we then recursively define: 
	        \begin{eqnarray}
    	        \label{eqn:cholesky_algorithm_final2}
    	        S_{n,j,k} &=& \phi'_{n,j}\phi'_{n,k}\left[S_{n-1, j, k} + D_{n-1, n-1}\tilde{W}_{n-1, j}\tilde{W}_{n-1,k}\right], \cr
    	        D_{n, n} &=& A'_{n, n} - \sum_{j=1}^{P}\sum_{k=1}^{P}\tilde U'_{n, j}S_{n, j, k}\tilde U'_{n, k}, \\
    	        \tilde W_{n, j} &=& \frac{1}{D_{n, n}}\left[\tilde V'_{n, j}-\sum_{k=1}^{P}\tilde U'_{n, k}S_{n, j, k}\right], \nonumber
    	    \end{eqnarray}
    	    for $n=2,...,N'$, $N' = N M$, with $P=2JM$ the number of rows in $\tilde U'$ and 
    	    $\tilde V'$. This additional factor of $M$ accounts for 
    	    the relatively poorer scaling of the method for 
    	    arbitrary $R$ over the outer-product case. For 
    	    arbitrary definitions of $R$, $P=2JM$ and the Cholesky decomposition thus
    	    scales as $\mathcal{O}(NJ^2M^3)$. 
    	    
\section{Computing the log-likelihood}
        \label{sec:likelihood}
    	The log-likelihood is given by 
    	\begin{equation}
            \label{eqn:simple_logL_appendix}
		    \ln\ \mathcal{L} = 
	    	-\frac{1}{2}(\bvec{y}-\bvec{\mu})^\T K^{-1}(\bvec{y}-\bvec{\mu})
		    -\frac{1}{2}\ln\ \mathrm{det}(K) - \frac{N'}{2}\ln(2\pi),
        \end{equation}
    	which incorporates both
    	the inverse and log-determinant of the covariance matrix, $K$. We therefore begin by 
    	describing the algorithms for each of these computations 
    	separately. The following algorithm comes directly 
    	from the original \celerite paper, but with our modified definitions of the semi-separable matrix components, $\tilde U^\prime, \tilde V^\prime$ and $\tilde W$, and $\phi_{n,j}^\prime$ rather than $\phi_{n,j}$ (see \S \ref{sec:outer_product_method}).
    	
    	The product of the inverse covariance matrix with 
    	a vector, $\bvec{z} = K^{-1}\bvec{y}$, is computed 
    	with a two-part algorithm. 
    	We first compute the intermediary $\bvec{z}'$, setting $z_1^\prime = y_1$, and then using the recursion relation 
    	\begin{eqnarray}
    	    \label{eqn:inv1}
    	    f_{n,j} &=& \phi_{n,j}^\prime\left[f_{n-1, j}+\tilde W_{n-1, j}z'_{n-1}\right] \\
    	    z'_{n} &=& y_n - \sum_{j=1}^{P}\tilde U'_{n, j}f_{n,j},
    	\end{eqnarray}
    	for $n=2,...,N'$, where $N' = NM$ and $f_{0,j}=0$ for all $j$.  We then use $\bvec{z}^\prime$ to compute $\bvec{z}$ in the second step of the algorithm, first setting $z_{N'} = z_{N'}^\prime/D_{N',N'}$, and then using downward recursion 
    	\begin{eqnarray}
    	    g_{n,j} &=& \phi_{n+1,j}^\prime\left[g_{n+1, j} + \tilde U'_{n+1, j}z_{n+1}\right] \\
    	    \label{eqn:inv2}
    	    z_n &=& \frac{z'_n}{D_{n,n}} - \sum_{j=1}^{P}\tilde W_{n,j}g_{n,j}
    	\end{eqnarray}
    	for $n = N'-1,..., 1$, where $g_{N', j} = 0$ for all $j$ and $P$ is the number of 
    	columns in $\tilde U'$, $\tilde V'$, and $\tilde W$. 
    	
    	The log-determinant of $K$ is given by 
    	\begin{equation}
    	    \ln{(\det K)} = \sum_{n=1}^{N'}\ln{(D_{n,n})}.
    	\end{equation}
    	Putting these two steps together we can compute the 
    	log-likelihood. Because the algorithm for taking products of the 
    	inverse requires $\mathcal{O}(NMP)$ operations, whereas the log-determinant 
    	can be computed in only $\mathcal{O}(NM)$ operations, the log-likelihood 
    	computation as a whole scales as $\mathcal{O}(NMP)$.
    	In practice, the 
    	bottleneck for applications such as maximizing the 
    	likelihood or MCMC 
    	is computing the Cholesky factor rather than computing the log-likelihood, 
    	since the log-likelihood computation itself 
    	is faster by $\mathcal{O}(P)$. Again we have $P=2J$ when 
    	$R$ is an outer product and $P=2JM$ when $R$ is any 
    	arbitrary covariance matrix. 
    	
\section{Prediction Algorithm}
    \label{sec:prediction}
    
    A Gaussian process prediction is an interpolation or 
    extrapolation of the observed data using with the GP model. 
    A prediction evaluated at each datapoint can also be 
    thought of as a smoothing operation as it yields an 
    estimate of the function with white noise removed. 
    
    The predictive distribution of a Gaussian process is 
    a multivariate normal with a mean $\mu^*$ and covariance $K^*$ evaluated at 
    the input coordinates $\bvec{x}^*$. For a 
    GP with no white noise component the mean is constrained 
    to pass directly through each observation of the data 
    points $\bvec{y}$. For a GP with a non-zero white noise 
    component the GP will act as a filter such that when the mean 
    is subtracted from the data the residuals will be distributed 
    according to 
    a Gaussian distribution whose width is given by the 
    GP white noise. 
    
    The predictive mean and covariance are computed as follows 
    	\begin{eqnarray}
    	\label{eqn:predictive_mean}
    	    \bvec{\mu^*} &=& \bvec{\mu_\theta}(\bvec{x^*}) + K(\bvec{x^*}, \bvec{x})K(\bvec{x}, \bvec{x})^{-1}[\bvec{y}-\bvec{\mu_\theta}(\bvec{x})] \\
    	    K^* &=& K(\bvec{x^*}, \bvec{x^*}) - K(\bvec{x^*}, \bvec{x})K(\bvec{x}, \bvec{x})^{-1}K(\bvec{x}, \bvec{x^*})
    	\end{eqnarray}
    	where $K(\bvec{x^*}, \bvec{x})$ and $K(\bvec{x}, \bvec{x^*})$ are the covariance 
    	kernel evaluated between the input coordinates and the data coordinates. If 
    	the input coordinates consist of $N^*$ points in the first dimension and 
    	$M^*$ points in the second then these matrices have dimensions $(M^*N^*{\times} NM)$ 
    	and $(NM{\times} N^*M^*)$ respectively. 
    	
    	For the 2D Kronecker-structured covariance matrix 
    	$K = T\otimes R$, we can rewrite equation \ref{eqn:predictive_mean} 
    	as 
    	\begin{eqnarray}
    	    \label{eqn:prediction_kron}
    	    \bvec{\mu}^* &=& \bvec{\mu_\theta}(\bvec{x}^*) + \left[T(\bvec{x}^*, \bvec{x})\otimes R(\bvec{x}^*, \bvec{x})\right]K(\bvec{x}, \bvec{x})^{-1}[\bvec{y}-\bvec{\mu}_\bvec{\theta}(\bvec{x})] \\
    	    &=& \bvec{\mu_\theta}(\bvec{x}^*) + \left[T(\bvec{x}^*, \bvec{x})\otimes R(\bvec{x}^*, \bvec{x})\right]\bvec{z}
    	\end{eqnarray}
    	where $\bvec{z} = K(\bvec{x}, \bvec{x})^{-1}\left[\bvec{y}-\bvec{\mu}_\bvec{\theta}(x)\right]$
    	Writing the second term of equation \ref{eqn:prediction_kron} in terms 
    	of the vectorization operator we have
    	\begin{equation}
    	    \label{eqn:prediction_vec}
    	    \left[T(\bvec{x}^*, \bvec{x})\otimes R(\bvec{x}^*, \bvec{x})\right]\bvec{z}
    	     = \left[T(\bvec{x}^*, \bvec{x})\otimes R(\bvec{x}^*, \bvec{x})\right]\mvec{Z}
    	\end{equation}
    	where $Z = Y - \mu_\theta(X)$ with $X$ and $Y$ matrices of size
    	$N{\times} M$ defined by $\bvec{x} = \mathrm{vec}(X)$ and 
    	$\bvec{y} = \mathrm{vec}(Y)$ respectively. 
    	For matrices $A$, $B$, and $C$ of sizes 
    	$(n{\times} m)$, $(m{\times} p)$, and $(p{\times} q)$ 
    	respectively there is an identity that 
    	states the following: 
    	\begin{equation}
    	    \mvec{ABC} = (A\otimes C^\T)\mvec{B}.
    	\end{equation} Applying this to 
    	equation \ref{eqn:prediction_vec} gives 
    	\begin{equation}
    	    \left[T(\bvec{x}^*, \bvec{x})\otimes R(\bvec{x}^*, \bvec{x})\right]\bvec{z} = \mvec{TZR}.
    	\end{equation}
    	The full expression for the predictive mean is now 
    	\begin{equation}
    	    \bvec{\mu}^* = \bvec{\mu}_\bvec{\theta}(\bvec{x}^*) + \mvec{TZR}.
    	\end{equation} 
    	
    	The matrix product $TZR$ can be computed via a modified version of 
    	the \celerite prediction algorithm presented in \citep{Foreman-Mackey2017}. 
    	
    	First, we compute the product $ZR$ at a computational cost 
    	of $\mathcal{O}(NM)$ when $R$ is outer product and 
    	$\mathcal{O}(NM^2)$ for arbitrary $R$. We then 
    	compute 
    	\begin{equation}
            \mu^*_{p, m} = \sum_{n=1}^N\sum_{j=1}^J e^{-c_j|t^*_p-t_n|}\left[
            a_j\mathrm{cos}(d_j|t_p^*-t_n|)+b_j\mathrm{sin}(d_j|t_p^*-t_n|)\right][ZR]_{n, m}.
        \end{equation}
        in two parts. Here $p$ and $m$ index the 
        elements of the predicted mean matrix. The first part consists of a forward pass 
        through $n_0=1, \dots, N$ where we define:
        \begin{eqnarray}
            G^-_{n, m, k} &=& \left[G^-_{n-1, p, k} +[ZR]_{n, m}\tilde{V}'_{n, k}\right]
            e^{-c_{k//2}(t_{n+1}-t_n)} \\
            H^-_{p, n, k} &=& e^{-c_{k//2}(t_p^*-t_{n+1})}\tilde{U}'^*_{p, k},
        \end{eqnarray} \footnote{$k//2$ denotes integer division of $k$ by 2. In other words, $k//2 = \mathrm{floor}(k/2)$.}
        and the second consisting of a backward pass through $n_0=N, ..., 1$ where we define
        \begin{eqnarray}
            G^+_{n, m, k} &=& \left[G^+_{n+1, p, k} + [ZR]_{n, m}\tilde{U}'_{n, k}\right]
            e^{-c_{k//2}(t_n-t_{n-1})} \\
            H^+_{p, n, k} &=& e^{-c_{k//2}(t_{n-1}^*-t_p)}\tilde{V}'^*_{p, k},
        \end{eqnarray}
        where $t_0=t_1$, $t_{N+1}=t_N$, $G^-_{0, m, k} = 0$, 
        and $G^+_{N+1, m, k}=0$ for $k=1, ..., 2J$ and for all 
        $m$. The expressions for $\tilde{U}'^*_{p, i}$ and $\tilde{V}'^*_{p, i}$ are evaluated at $t^*_p$. 
        For each value of $p$, $G^{\pm}$ are evaluated recursively 
        from $n$ to $n_0$ and then the prediction $\mu^*_{p, m}$ 
        is computed from 
        \begin{equation}
           \mu^*_{p, m} = \sum_{k=1}^P\left[G^-_{n_0, m, k}H^-_{p, n_0, k} 
            + G^+_{n_0+1, p, k}H^+_{p, n_0+1, k}\right].
        \end{equation}
        This two part computation scales as $\mathcal{O}(nN + n^*N^*)$ 
        where $n$ and $n^*$ are constants. The overall scaling is 
        therefore determined by the cost of the matrix multiplication step. 

\section{Sampling from the GP}
    \label{sec:sampling}
    A sample $\bvec{y}$ can be drawn from a Gaussian process by 
    computing 
    \begin{equation}
        \bvec{y} = \bvec\mu + L\bvec{n}
    \end{equation}
    where $\bvec\mu$ is the mean function and $\bvec{n}$ is a 
    vector of draws from a normal distribution 
    \begin{equation}
        n_i \sim \mathcal{N}(0, D_{i,i}^{1/2})
    \end{equation}
    for each entry $n_i$ in $\bvec{n}$. The ordering of 
    entries in $\bvec\mu$ and consequently $\bvec{y}$ is 
    determined by the structure of $K$. For the 
    Kronecker structured covariance matrix given 
    in equation \ref{eqn:kronecker_cov}, $\bvec\mu$ is 
    the concatenation of the $N$ length-$M$ vectors 
    containing the mean function evaluated at each 
    point in the second dimension at a given point 
    in the first. In other words, 
    \begin{equation}
        \bvec\mu = \left(\bvec\mu_1, \bvec\mu_2, \dots \bvec\mu_N \right)
    \end{equation}
    where $\bvec\mu_i = \left(\mu_{i, 1}, \mu_{i, 2}, \dots \mu_{i, M} \right)$ is the mean function evaluated at the 
    $i^\mathrm{th}$ point in the first dimension. 
    
    Thus $\bvec\mu$ is a one-dimensional 
    vector of length $N' = NM$ where $N$ is the size of the first dimension 
    and $M$ the size of the second. The sample vector $\bvec{y}$ then 
    has the same structure. Most users will wish to either unpack the sample 
    into $M$ separate vectors obtained by taking every $M^\mathrm{th}$ 
    entry in $\bvec{y}$ or reshape it into an $N{\times} M$ array 
    before displaying or examining the sample.
        
    \begin{figure*}
	        \plottwo{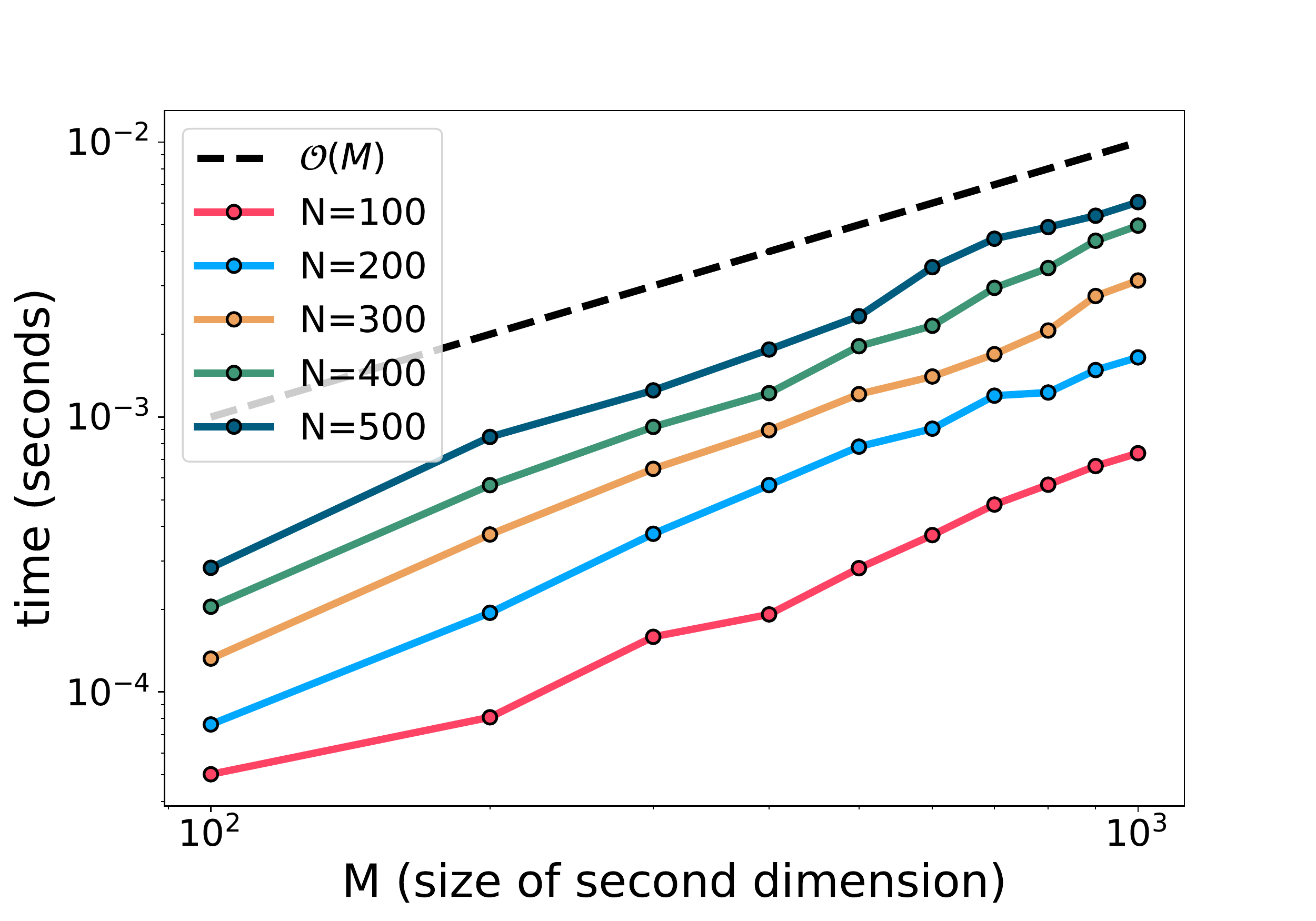}{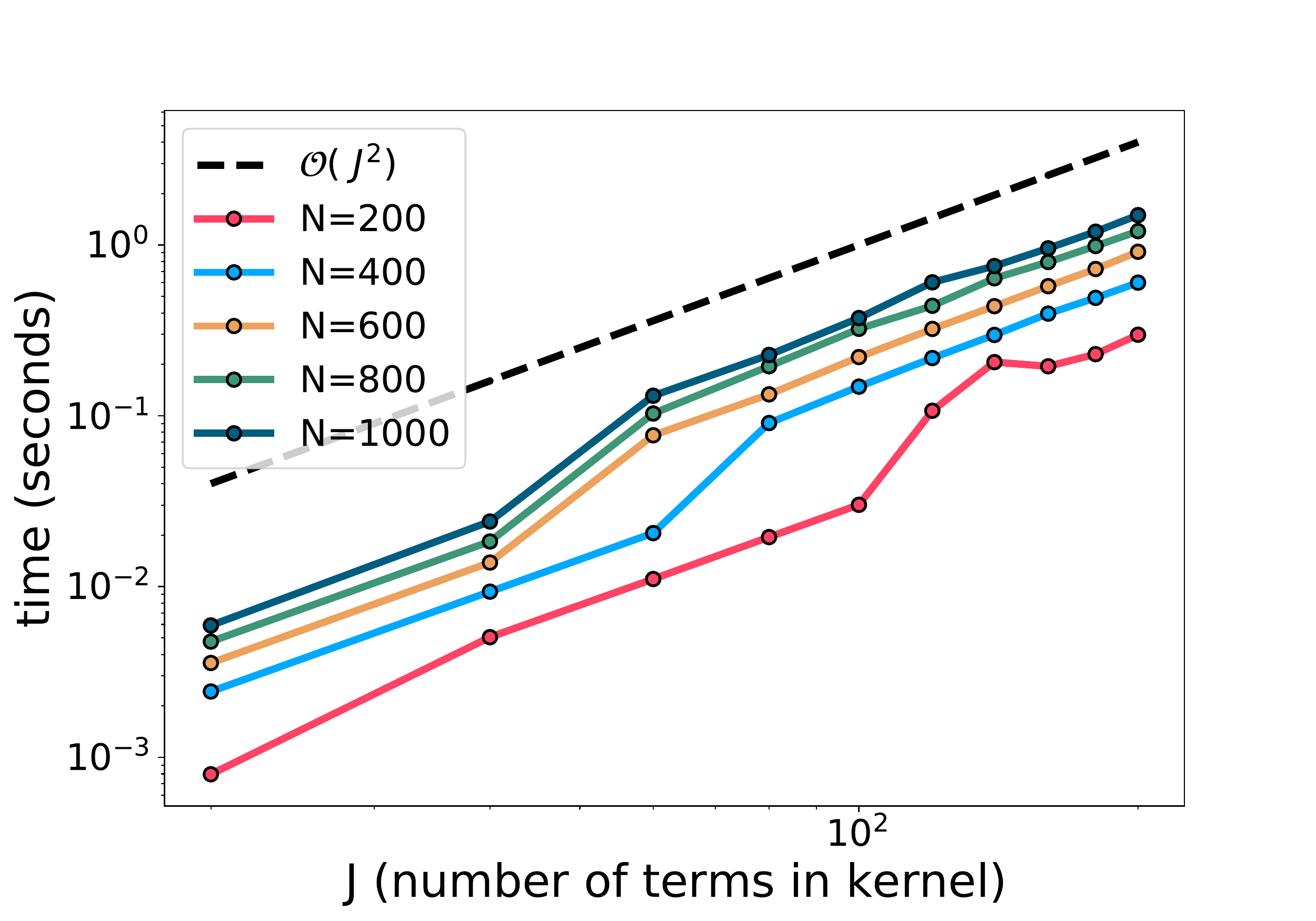}
	        \plottwo{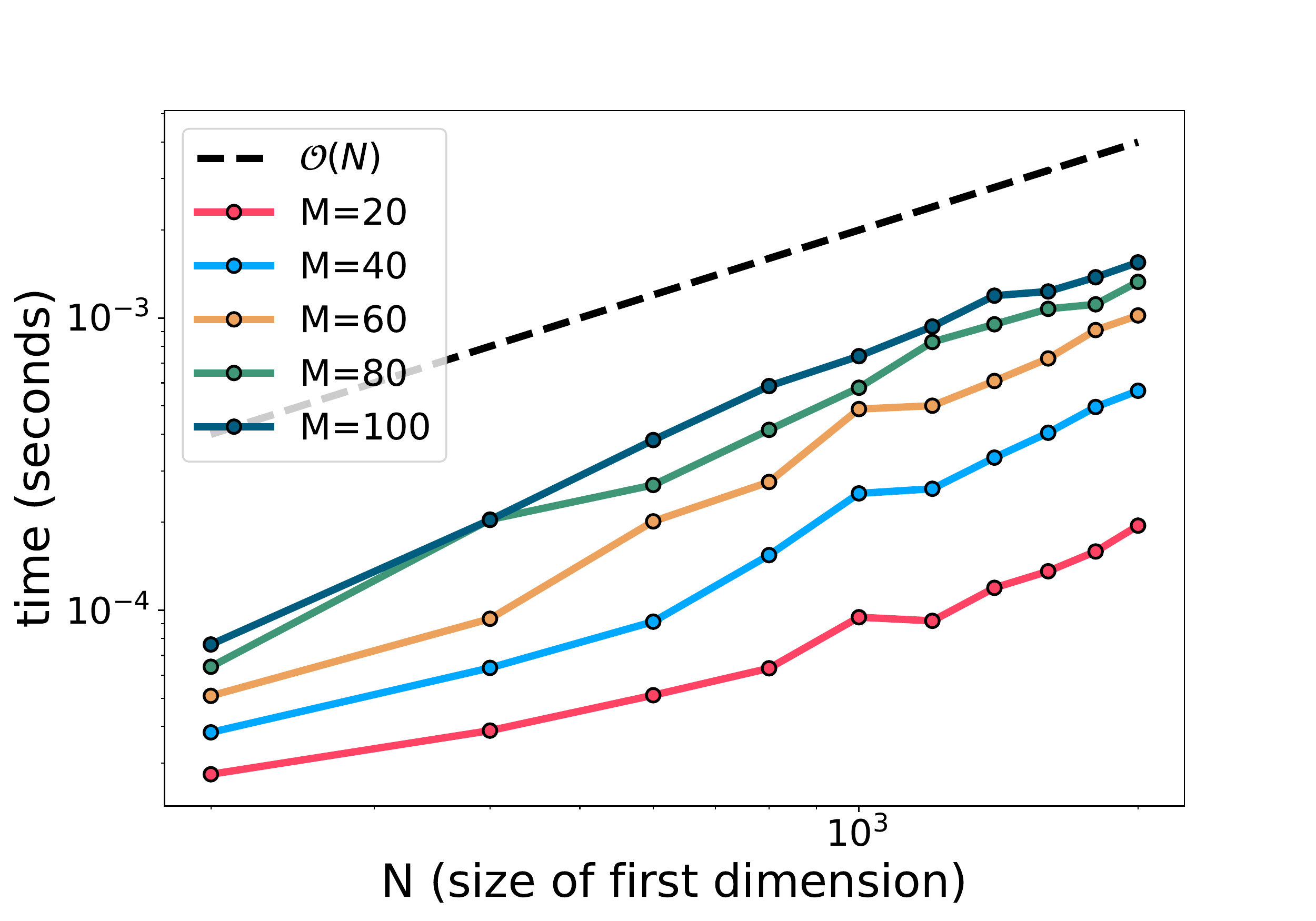}{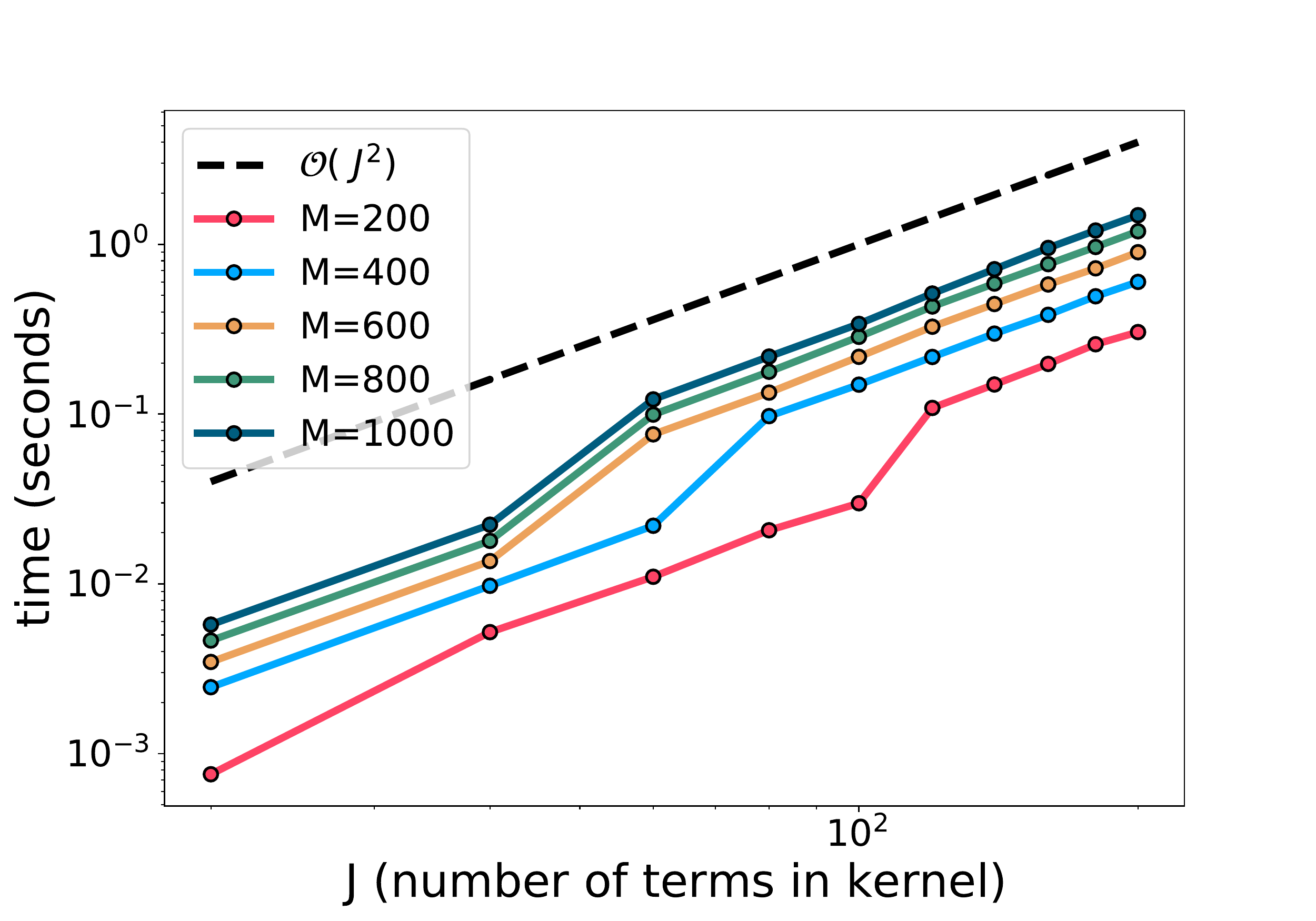}
	        \plottwo{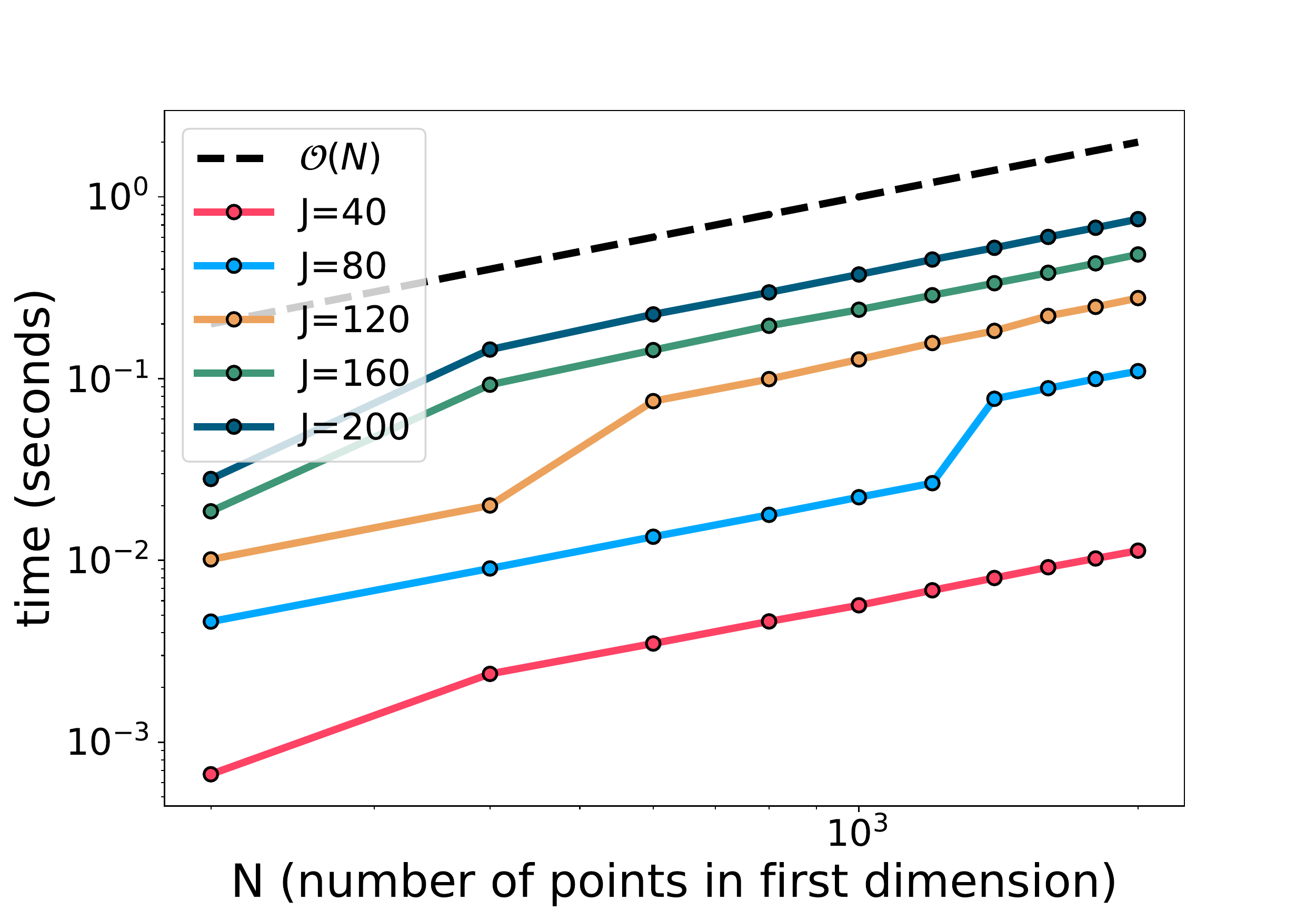}{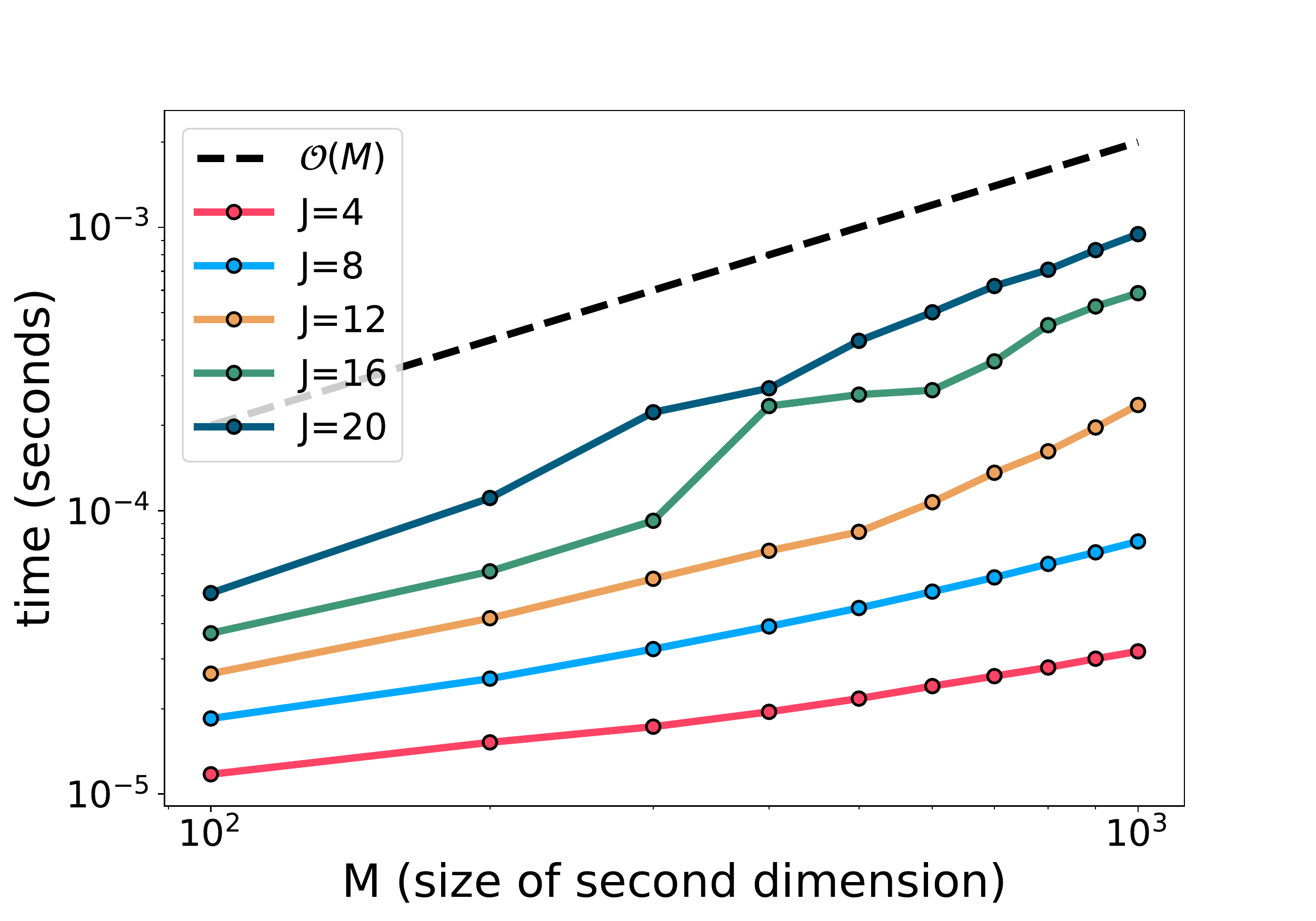}
            \caption{Benchmarks for the two-dimensional celerite 
            implementation with outer-product covariance in the 
            second dimension. We recover the anticipated linear 
            scaling with respect to both $N$ and $M$, and 
            the quadratic scaling with respect to $J$.}
            \label{fig:benchmarks}
        \end{figure*}
        
\section{Notation}
    \label{sec:notation}
    Notation and symbols, in order of appearance:
    \begin{itemize}
        \item[] $K$: covariance matrix 
        \item[] $k$: kernel function corresponding to $K$
        \item[] $x$: general independent variable for the GP
        \item[] $\mathcal{L}$: GP Likelihood
        \item[] $\bvec{\mu}$: GP mean vector 
        \item[] $\bvec{y}$: vector of observations
        \item[] $N'$: number of observations corresponding to the length of vector $\bvec{y}$
        \item[] $\omega_0$: characteristic frequency of simple harmonic oscillator (SHO) term
        \item[] $\epsilon$: stochastic force term, driving force of SHO
        \item[] $t$: an independent variable used to represent time
        \item[] $Q$: quality factor of SHO
        \item[] $\omega$: an independent variable used to represent frequency in expressions for the power spectral density of a process
        \item[] $S_0$: Amplitude of the SHO 
        \item[] $\tau$: an independent variable used to represent time lag, as in $\tau_{|i-j|} = |t_i-t_j|$
        \item[] $n, m, i, j, p, q$: integers used to index independent variables and matrices
        \item[] $x_h, x_c$: covering fractions of a hot and cold component of the stellar photosphere
        \item[] $R_*$: stellar radius 
        \item[] $d$: distance from star to observer
        \item[] $F$: flux 
        \item[] $B_n$: $n^\mathrm{th}$ spectral band
        \item[] $\lambda$: independent variable representing wavelength 
        \item[] $S_c(\lambda), S_h(\lambda)$: spectra of the hot and 
        cold components of a stellar photosphere 
        \item[] $\mathcal{R}_{B_i}(\lambda)$: Response curve for band $B_i$
        \item[] $\alpha_i$: variability amplitude 
        integrated over band $B_i$
        \item[] $\sigma_c$, $\sigma_h$: $\mathrm{var}(x_c)^{1/2}$, $\mathrm{var}(x_h)^{1/2}$  respectively; the RMS of the cold and hot covering fraction.
        \item[] $\Sigma_i$: diagonal matrix containing 
        the white noise variances for each wavelength at the  $i$th 
        time index.
        \item[] $T$: covariance matrix representing the first dimension or time dimension of the 2D GP. $T$ will always be described by a \celerite kernel function.
        \item[] $R$: covariance matrix representing 
        the second dimension or wavelength dimension of the 2D GP. $R$ may be an 
        arbitrary covariance matrix or an outer-product.
        \item[] $N$: Length of the first dimension, equal to the number of times in our example application of multiband time series 
        \item[] $M$: Length of second dimension, equal to the number of bands in our example application of multiband time series
        \item[] $J$: number of \celerite terms in kernel function
        \item[] $P$: rank of \celerite generator matrices
        \item[] $\bvec{\alpha}$: Vector of correlated noise amplitudes in the second dimension.
        \item[] $\Sigma$: diagonal matrix containing 
        the white noise variances for each observation; the white noise component of 
        the GP covariance matrix
        \item[] $\sigma_i^2$: white noise variance 
        for $i^\mathrm{th}$ datapoint
        \item[] $R_p$: planetary radius 
         \item[] $t_0$: time of center of transit 
        \item[] $\delta_\mathrm{in}$: duration of transit ingress/egress
        \item[] $\delta$: Transit duration (mid-ingress to mid-egress).
        \item[] $\bvec{\theta}$: Vector of transit parameters.
        \item[] $\mu_\mathrm{trap}$: Transit mean model.
        \item[] $f_0$:  Characteristic frequency of the correlated noise model.
        \item[] $\bar \alpha$: Weighted mean of $\bvec{\alpha}$ used to represent the total amplitude of 
        the correlated variability component of the GP summed over all bands (``monochromatic").
        \item[] $\sigma$: Mean of $\bvec{\sigma}$, the vector of white noise terms; used to represent the total amplitude of the uncorrelated variability component of the GP
        \item[] $\mathcal{I}$: Information matrix.
        \item[] $N_\theta$: Number of mean-model parameters, equal to the length of $\bvec{\theta}$
        \item[] $\sigma_{R_p^2}$:  Uncertainty on the transit depth (with ``poly" and ``mono" to indicate the polychromatic and monochromatic values).
        \item[] $\bvec{\beta}$: Vector of coefficients used in defining the \celerite kernel (\citet{Foreman-Mackey2017} use $\bvec{\alpha}$).
        \item[] $a, b, c, d$: \celerite 
        coefficients
        \item[] $A$: diagonal component of full 
        kernel function; $K = A + \mathrm{tril}(UV^\T) + \mathrm{triu}(VU^\T)$
        \item[] $U, V$: \celerite generator 
        matrices
        \item[] $L$: Lower triangaular matrix used in LDLT Cholesky
        decomposition.
        \item[] tril, triu: lower and upper triangular matrix operators
        \item[] $D$: Diagonal matrix used in decomposition.
        \item[] $W$: Matrix used in semi-separable LDLT Cholesky decomposition.
        \item[] $I$: Identity matrix.
        \item[] $S$: intermediary matrix used in the \celerite decomposition algorithm
        \item[] $D$: diagonal matrix in the Cholesky 
        decomposition of $K$
        \item[] $A_0$: diagonal component of $K$ with 
        white noise amplitude set to zero; $A_0 = A - \Sigma$
        \item[] $U', V', A'$: Kronecker products 
        of $U, V,$ and $A$ taken with $\vec{\alpha}$ or $R$ and $I_M$ 
        \item[] $\tilde U, \tilde V, \tilde W$: refactored \celerite matrices corresponding to $U$, $V$, and $W$
        \item[] $\tilde U', \tilde V', \tilde W'$: refactored \celerite matrices 
        corresponding to $U'$, $V'$, and $A'$.
        \item[] $\phi, \phi'$: matrices used in the refactored version of \celerite
        \item[] $F^\pm, G^\pm$: intermediary matrices for prediction algorithm 
        \item[] $f_{n, j}, g_{n, j}$: intermediary vectors used to compute the likelihood of the GP model
        \item[] $\bvec{\mu}^*$: predictive mean model
        \item[] $K^*$: predictive covariance
        \item[] $\bvec{x}^*$ independent variable used to represent the points 
        at which the predictive mean and covariance of the GP are evaluated
        \item[] $\bvec{z}$: the product between the observed vector $\bvec{y}$ and the inverse of the 
        covariance matrix $K^{-1}$ used to compute the GP likelihood
        \item[] $\bvec{z}'$: intermediary vector used to compute $\bvec{z}$
        \item[] $X, Y, Z$: $\bvec{x} = \mathrm{vec}(X)$, $\bvec{y} = \mathrm{vec}(Y)$, and $\bvec{z} = \mathrm{vec}(Z)$ 
        respectively; the matrix versions of $\bvec{x}$, $\bvec{y}$, and $\bvec{z}$ for 
        the two-dimensional GP.
        \item[] $t^*$: independent variable used to represent the points at which the predictive mean and covariance of the GP are evaluated; same as $x^*$ when the 
        independent variable is time.
        \item[] $H^\pm$: intermediary matrix used to compute the GP prediction ($Q^\pm$ in 
        \cite{Foreman-Mackey2017})
        \item[] $N^*$: the number of points at which the prediction is evaluated in 
        the first dimensions
        \item[] $n^*$: constant on which the computational scaling of the prediction algorithm depends
        \item[] $\bvec{n}$: vector of random draws from a standard normal 
        distribution used to draw a sample from the GP
    \end{itemize}
     
\bibliography{main}

\begin{thebibliography}{}
\expandafter\ifx\csname natexlab\endcsname\relax\def\natexlab#1{#1}\fi
\providecommand{\url}[1]{\href{#1}{#1}}
\providecommand{\dodoi}[1]{doi:~\href{http://doi.org/#1}{\nolinkurl{#1}}}
\providecommand{\doeprint}[1]{\href{http://ascl.net/#1}{\nolinkurl{http://ascl.net/#1}}}
\providecommand{\doarXiv}[1]{\href{https://arxiv.org/abs/#1}{\nolinkurl{https://arxiv.org/abs/#1}}}

\bibitem[{Agol \& Fabrycky(2018)}]{Agol2018}
Agol, E., \& Fabrycky, D.~C. 2018, in Handbook of Exoplanets (Cham: Springer
  International Publishing), 797--816.
\newblock \url{http://link.springer.com/10.1007/978-3-319-55333-7{\_}7}

\bibitem[{{Agol} {et~al.}(2005){Agol}, {Steffen}, {Sari}, \&
  {Clarkson}}]{Agol2005}
{Agol}, E., {Steffen}, J., {Sari}, R., \& {Clarkson}, W. 2005, \mnras, 359,
  567, \dodoi{10.1111/j.1365-2966.2005.08922.x}

\bibitem[{{Ambikasaran} {et~al.}(2015){Ambikasaran}, {Foreman-Mackey},
  {Greengard}, {Hogg}, \& {O'Neil}}]{Ambikasaran2015}
{Ambikasaran}, S., {Foreman-Mackey}, D., {Greengard}, L., {Hogg}, D.~W., \&
  {O'Neil}, M. 2015, IEEE Transactions on Pattern Analysis and Machine
  Intelligence, 38, 252, \dodoi{10.1109/TPAMI.2015.2448083}

\bibitem[{Anderson \& Jefferies(1990)}]{Anderson1990}
Anderson, E.~R., \& Jefferies, S.~M. 1990, 364, 699

\bibitem[{Bagnasco {et~al.}(2007)Bagnasco, Kolm, Ferruit, Honnen, Koehler,
  Lemke, Maschmann, Melf, Noyer, Rumler, Salvignol, Strada, \&
  Plate}]{Bagnasco2007}
Bagnasco, G., Kolm, M., Ferruit, P., {et~al.} 2007, in Cryogenic Optical
  Systems and Instruments {XII}, ed. J.~B. Heaney \& L.~G. Burriesci ({SPIE}).
\newblock \url{https://doi.org/10.1117/12.735602}

\bibitem[{Barclay {et~al.}(2015)Barclay, Endl, Huber, Foreman-Mackey, Cochran,
  Macqueen, Rowe, \& Quintana}]{Barclay2015}
Barclay, T., Endl, M., Huber, D., {et~al.} 2015, The Astrophysical Journal,
  800, 46, \dodoi{10.1088/0004-637X/800/1/46}

\bibitem[{Barros {et~al.}(2020)Barros, Demangeon, D{\'{\i}}az, Cabrera, Santos,
  Faria, \& Pereira}]{Barros2020}
Barros, S. C.~C., Demangeon, O., D{\'{\i}}az, R.~F., {et~al.} 2020, Astronomy
  {\&} Astrophysics, 634, A75, \dodoi{10.1051/0004-6361/201936086}

\bibitem[{Beichman {et~al.}(2014)Beichman, Benneke, Knutson, Smith, Lagage,
  Dressing, Latham, Lunine, Birkmann, Ferruit, Giardino, Kempton, Carey, Krick,
  Deroo, Mandell, Ressler, Shporer, Swain, Vasisht, Ricker, Bouwman,
  Crossfield, Greene, Howell, Christiansen, Ciardi, Clampin, Greenhouse,
  Sozzetti, Goudfrooij, Hines, Keyes, Lee, McCullough, Robberto, Stansberry,
  Valenti, Rieke, Rieke, Fortney, Bean, Kreidberg, Ehrenreich, Deming, Albert,
  Doyon, \& Sing}]{Beichman2014}
Beichman, C., Benneke, B., Knutson, H., {et~al.} 2014, Publications of the
  Astronomical Society of the Pacific, 126, 1134, \dodoi{10.1086/679566}

\bibitem[{{Berta} {et~al.}(2012){Berta}, {Charbonneau}, {D{\'e}sert},
  {Miller-Ricci Kempton}, {McCullough}, {Burke}, {Fortney}, {Irwin}, {Nutzman},
  \& {Homeier}}]{Berta2012}
{Berta}, Z.~K., {Charbonneau}, D., {D{\'e}sert}, J.-M., {et~al.} 2012, \apj,
  747, 35, \dodoi{10.1088/0004-637X/747/1/35}

\bibitem[{Boone(2019)}]{Boone2019}
Boone, K. 2019, The Astronomical Journal, 158, 257,
  \dodoi{10.3847/1538-3881/ab5182}

\bibitem[{Carter {et~al.}(2008)Carter, Yee, Eastman, Gaudi, \&
  Winn}]{Carter2008}
Carter, J.~A., Yee, J.~C., Eastman, J., Gaudi, B.~S., \& Winn, J.~N. 2008, The
  Astrophysical Journal, 689, 499, \dodoi{10.1086/592321}

\bibitem[{{Chakrabarty} \& {Sengupta}(2019)}]{Chakrabarty2019}
{Chakrabarty}, A., \& {Sengupta}, S. 2019, \aj, 158, 39,
  \dodoi{10.3847/1538-3881/ab24dd}

\bibitem[{{Dawson} {et~al.}(2014){Dawson}, {Johnson}, {Fabrycky},
  {Foreman-Mackey}, {Murray-Clay}, {Buchhave}, {Cargile}, {Clubb}, {Fulton},
  {Hebb}, {Howard}, {Huber}, {Shporer}, \& {Valenti}}]{Dawson2014}
{Dawson}, R.~I., {Johnson}, J.~A., {Fabrycky}, D.~C., {et~al.} 2014, \apj, 791,
  89, \dodoi{10.1088/0004-637X/791/2/89}

\bibitem[{Deisentroth \& Ng(2015)}]{Deisenroth2015}
Deisentroth, M.~P., \& Ng, J.~W. 2015

\bibitem[{Foreman-Mackey {et~al.}(2017)Foreman-Mackey, Agol, Ambikasaran, \&
  Angus}]{Foreman-Mackey2017}
Foreman-Mackey, D., Agol, E., Ambikasaran, S., \& Angus, R. 2017, The
  Astronomical Journal, 154, 220, \dodoi{10.3847/1538-3881/aa9332}

\bibitem[{Foreman-Mackey {et~al.}(2019)Foreman-Mackey, Czekala, Luger, Agol,
  Barentsen, \& Barclay}]{exoplanet:exoplanet}
Foreman-Mackey, D., Czekala, I., Luger, R., {et~al.} 2019, dfm/exoplanet
  v0.2.3, \dodoi{10.5281/zenodo.1998447}.
\newblock \url{https://doi.org/10.5281/zenodo.1998447}

\bibitem[{Frohlich {et~al.}(1995)Frohlich, Romero, Roth, Wehrli, Andersen,
  Appourchaux, Domingo, Telljohann, Berthomieu, Delache, Provost, Toutain,
  Crommelynck, Chevalier, Däppen, Gough, Hoeksema, Jiménez, Gómez, Herreros,
  Cortés, Jones, Pap, \& Willson}]{Frohlich1995}
Frohlich, C., Romero, J., Roth, H., {et~al.} 1995, Solar Physics, 162, 101,
  \dodoi{10.1007/BF00733428}

\bibitem[{Gardner {et~al.}(2018)Gardner, Pleiss, Bindel, Weinberger, \&
  Wilson}]{Gardner2018}
Gardner, J.~R., Pleiss, G., Bindel, D., Weinberger, K.~Q., \& Wilson, A.~G.
  2018, CoRR, abs/1809.11165.
\newblock \doarXiv{1809.11165}

\bibitem[{{Gillon} {et~al.}(2017){Gillon}, {Triaud}, {Demory}, {Jehin}, {Agol},
  {Deck}, {Lederer}, {de Wit}, {Burdanov}, {Ingalls}, {Bolmont}, {Leconte},
  {Raymond}, {Selsis}, {Turbet}, {Barkaoui}, {Burgasser}, {Burleigh}, {Carey},
  {Chaushev}, {Copperwheat}, {Delrez}, {Fernand es}, {Holdsworth}, {Kotze},
  {Van Grootel}, {Almleaky}, {Benkhaldoun}, {Magain}, \& {Queloz}}]{Gillon2017}
{Gillon}, M., {Triaud}, A. H.~M.~J., {Demory}, B.-O., {et~al.} 2017, \nat, 542,
  456, \dodoi{10.1038/nature21360}

\bibitem[{{Grimm} {et~al.}(2018){Grimm}, {Demory}, {Gillon}, {Dorn}, {Agol},
  {Burdanov}, {Delrez}, {Sestovic}, {Triaud}, {Turbet}, {Bolmont}, {Caldas},
  {de Wit}, {Jehin}, {Leconte}, {Raymond}, {Van Grootel}, {Burgasser}, {Carey},
  {Fabrycky}, {Heng}, {Hernandez}, {Ingalls}, {Lederer}, {Selsis}, \&
  {Queloz}}]{Grimm2018}
{Grimm}, S.~L., {Demory}, B.-O., {Gillon}, M., {et~al.} 2018, \aap, 613, A68,
  \dodoi{10.1051/0004-6361/201732233}

\bibitem[{Hey {et~al.}(2020)Hey, Murphy, Foreman-Mackey, Bedding, Pope, \&
  Hogg}]{Hey2020}
Hey, D.~R., Murphy, S.~J., Foreman-Mackey, D., {et~al.} 2020, The Astronomical
  Journal, 159, 202, \dodoi{10.3847/1538-3881/ab7d38}

\bibitem[{{Hippke} {et~al.}(2019){Hippke}, {David}, {Mulders}, \&
  {Heller}}]{Hippke2019}
{Hippke}, M., {David}, T.~J., {Mulders}, G.~D., \& {Heller}, R. 2019, \aj, 158,
  143, \dodoi{10.3847/1538-3881/ab3984}

\bibitem[{Hoffman \& Gelman(2014)}]{Hoffman2014}
Hoffman, M.~D., \& Gelman, A. 2014, J. Mach. Learn. Res., 15, 1593

\bibitem[{Hojjati {et~al.}(2013)Hojjati, Kim, \& Linder}]{Hojjati2013}
Hojjati, A., Kim, A.~G., \& Linder, E.~V. 2013, Physical Review D, 87,
  \dodoi{10.1103/physrevd.87.123512}

\bibitem[{Hojjati \& Linder(2014)}]{Hojjati2014}
Hojjati, A., \& Linder, E.~V. 2014, Physical Review D, 90,
  \dodoi{10.1103/physrevd.90.123501}

\bibitem[{Holman(2005)}]{Holman2005}
Holman, M.~J. 2005, Science, 307, 1288, \dodoi{10.1126/science.1107822}

\bibitem[{Kallinger {et~al.}(2014)Kallinger, {De Ridder}, Hekker, Mathur,
  Mosser, Gruberbauer, Garc{\'{i}}a, Karoff, \& Ballot}]{Kallinger2014}
Kallinger, T., {De Ridder}, J., Hekker, S., {et~al.} 2014, Astronomy {\&}
  Astrophysics, 570, A41, \dodoi{10.1051/0004-6361/201424313}

\bibitem[{Kelly {et~al.}(2009)Kelly, Bechtold, \& Siemiginowska}]{Kelly2009}
Kelly, B.~C., Bechtold, J., \& Siemiginowska, A. 2009, The Astrophysical
  Journal, 698, 895, \dodoi{10.1088/0004-637x/698/1/895}

\bibitem[{Kelly {et~al.}(2014)Kelly, Becker, Sobolewska, Siemiginowska, \&
  Uttley}]{Kelly2014}
Kelly, B.~C., Becker, A.~C., Sobolewska, M., Siemiginowska, A., \& Uttley, P.
  2014, The Astrophysical Journal, 788, 33, \dodoi{10.1088/0004-637X/788/1/33}

\bibitem[{Kim {et~al.}(2013)Kim, Thomas, Aldering, Antilogus, Aragon, Bailey,
  Baltay, Bongard, Buton, Canto, Cellier-Holzem, Childress, Chotard, Copin,
  Fakhouri, Gangler, Guy, Kerschhaggl, Kowalski, Nordin, Nugent, Paech, Pain,
  Pecontal, Pereira, Perlmutter, Rabinowitz, Rigault, Runge, Saunders, Scalzo,
  Smadja, Tao, Weaver, \& Wu}]{Kim2013}
Kim, A.~G., Thomas, R.~C., Aldering, G., {et~al.} 2013, The Astrophysical
  Journal, 766, 84, \dodoi{10.1088/0004-637x/766/2/84}

\bibitem[{{Kipping} {et~al.}(2013){Kipping}, {Hartman}, {Buchhave}, {Schmitt},
  {Bakos}, \& {Nesvorn{\'y}}}]{Kipping2013}
{Kipping}, D.~M., {Hartman}, J., {Buchhave}, L.~A., {et~al.} 2013, \apj, 770,
  101, \dodoi{10.1088/0004-637X/770/2/101}

\bibitem[{{Kreidberg} {et~al.}(2019){Kreidberg}, {Luger}, \&
  {Bedell}}]{Kreidberg2019}
{Kreidberg}, L., {Luger}, R., \& {Bedell}, M. 2019, \apjl, 877, L15,
  \dodoi{10.3847/2041-8213/ab20c8}

\bibitem[{{Loper} {et~al.}(2020){Loper}, {Blei}, {Cunningham}, \&
  {Paninski}}]{Loper2020}
{Loper}, J., {Blei}, D., {Cunningham}, J.~P., \& {Paninski}, L. 2020, arXiv
  e-prints, arXiv:2003.05554.
\newblock \doarXiv{2003.05554}

\bibitem[{MacLeod {et~al.}(2010)MacLeod, Ivezi{\'{c}}, Kochanek, Koz{\l}owski,
  Kelly, Bullock, Kimball, Sesar, Westman, Brooks, Gibson, Becker, \&
  de~Vries}]{MacLeod2010}
MacLeod, C.~L., Ivezi{\'{c}}, {\v{Z}}., Kochanek, C.~S., {et~al.} 2010, The
  Astrophysical Journal, 721, 1014, \dodoi{10.1088/0004-637x/721/2/1014}

\bibitem[{Mahadevan {et~al.}(2019)Mahadevan, Bender, Hambleton, Fleming,
  Deshpande, Conroy, Matijevi{\v{c}}, Hebb, Roy, Ak, Leban, \&
  Pr{\v{s}}a}]{Mahadevan2019}
Mahadevan, S., Bender, C.~F., Hambleton, K., {et~al.} 2019, The Astrophysical
  Journal, 884, 126, \dodoi{10.3847/1538-4357/ab3793}

\bibitem[{{Mandell} {et~al.}(2013){Mandell}, {Haynes}, {Sinukoff},
  {Madhusudhan}, {Burrows}, \& {Deming}}]{Mandell2013}
{Mandell}, A.~M., {Haynes}, K., {Sinukoff}, E., {et~al.} 2013, \apj, 779, 128,
  \dodoi{10.1088/0004-637X/779/2/128}

\bibitem[{{Mazeh} \& {Faigler}(2010)}]{Mazeh2010}
{Mazeh}, T., \& {Faigler}, S. 2010, \aap, 521, L59,
  \dodoi{10.1051/0004-6361/201015550}

\bibitem[{Morris {et~al.}(2020)Morris, Bobra, Agol, Lee, \&
  Hawley}]{Morris2020}
Morris, B.~M., Bobra, M.~G., Agol, E., Lee, Y.~J., \& Hawley, S.~L. 2020,
  Monthly Notices of the Royal Astronomical Society, 493, 5489,
  \dodoi{10.1093/mnras/staa618}

\bibitem[{Pancoast {et~al.}(2014)Pancoast, Brewer, Treu, Park, Barth, Bentz, \&
  Woo}]{Pancoast2014}
Pancoast, A., Brewer, B.~J., Treu, T., {et~al.} 2014, Monthly Notices of the
  Royal Astronomical Society, 445, 3073, \dodoi{10.1093/mnras/stu1419}

\bibitem[{Pereira {et~al.}(2019)Pereira, Campante, Cunha, Faria, Santos,
  Barros, Demangeon, Kuszlewicz, \& Corsaro}]{Pereira2019}
Pereira, F., Campante, T.~L., Cunha, M.~S., {et~al.} 2019, Monthly Notices of
  the Royal Astronomical Society, 489, 5764, \dodoi{10.1093/mnras/stz2405}

\bibitem[{Peters {et~al.}(2015)Peters, Richards, Myers, Strauss, Schmidt,
  Ivezi\'c, Ross, MacLeod, \& Riegel}]{Peters2015}
Peters, C.~M., Richards, G.~T., Myers, A.~D., {et~al.} 2015, The Astrophysical
  Journal, 811, 95, \dodoi{10.1088/0004-637x/811/2/95}

\bibitem[{Press {et~al.}(1992)Press, Rybicki, \& Hewitt}]{Press1992}
Press, W.~H., Rybicki, G.~B., \& Hewitt, J.~N. 1992, The Astrophysical Journal,
  385, 404, \dodoi{10.1086/170951}

\bibitem[{Rajpaul {et~al.}(2015)Rajpaul, Aigrain, Osborne, Reece, \&
  Roberts}]{Rajpaul2015}
Rajpaul, V., Aigrain, S., Osborne, M.~A., Reece, S., \& Roberts, S. 2015,
  MNRAS, 452, 2269, \dodoi{10.1093/mnras/stv1428}

\bibitem[{Rasmussen \& Williams(2006)}]{Rasmussen2006}
Rasmussen, C.~E., \& Williams, C. K.~I. 2006, {Gaussian processes for machine
  learning} (MIT Press), 248.
\newblock \url{http://www.gaussianprocess.org/gpml/}

\bibitem[{Rybicki \& Press(1992)}]{Rybicki1992}
Rybicki, G.~B., \& Press, W.~H. 1992, The Astrophysical Journal, 398, 169,
  \dodoi{10.1086/171845}

\bibitem[{Rybicki \& Press(1995)}]{Rybicki1995}
---. 1995, Physical Review Letters, 74, 1060,
  \dodoi{10.1103/physrevlett.74.1060}

\bibitem[{{Sarkar} {et~al.}(2018){Sarkar}, {Argyriou}, {Vandenbussche},
  {Papageorgiou}, \& {Pascale}}]{Sarkar2018}
{Sarkar}, S., {Argyriou}, I., {Vandenbussche}, B., {Papageorgiou}, A., \&
  {Pascale}, E. 2018, \mnras, 481, 2871, \dodoi{10.1093/mnras/sty2453}

\bibitem[{Sarkar {et~al.}(2019)Sarkar, Madhusudhan, \&
  Papageorgiou}]{Sarkar2019}
Sarkar, S., Madhusudhan, N., \& Papageorgiou, A. 2019, Monthly Notices of the
  Royal Astronomical Society, 491, 378, \dodoi{10.1093/mnras/stz2958}

\bibitem[{Sulis {et~al.}(2020)Sulis, Lendl, Hofmeister, Veronig, Fossati,
  Cubillos, \& Grootel}]{Sulis2020}
Sulis, S., Lendl, M., Hofmeister, S., {et~al.} 2020, Astronomy {\&}
  Astrophysics, 636, A70, \dodoi{10.1051/0004-6361/201937412}

\bibitem[{{Teachey} {et~al.}(2019){Teachey}, {Kipping}, {Burke}, {Angus}, \&
  {Howard}}]{Teachey2019}
{Teachey}, A., {Kipping}, D., {Burke}, C.~J., {Angus}, R., \& {Howard}, A.~W.
  2019, arXiv e-prints, arXiv:1904.11896.
\newblock \doarXiv{1904.11896}

\bibitem[{{Teachey} \& {Kipping}(2018)}]{Teachey2018}
{Teachey}, A., \& {Kipping}, D.~M. 2018, Science Advances, 4, eaav1784,
  \dodoi{10.1126/sciadv.aav1784}

\bibitem[{{Tracey} \& {Wolpert}(2018)}]{Tracey2018}
{Tracey}, B.~D., \& {Wolpert}, D.~H. 2018, arXiv e-prints, arXiv:1801.06147.
\newblock \doarXiv{1801.06147}

\bibitem[{Uttley {et~al.}(2005)Uttley, McHardy, \& Vaughan}]{Uttley2005}
Uttley, P., McHardy, I.~M., \& Vaughan, S. 2005, Monthly Notices of the Royal
  Astronomical Society, 359, 345, \dodoi{10.1111/j.1365-2966.2005.08886.x}

\bibitem[{Vallisneri(2008)}]{Vallisneri2008}
Vallisneri, M. 2008, Physical Review D, 77, \dodoi{10.1103/physrevd.77.042001}

\bibitem[{{Wilson} \& {Nickisch}(2015)}]{Wilson2015}
{Wilson}, A.~G., \& {Nickisch}, H. 2015, arXiv e-prints, arXiv:1503.01057.
\newblock \doarXiv{1503.01057}

\bibitem[{Zu {et~al.}(2011)Zu, Kochanek, \& Peterson}]{Zu2011}
Zu, Y., Kochanek, C.~S., \& Peterson, B.~M. 2011, The Astrophysical Journal,
  735, 80, \dodoi{10.1088/0004-637x/735/2/80}

\end{thebibliography}

\end{document}